\documentclass{article}
\usepackage{arxiv}

\usepackage[utf8]{inputenc} % allow utf-8 input
\usepackage[T1]{fontenc}    % use 8-bit T1 fonts
\usepackage{hyperref}       % hyperlinks
\usepackage{url}            % simple URL typesetting
\usepackage{booktabs}       % professional-quality tables
\usepackage{amsfonts}       % blackboard math symbols
\usepackage{nicefrac}       % compact symbols for 1/2, etc.
\usepackage{microtype}      % microtypography
\usepackage{lipsum}		% Can be removed after putting your text content
\usepackage{graphicx}
\usepackage[round]{natbib}
\usepackage{doi}
\usepackage[toc,page]{appendix}
\usepackage{latexsym}
\usepackage{amsmath}
\usepackage{amsfonts}
\usepackage{amssymb}
\usepackage{mathrsfs}  
\usepackage{makecell}
\usepackage{footmisc}
\usepackage{float}
\usepackage{array}
\usepackage{pdfpages}
\newcolumntype{C}[1]{>{\centering\arraybackslash}m{#1}}
\newcolumntype{L}[1]{m{#1}}
\usepackage{multicol}
\usepackage{multirow}
\usepackage{bm}
\usepackage{enumitem}
\usepackage[figuresright]{rotating}
\usepackage{siunitx}
\usepackage{verbatim} 
\usepackage{soul}
\usepackage{color}
\setul{0.5ex}{0.15ex}
\setulcolor{blue}
\setstcolor{red}
\usepackage{setspace}
\usepackage{appendix}

\usepackage{enumitem}
\setlist{nolistsep}

\title{Ensemble methods for survival function estimation with time-varying covariates}

\author{Weichi Yao \\
	Department of Technology, Operations, and Statistics \\
	Stern School of Business\\
	New York University \\
	New York, NY 10012, USA\\
	\texttt{wyao@stern.nyu.edu} \\
    \And
	Halina Frydman\\
	Department of Technology, Operations, and Statistics \\
	Stern School of Business\\
	New York University \\
	New York, NY 10012, USA\\
	\texttt{hfrydman@stern.nyu.edu } \\
	\And
	Denis Larocque \\
	Department of Decision Sciences \\
	HEC Montr\'{e}al\\
    Montr\'{e}al, Qu\'{e}bec, Canada H3T 2A7\\
	\texttt{denis.larocque@hec.ca} \\
	\And
	Jeffrey S. Simonoff\\
	Department of Technology, Operations, and Statistics \\
	Stern School of Business\\
	New York University \\
	New York, NY 10012, USA\\
	\texttt{jsimonof@stern.nyu.edu} \\
}

\date{}

\renewcommand{\headeright}{Research article}
\renewcommand{\undertitle}{Research article}
\renewcommand{\shorttitle}{YAO, FRYDMAN, LAROCQUE and SIMONOFF}

\begin{document}
\maketitle

\begin{abstract}
    Survival data with time-varying covariates are common in practice. If relevant, they can improve on the estimation of a survival function. However, the traditional survival forests - conditional inference forest, relative risk forest and random survival forest - have accommodated only time-invariant covariates. We generalize the conditional inference and relative risk forests to allow time-varying covariates. We also propose a general framework for estimation of a survival function in the presence of time-varying covariates. We compare their performance with that of the Cox model and transformation forest, adapted here to accommodate time-varying covariates, through a comprehensive simulation study in which the Kaplan-Meier estimate serves as a benchmark, and performance is compared using the integrated $L_2$ difference between the true and estimated survival functions. In general, the performance of the two proposed forests substantially improves over the Kaplan-Meier estimate. Taking into account all other factors, under the proportional hazard (PH) setting, the best method is always one of the two proposed forests, while under the non-PH setting, it is the adapted transformation forest. $K$-fold cross-validation is used as an effective tool to choose between the methods in practice.    \end{abstract}

\section{Introduction}
Methodology for survival data often assumes that covariate information is time-invariant; that is, only values measured at time 0 are used. In this situation, survival analysis models can provide an estimate of the survival function (the probability of surviving past time $t$) for a subgroup of the population (i.e., a subpopulation) with a specific set of values for the covariates. Time-varying covariates, however, are common in practice and play an important role in the analysis of censored time-to-event data. For example, in a study of the effect of heart transplant on survival for heart patients, the occurrence of a transplant can be modeled as a time-varying binary covariate \citep{exampletvdata1},
and in a study of the effect of $\mathrm{CD}4+$ T-cell counts on the occurrence of AIDS or death for HIV-infected patients, the cell count was used as a time-varying numerical covariate, measured longitudinally \citep{exampletvdata2}.

The Cox proportional hazards model \citep{Cox} has a long history of being used to model and analyze censored survival data. As a semi-parametric model, it assumes that the time-invariant covariates have a proportional effect on the hazard function. The Cox model was extended to fit time-varying covariates using a counting process formulation as follows \citep{timevarying1}. Consider continuous-time survival data with time-varying covariates, where each subject may have multiple records of measurements of risk factors at multiple time points. 
In practice, as the subjects are observed intermittently, the time-varying covariates are assumed constant between observation times. 
One can then reformat the data structure using the counting process approach by which a data record of a subject becomes a list of pseudo-subjects, that are treated as being independent, left-truncated and right-censored observations.

It is important to recognize that survival regression models, like all regression models, can be used for the two distinct purposes of estimation and prediction. These two purposes, while related (prediction almost always involves estimation of some kind as a first step), are distinctly different from each other. Prediction is meaningful at the level of an individual. In the context of survival data, this would correspond to an estimated survival function for a particular individual. For such an individual, if a change in the value of a covariate at time $t^{*}$ can potentially impact the future probabilities of survival of that individual, then the estimated survival function for that individual must be 1 for any time $t < t^{*}$ (since the covariate couldn't impact the future probabilities of survival unless the individual was alive at that time). A related point is the existence of so-called internal covariates, in which a variable (for example, blood pressure) can only be measured when an individual is alive, meaning that the act of measuring the covariate implies that the survival function value for that individual must equal 1. These facts, and their implications for prediction, are well-known, and are the reason for the prominence of joint modeling methods \citep{jointmodelbook} in the prediction of survival data with time-varying covariates.

The focus of this paper, however, is exclusively on estimation, rather than prediction. The goal here is not prediction of an individual's survival function, but rather estimation of a population-level survival function. This is an important problem in any situation that involves strategic decision making, such as public health policy or business operations and planning. Civic policymakers are interested in developing strategies that change over time, in response to changing conditions; this does not involve prediction at the individual level, but rather estimation of population probabilities. Such estimation can answer ``what if'' questions, in order to either be prepared for what might happen, or to try to control what will happen. Recognizing that various aspects of society can change, it can also examine what the effects of such changes might be. Examples of this could include modeling the incidence of COVID-19 infection in the population as vaccination and testing rates change, or estimating what proportion of deliveries will be made by Christmas as various characteristics of the supply chain change. These problems are fundamentally different from the problem of making a prediction for a particular individual, and issues related to joint modeling and the distinction between internal and external covariates are no longer relevant.  
 
When time-varying covariates are available, it is important to use the updated covariate values to dynamically adjust the estimated survival function. We use a simple hypothetical example to fix ideas and illustrate the mechanics of the survival function adjustment. Consider the following simple example, based on the COVID-19 problem noted earlier. The time-to-event is the time to a positive COVID test, and there is a time-varying covariate $X(t)$ that describes the vaccination status of a subject, assuming a two-shot vaccine (unvaccinated, vaccinated with one dose, vaccinated with two doses, vaccinated with two doses and a booster). Starting with a sample of unvaccinated and COVID-free subjects, the changes of $X(t)$ define distinct subgroups of the sample (and hence the population). Estimation of the population survival function for subjects who have gotten one dose of the vaccine would be based only on those subjects who survived to get that dose, but the overall population survival function estimate would have to be adjusted to account for the probability that unvaccinated subjects did not survive to that time. This same adjustment would be applied at the other times that vaccination status changes, resulting in a survival function estimate that is appropriate for all members of the population who followed the specified temporal vaccination pattern. In the next section we describe how such a survival function estimate can be constructed, and in Appendix A we illustrate the detailed calculations of the dynamically adjusted survival function for this simple example.

The Cox proportional hazards model with time-varying covariates, hereafter referred to as the extended Cox model, relies on restrictive assumptions such as proportional hazards and a log-linear relationship between the hazard function and covariates. Tree-based methods and their ensembles, which are useful non-parametric alternatives to the extended Cox model, also can incorporate time-varying covariates. 
Recently, two types of survival trees were proposed as extensions of the relative risk tree \citep{LeBlanc} and of the conditional inference tree \citep{ctree}, respectively, to left-truncated right-censored (LTRC) data, referred to as LTRC trees  \citep{LTRCtrees}. The proposed LTRC tree algorithms allow for time-varying covariate data after the data structure is reformatted using the counting process approach.
Another tree-based method that can handle LTRC survival data and therefore potentially be applied to time-varying covariate data is the novel ``transformation tree,'' and the corresponding ensemble is the ``transformation forest'' \citep{transformationforest}. These two algorithms are based on a parametric family of distributions characterized by their transformation function and developed to detect distributional alternatives to proportional hazards. 
None of the above methods have considered the estimation of the survival function. Similarly, recently developed methods for hazard function estimation in the presence of time-varying covariates haven't dealt with survival function estimation in general \citep{dynamicHazardROC,dynamicHazardRFSLAM}. 
There exist other survival trees and forest methods that can handle time-varying covariate data, but only for discrete-time survival data \citep{timevarying5discrete,timevarying6discrete,timevarying7discrete,timevarying8discrete, timevarying9discrete}. 

In this paper, we focus on forest algorithms for dynamic estimation of the survival function for continuous-time survival data. Ensemble methods like forest algorithms are known to preserve low bias while reducing variance and therefore can substantially improve prediction accuracy, compared to tree algorithms \citep{RF}. 
The most well-known ensemble methods for survival analysis are perhaps the relative risk forest \citep{CARTFOREST}, random survival forest \citep{RSF} and conditional inference forest \citep{SE}. 
These forest methods provide estimates of survival functions, but only for right-censored survival data with time-invariant covariates.
We propose to extend the relative risk and conditional inference forests, as well as the transformation forest, to allow time-varying (TV) covariates. We refer to them as RRF-TV, CIF-TV, and TSF-TV, respectively. 

The proposed methods by design can handle survival data with all combinations of left-truncation and right-censoring in the survival outcome, and with both time-invariant and time-varying covariates. 
In this paper we focus on survival data with time-varying covariates. Similar analysis for LTRC data with time-invariant covariates is provided in Section S2 in the Supplemental Material.

\section{Proposed forests for time-varying covariate data}
Assume $p$ covariates $\bm{X}= (X_1,X_2,\ldots, X_p)$ are available, some of which are TV covariates and the others are time-invariant (TI). For example, assume $X_1$ is the only time-invariant covariate among all $p$ covariates, then at time $t$, 
$\bm{X}(t) = (X_1, X_2(t), \ldots, X_p(t))$. For ease of exposition, we write $\bm{X}(t) = (X_1(t), X_2(t), \ldots, X_p(t))$ with $X_1(t) \equiv X_1$ for all $t$. 
Observations are obtained from $N$ subjects. Note that the subjects are observed only intermittently, for example, $J^{(i)}$ times for subject $i$, initially observed at $t_0^{(i)}$, and then at $t_j^{(i)}$, $j=1,2,\ldots,J^{(i)}-1$, for followup visits, with corresponding observed values $\bm{x}^{(i)}_j = (x_{j,1}^{(i)},\ldots,x^{(i)}_{j,p})$.  
Let $t_{J^{(i)}}^{(i)}$ denote $\tilde{T}^{(i)}= \min(T^{(i)}, C^{(i)})$, the minimum value of the true survival time $T^{(i)}$ and censoring time $C^{(i)}$. 
We assume non-informative censoring conditional on the
covariates.
Denote $\Delta = \mathbb{I}\{T^{(i)} \le C^{(i)}\}$, which indicates whether a subject experienced an event ($\Delta = 1$) or was right-censored ($\Delta = 0$). If $t_0^{(i)}\neq 0$, then we say the survival time is left-truncated. The outcome of interest is the time to the event.  

At any time $t$, let $\mathcal{I}(t)$ denote an arbitrary set of time values up to time $t$, that is, $\mathcal{I}(t)\subseteq [0,t]$.
It could be a finite number of time points, a finite number of intervals, or a disjoint set of time intervals and/or time points.  
Given the historical data for $N$ subjects observed up to the death or censoring time, the goal is to estimate the conditional survival function $S(t\mid \bm{X}(u) = \bm{x}(u), u\in\mathcal{I}(t))$ where $\{\bm{x}(u), u\in\mathcal{I}(t)\}$ is a particular stream of covariate values.
This true survival function is defined as the population proportion of subjects who have the specified covariate values at the specified times up to either time $t$ or their event time (whichever comes first) that are alive at time $t$. 
Note that the population includes all subjects for whom the specified conditions hold up until their time of event if that occurs before the evaluation time $t$, even if the conditions do not hold after the time of event. This is true if measurement of the covariate is no longer meaningful after the event occurs (as might be the case for a so-called internal covariate, such as blood pressure), or it is meaningful and available but no longer satisfies the conditions (a so-called external covariate, such as pollutant level). The reason is that the influence of the covariate on survival in either case is irrelevant for a subject for whom the event has occurred.

The proposed forest methods provide the survival function estimate by following three steps. First, we adopt the counting process approach to reformat the data structure. 
This approach assumes that the time-varying covariates are constant between the observed time points, that is,
\begin{align*}
    \bm{X}^{(i)}(t) = \bm{x}_j^{(i)},\quad t\in \big[t_j^{(i)}, t_{j+1}^{(i)}\big), \quad j=0,1,\ldots, J^{(i)}-1.
\end{align*}
It then splits the $i$-th subject observation into $J^{(i)}$ pseudo-subject observations: $(t_j^{(i)}, t_{j+1}^{(i)}, \delta_j^{(i)}, \bm{x}_j^{(i)})$ with LTRC times $t_j^{(i)}$, $t_{j+1}^{(i)}$, and event indicator $\delta_j^{(i)} = \Delta\mathbb{I}\{j= J^{(i)}-1\}$, $j=0,1,\ldots,J^{(i)}-1$. 
The multiple records from $N$ subjects now become a list of pseudo-subjects,
\begin{align*}
    \big\{\big\{\big(t_j^{(i)}, t_{j+1}^{(i)}, \delta_j^{(i)}, \bm{x}_j^{(i)}\big)\big\}_{j=0}^{J^{(i)}-1}\big\}_{i=1}^N.
\end{align*}
The set of pseudo-subjects is treated as if they were independent in the following form
\begin{align}
    \big\{\big(L_{l}^{\prime}, R_{l}^{\prime}, \delta_l^{\prime}, \bm{x}_l^{\prime}\big)\big\}_{l=1}^n,\quad n = \sum_{i=1}^NJ^{(i)},
    \label{eq:dataset_pseudosubject_indep}
\end{align}
where $\bm{x}_l^\prime = (x_{l,1}^\prime,\ldots,x_{l,p}^\prime)$ is the vector of the observed values of $p$ covariates from the $l$-th pseudo-subjects in the reformatted dataset. 
The second step is to apply the forest algorithms on the reformatted dataset given in (\ref{eq:dataset_pseudosubject_indep}), to fit a model. 
Finally, in the third step, given a particular stream of covariate values $\bm{x}_j^\ast$ at the corresponding time values  $t_j^\ast$, $j=0,1,\ldots$, a survival function estimate is  constructed based on the outputs of the proposed forest algorithms. More specifically, at any time $t$, with $\mathcal{X}^{\ast}(t)$ denoting the covariate information up to time $t$,
\begin{align}
    \mathcal{X}^{\ast}(t) = \big\{\bm{x}_j^{\ast}, \forall j:0\le t_j^{\ast}\le t\big\},
    \label{eq:covhist}
\end{align}
we compute the estimated survival probability $\widehat{S}(t\mid \mathcal{X}^\ast(t))$.

\subsection{Extending right-censored TI survival forests to the proposed TV forests}
The conditional inference forest and the relative risk forest are both tree-based ensemble methods, where $B$ individual trees are grown from $B$ bootstrap samples drawn from the original data. 
Randomness is induced into each node of each individual tree when selecting a variable to split on. Only a random subset $I$ of the total $p$ covariates is considered for splitting at each node. 
The node is then split using the candidate covariates based on different criteria for different forest methods. To extend the two forest methods for right-censored survival data with time-invariant covariates to the forests for (left-truncated) right-censored survival data with time-varying covariates, the splitting criteria are modified. 

\subsubsection{Recursive partitioning in the proposed CIF-TV forest}
Consider right-censored survival time data of the form  $\big(\tilde{T},\Delta,\bm{X}\big)$, with survival/censored time $\tilde{T}$, event indicator $\Delta$ ($\tilde{T}$ denotes the survival time if $\Delta = 1$, or censored time if $\Delta = 0$), and $p$ time-invariant covariates $\bm{X} = \big(X_1,\ldots, X_p\big)$. 
In each node, the recursive partitioning in the conditional inference forest algorithm is based on a test of the global null hypothesis of independence between the response variable in the right censored case $\bm{V} = \big(\tilde{T},\Delta\big)$ and any of the covariates in the random subset $I$. It is formulated in terms of $\vert I\vert$ partial hypotheses, $H_0 = \cap_{k=1}^{\vert I\vert}H_0^k$ with
\begin{align}
    H_0^k : D\big(\bm{V}\mid X_k\big) = D(\bm{V}), \quad k=1,\ldots,\vert I\vert,
    \label{eq:CIpartialH0}
\end{align}
where $D(\bm{V}\mid X_k)$ denotes the conditional distribution of $\bm{V}$ given the covariate $X_k$. 
The independence is measured by linear statistics incorporating the log-rank scores that take censoring into account. 
In the extension of conditional inference tree to LTRC conditional inference tree, the log-rank score can be modified as follows for LTRC data \citep{LTRCtrees}. 

Given the list of pseudo-subject observations with LTRC survival times as in (\ref{eq:dataset_pseudosubject_indep}), the response variable now becomes $\bm{V} = \big(L_l^\prime, R_{l}^\prime, \delta_l^\prime\big)$ in the test of partial null hypothesis of independence (\ref{eq:CIpartialH0}) for the $l$-th observation $\big(L_l^{\prime}, R_{l}^{\prime}, \delta_l^{\prime},\bm{x}_l^\prime\big)$.
The corresponding log-rank score is defined as  
\begin{equation}
    U_l= \begin{cases}
    1+\log\widehat{S}(R_{l}^\prime)-\log\widehat{S}(L_l^\prime),\quad
	&\text{if }\delta_l^\prime=1,\\
	\log\widehat{S}(R_{l}^\prime)-\log\widehat{S}(L_l^\prime),\quad
	& \text{otherwise}.
	\end{cases}\label{eq:logrank_score_ltrc}
\end{equation}
	
Note that $\widehat{S}$ is the nonparametric maximum likelihood estimator (NPMLE) of the survival function which takes
into account left-truncation. 
Such an estimator can be constructed using the product-limit estimator, i.e. Kaplan-Meier estimator with pseudo-subjects that fall into the current node \citep{nonparametricLTRC1996Shulamith, ProductLimitEstimator1987Tsai}. 
We similarly use the log-rank score $U_l$ in the proposed extension of conditional inference forest to LTRC conditional inference forest.

\subsubsection{Recursive partitioning in the proposed RRF-TV forest}
The relative risk forest combines the use of relative risk trees \citep{LeBlanc} with random forest methodology \citep{RF} as a way to reliably estimate relative risk values. 
The Classification and Regression Tree (CART) paradigm \citep{CART} is used to produce a relative risk forest by exploiting an equivalence with Poisson tree likelihoods.

The splitting criterion under the relative risk framework is to maximize the reduction in the one-step deviance between the log-likelihood of the saturated model and the maximized log-likelihood. 
At node $h$, let $\mathcal{R}_h$ denote the set of labels of those observations that fall into the region corresponding to node $h$, and let $\lambda_h(t)$ and $\Lambda_h$ denote the corresponding hazard and cumulative hazard function, respectively. 
Under the assumption of proportional hazards, 
\begin{align*}
	\lambda(t) = \lambda_0(t)\varphi_h,
\end{align*}
where $\lambda_0$ is the baseline hazard and $\varphi_h$ is the nonnegative relative risk of the node $h$. Given the right-censored observations $\big(\tilde{t}_l, \delta_l\big)$, $l\in \mathcal{R}_h$, the maximum likelihood estimate of $\varphi_h$ is 
\begin{align*}
    \widehat{\varphi}_h=\dfrac{\sum_{l\in \mathcal{R}_h}\delta_l}{\sum_{l\in \mathcal{R}_h}\Lambda_0(\tilde{t}_l)},
\end{align*}
where the Nelson-Aalen estimator using all of the data at the root node $\widehat{\Lambda}_0$ is used for $\Lambda_0$ \citep{LeBlanc}. The full likelihood deviance residual for node $h$ is defined as 
\begin{align}
	d_h=\sum_{l\in \mathcal{R}_h}2\Big[\delta_l\log\big(\dfrac{\delta_l}{\widehat{\Lambda}_0(\tilde{t}_l)
	\widehat{\varphi}_h}\big)-\big(\delta_l-\widehat{\Lambda}_0(\tilde{t}_l)\widehat{\varphi}_h\big)\Big].
	\label{eq:dev_leblanc}
\end{align} 
For a Poisson regression model, let $\varrho_h$ denote the event rate, $s_l$ and $c_l$ be the exposure time and the event count for observation $l$, respectively, then (\ref{eq:dev_leblanc}) is equivalent in form to the deviance residual based on the Poisson regression model,
\begin{align}
    d_h^{\mathrm{Pois}}=\sum_{l\in \mathcal{R}_h}2\Big[c_l\log\big(\dfrac{c_l}{s_l
	\widehat{\varrho}_h}\big)-\big(c_l-s_l\widehat{\varrho}_h\big)\Big]
	\label{eq:dev_poisson}
\end{align}
with $\widehat{\varrho}_h = \frac{\sum_{l\in \mathcal{R}_h}c_l}{\sum_{l\in \mathcal{R}_h}s_l}$,
by replacing $\widehat{\varrho}_h$ with $\widehat{\varphi}_h$,  $s_l$ with $\widehat{\Lambda}_0(\tilde{t}_l)$, and $c_l$ with $\delta_l$ \citep{LeBlanc}. 

To adapt the Poisson regression tree approach for left-truncated right-censored survival observations $\{(L_l^\prime,R_{l}^\prime,\delta_l^\prime)\}$, the key is to modify the estimated $\widehat{\Lambda}_0(\tilde{t}_l)$ and $\delta_l$ to replace $s_l$, $c_l$ and $\widehat{\varrho}_h$ in (\ref{eq:dev_poisson}).
First, compute the estimated cumulative hazard function $\widehat{\Lambda}_0(\cdot)$ based on all (pseudo-subject) observations. 
The exposure time $s_l$ and the event count $c_l$ for observation $l$ in (\ref{eq:dev_poisson}) are then replaced by $\widehat{\Lambda}_0(R_{l}^\prime)-\widehat{\Lambda}_0(L_l^\prime)$ and $\delta_l^\prime$ to obtain the deviance residual appropriate for LTRC data \citep{LTRCtrees}.

\subsubsection{Implementation of the proposed forests}
To implement the CIF-TV and RRF-TV algorithms, we make use of the fast algorithms provided in the packages \textsf{partykit} \citep{partykit} and \textsf{randomForestSRC} \citep{rfsrcpackage}, respectively. 
The RRF-TV building architecture is based on employing the fast \textsf{C} code from \textsf{randomForestSRC}. 
The Poisson splitting rule \citep{LeBlanc} is coded in \textsf{C} and is incorporated by exploiting the custom splitting rule feature in the \textit{rfsrc} function. 
The CIF-TV is built by extending the survival forest algorithms coded in the \textit{cforest} function from \textsf{partykit} with the log-rank score adapted for LTRC data.

\subsection{Bootstrapping subjects vs. bootstrapping pseudo-subjects}
In forest-like algorithms, bootstrapped samples are typically used to construct each individual tree to increase independence between these base learners. 
The nonparametric bootstrap approach is used in all three types of forests being considered here (CIF-TV, RRF-TV, and TSF-TV). 
It places positive integer weights that sum to the sample size on approximately 63\% of the observations in any given bootstrap sample, and the 37\% of the data excluded during this procedure is called out-of-bag data (OOB data). As we split each subject into several pseudo-subjects and treat these pseudo-subjects as independent observations on which to build the forests, we have two bootstrapping options: we can bootstrap subjects or bootstrap pseudo-subjects.

Bootstrapping pseudo-subjects is used for some discrete survival forest methods \citep{timevarying5discrete, timevarying6discrete}. Since all pseudo-subjects are treated as independent observations in the recursive partitioning process \citep{timevarying2,LTRCtrees}, bootstrapping pseudo-subjects is just bootstrapping ``independent'' observations as the first step of any forest algorithm. 
On the other hand, bootstrapping subjects is a natural approach, as it keeps all of the pseudo-subjects for each subject in the bootstrap sample. 
In fact, simulations have shown that the two different bootstrapping mechanisms do not result in fundamentally different levels of performance; see Section S1.5 in the Supplemental Material for more details. 
This paper will focus on forests based on bootstrapping subjects.

\subsection{Regulating the construction of individual trees in the proposed forests}
In a forest algorithm, only a random subset of covariates is considered for splitting at each node. The size of this random set is denoted by \textit{mtry}.
In addition to \textit{mtry}, many other parameters play an important role in establishing a split in the individual tree. 
In both the \textit{cforest}\citep{partykit} function for the conditional inference forest and the \textit{traforest}\citep{trtf} function for the transformation forest algorithms, \textit{minsplit} (the minimum sum of weights in a node in order to be considered for splitting), \textit{minprob} (the minimum proportion of observations needed to establish a terminal node) and \textit{minbucket} (the minimum sum of weights in a terminal node) control whether or not to implement a split; in the \textit{rfsrc}\citep{randomForest} function for the relative risk forest algorithms, \textit{nodesize} controls the average terminal node size. These tuning parameters thereby regulate the size of the individual trees. 
The recommended values for these parameters are usually given as defaults to the algorithm. 
For example, \textit{mtry} is usually set to be $\sqrt{p}$, 
where $p$ is the total number of covariates \citep{SE,RSF}, 
\textit{nodesize} to be $15$ in the \textit{rfsrc} function, 
(\textit{minsplit}, \textit{minbucket}) to be $(20, 7)$ in the 
\textit{cforest} function and the \textit{traforest} function, 
which we refer to as the \textit{default parameter settings}.
 
The best values for these parameters would be expected to depend on the problem and they should be treated as tuning parameters \citep{book}. 
It has been shown for conditional inference forests for interval-censored data \citep{ICcforest} that these parameters have a non-negligible effect on the overall performance of the forest algorithm.
As we extend the forest framework to allow for left truncation, and from time-invariant covariate data to time-varying covariate data, we should also consider rules for choosing tuning parameters. 

The algorithm designed in survival forests for interval-censored data with time-invariant covariates \citep{ICcforest} tunes the value of \textit{mtry} on the ``out-of-bag observations''. 
To adapt the same idea to survival forests based on bootstrapping subjects on a dataset with time-varying covariates, we define the ``out-of-bag observations'' for the $b$-th tree to be the observations from those subjects that are left out of the $b$-th bootstrap sample and not used in the construction of the $b$-th tree. 
The survival curve can be estimated by using each of the $B$ trees in which that subject was ``out-of-bag,'' denoted as $\widehat{S}^{\mathrm{OOB}}$. 
To evaluate the fit of the out-of-bag estimate $\widehat{S}^{\mathrm{OOB}}$ with a specific value of \textit{mtry}, we compute the estimation error defined as the integrated Brier score designed for time-invariant covariate data \citep{brier}, adapted here for time-varying covariate data as follows.

For a given dataset $\mathcal{D}$, define the integrated Brier score $\widehat{\mathrm{IBS}}\big(\widehat{S};\mathcal{D}\big)$ for the estimated survival function $\widehat{S}$ as 
\small
\begin{equation}
	\widehat{\mathrm{IBS}}\big(\widehat{S};\mathcal{D}\big)=\frac{1}{\vert \mathcal{D}\vert}\sum_{i\in \mathcal{D}}\frac{1}{\tau^{(i)}}\int_0^{\tau^{(i)}}\widehat{W}^{(i)}(t)\Big[\tilde{Y}^{(i)}(t)-\widehat{S}\big(t\mid\mathcal{X}^{(i)}(t)\big)\Big]^2dt 
	\label{eq:IBSrewrite} 
\end{equation}
\normalsize
where $\tau^{(i)}$ determines the length of difference evaluation time span for subject $i$, $\tilde{Y}^{(i)}(t) = \mathbb{I}\{\tilde{T}^{(i)}>t\}$ is the observed status ($\tilde{T}^{(i)}$ is the survival/censored time), $\widehat{W}^{(i)}(t)$ is the inverse probability of censoring weights,
\small
\begin{equation*}
	\widehat{W}^{(i)}(t) = \frac{\big(1-\tilde{Y}^{(i)}(t)\big)\Delta^{(i)}}{\widehat{G}\big(\tilde{T}^{(i)}\big)}+
	\frac{\tilde{Y}^{(i)}(t)}{\widehat{G}(t)} 
\end{equation*}
\normalsize
with $\widehat{G}$ the Kaplan-Meier estimate of the censoring distribution based on $\{(\tilde{T}^{(i)},1-\Delta^{(i)})\}_{i\in\mathcal{D}}$ \citep{brier}.
The corresponding Brier score $\widehat{\mathrm{BS}}(t,\widehat{S};\mathcal{D}) $ at time $t$ is defined as
\small
\begin{align}
	\widehat{\mathrm{BS}}(t,\widehat{S};\mathcal{D}) = \frac{1}{\vert \mathcal{D}\vert} \sum_{i\in \mathcal{D}}\widehat{W}^{(i)}(t)\Big[\tilde{Y}^{(i)}(t)-\widehat{S}\big(t\mid\mathcal{X}^{(i)}(t) \big)\Big]^2.
 	\label{eq:BS}
\end{align}
\normalsize
The resulting estimation error for the ensemble method with a specific value of \textit{mtry} 
can then be computed by setting $\widehat{S}=\widehat{S}^{\mathrm{OOB}}$ in (\ref{eq:IBSrewrite}). 
An appropriate value of \textit{mtry} is the one that minimizes the ``out-of-bag'' estimation error. 

Regarding the values of other tuning parameters, the optimal values that determine the split vary from case to case. 
As fixed numbers, the default values may not affect the splitting at all 
when the sample size is large, while having a noticeable effect in smaller 
data sets. This inconsistency can potentially result in good performance in 
some data sets and poor performance in others. 
In the simulations, we set \textit{minsplit}, \textit{minbucket} and \textit{nodesize} 
to be the maximum of the default value and the square root of the number of pseudo-subject observations $n$. 
This set of values can automatically adjust to the change in size of the data set. 
We refer to the above choice of \textit{mtry}, \textit{minsplit}, \textit{minbucket} and \textit{nodesize} 
as the \textit{proposed parameter settings}, as opposed to the default settings.

\subsection{Constructing a survival function estimate for time-varying covariate data}
Consider a particular stream of covariate values $\bm{x}_j^\ast$ at time $t_j^\ast$, for $j=0,1,\ldots,J-1$. Denote $\mathcal{X}^\ast(u)=\{\bm{x}_j^\ast:\forall j,\; t_j^\ast\le u \}$ the set of the covariate values up to time $u$. At time $t\in \big[t_j^\ast, t_{j+1}^\ast\big)$, 
we derive a recursive computation as follows
\small
\begin{equation}
    \widehat{S}(t\mid \mathcal{X}^\ast(t)\big)=
    \begin{cases} 
     1, &  t = t_0^\ast;\\
     \frac{\widehat{S}_{A,j}(t)}{\widehat{S}_{A,j}(t_j^\ast)}\widehat{S}\big(t_j^\ast\mid \mathcal{X}^\ast(t)\big),&  t\in[t_j^\ast,t_{j+1}^\ast),
    \label{eq:SurvEstimate_recursive}
    \end{cases}
\end{equation}
\normalsize
where $\widehat{S}_{A,j}(t)\triangleq\widehat{\mathbb{P}}(T>t\; | \;\bm{x}_j^\ast)$ denotes the output of the  algorithm for the input with time-invariant covariate value $\bm{x}_j^\ast$. See Appendix B for details of derivation.

Further, by expansion, (\ref{eq:SurvEstimate_recursive}) is equivalent to
\small
\begin{align}
    \widehat{S}\big(t\mid \mathcal{X}^\ast(t)\big) 
    = \frac{\widehat{S}_{A,j}(t)}{\widehat{S}_{A,j}(t_j^\ast)}\prod_{l=0}^{j-1}\frac{\widehat{S}_{A,l}(t_{l+1}^\ast)}{\widehat{S}_{A,l}(t_l^\ast)},
    \label{eq:SurvEstimate_recursive_expanded}
\end{align}
\normalsize
for $t\in \big[t_j^\ast,t_{j+1}^\ast\big)$. The formulation in (\ref{eq:SurvEstimate_recursive_expanded}) provides another perspective to view the resulting survival curve estimate -- it is constructed by combining the pseudo-subject-specific ensemble estimates of the survival function with multiplicative correction factors. These correction factors ensure monotonicity of the overall curve. 

Note that the construction in (\ref{eq:SurvEstimate_recursive_expanded}) coincides with what the function \textit{survfit} in the \texttt{R} package \textsf{survival} \citep{survival} uses to give a subject's survival function estimate from a \textit{coxph} fit using the same counting process approach.

\begin{figure}[t]
    \centering
    \includegraphics[scale = 0.5]{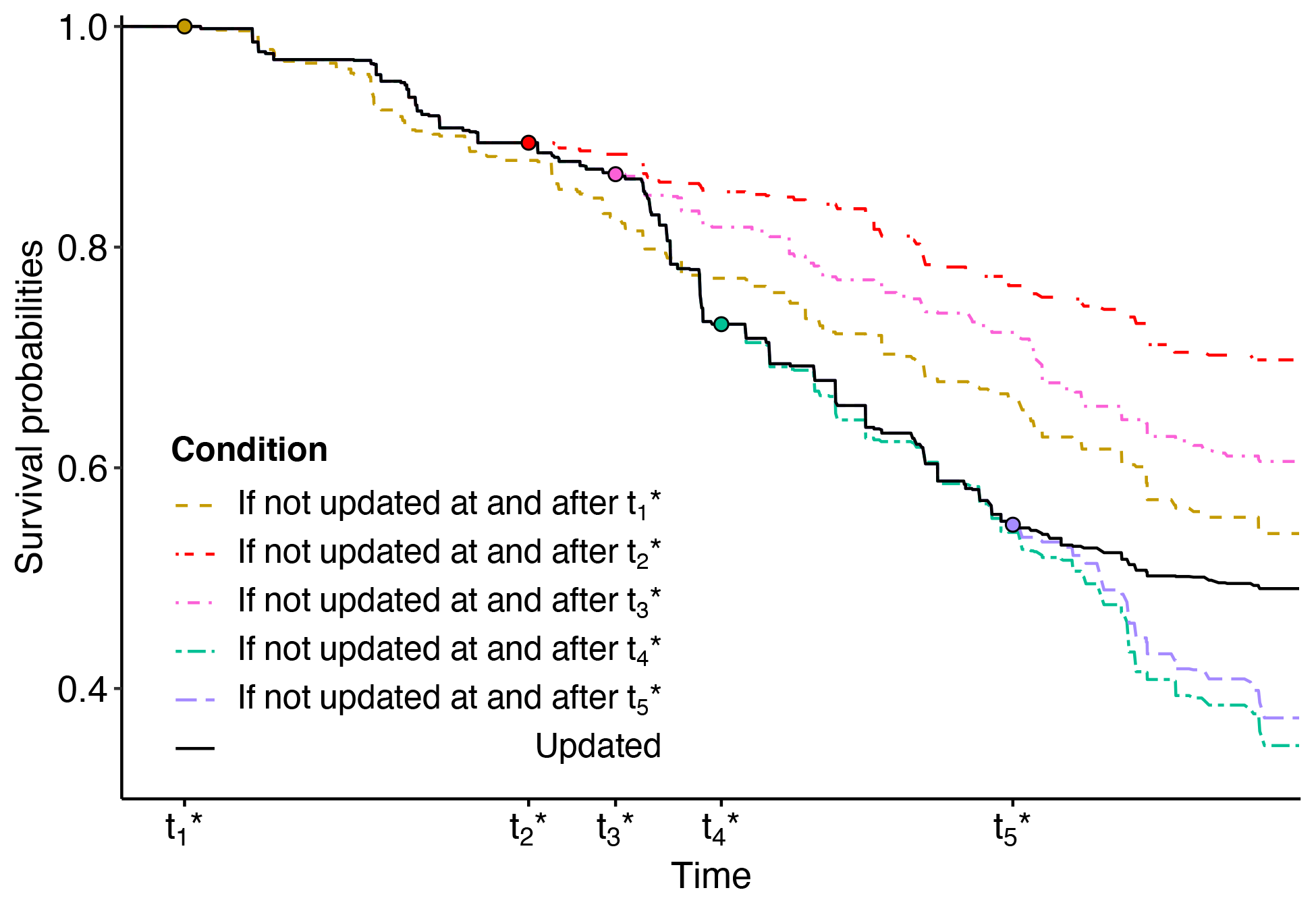}
    \caption{Illustration of estimated survival functions with or without changing the covariate values at $t_j^\ast$, $j=1,2,\ldots,5$.
    At $t_j^\ast$, the dot on the curve shows the estimated survival function at the time of change (having been updated at all of the previous time points $t_{1}^\ast,\ldots, t_{j-1}^\ast$). If not updated with the latest change, the estimated survival function at $t>t_j^\ast$ is shown as the dashed line with the same color as the dot.
    The solid black line shows the one estimated by CIF-TV and constructed as given in (\ref{eq:SurvEstimate_recursive}). It tracks all of the changes in covariate values and updates the estimated survival probabilities at each time step of change. }
    \label{fig:survCurv}
\end{figure}

Figure \ref{fig:survCurv} gives an illustration of the estimated survival functions with or without ``updating'' the covariate values at time $t_1^\ast,t_2^\ast,\ldots, t_5^\ast$. 
For each $j$, the ``update'' in the estimated survival probability starting from time $t_j^\ast$ for all the future time $t>t_j^\ast$ reflects the difference in the estimated surviving proportions of two subpopulations of subjects with their covariate trajectories diverging from the shared past before $t_j^\ast$. 
One can see that the change in covariate information at each time point of change can make a huge impact on the future path of the estimated survival function.

\section{Simulation study}
\subsection{Data generation scheme}
In the simulation study, observations from $N$ subjects are generated independently with $p$ covariates $\bm{X} = \big(X_1,\ldots,X_{p}\big)$. We set $p=20$. 
Eight of these covariates are time-invariant: $X_1,X_{11}\sim\mathrm{Bern}(0.5)$, $X_2, X_7, X_{10}\sim\mathrm{Unif}(0,1)$, $X_8\sim\mathrm{Unif}(1,2)$, $X_9$ follows a categorical distribution with possible outcomes $\{1,2,3,4,5\}$ with equal probability, $X_{12}$ follows a categorical distribution with possible outcomes $\{0,1,2\}$ with equal probability. 
The others are time-varying, whose values are obtained at $m$ randomly generated time points, different for each subject.   
In the simulations, we set $m=11$. At each of these preset time points, for some time-varying covariates, the value is randomly resimulated from its distribution: $X_3,X_{19}\sim\mathrm{Bern}(0.5)$, $X_4,X_{15},X_{17}\sim\mathrm{Unif}(0,1)$,
$X_5$ and $X_{14}$ both follow a categorical distribution with possible outcomes $\{1,2,3,4,5\}$ with equal probability; for other time-varying covariates, the value is resimulated following particular patterns: $X_6$, whose initial value is randomly generated from $\{0,1,2\}$ with equal probability, which will choose to stay at the original value or move one level up but the largest value can only be $2$; the changing pattern of $X_{13}$ is always $0\rightarrow 1$; the changing pattern of $X_{16}$ is either $0\rightarrow1$ or $1\rightarrow2$;
the changing pattern of $X_{18}$ is $0\rightarrow1\rightarrow2$; value of $X_{20}$ is a linear function of the left-truncated time point of the interval with slope and intercept follows $\mathrm{Unif}(0,1)$. 
Further details of the changing pattern of $X_6$, $X_{13}$, $X_{16}$, $X_{18}$ and $X_{20}$ can be found in Section S1.2 in the Supplemental Material.

After the time-varying covariates' values are generated at each of those $m$ preset time points, 
the true survival time $T$ is then computed under different model setups and the right-censoring time $C$ is generated independently. 
\subsection{Model setup}
We consider the following factors for different variations of data generating models:
\begin{enumerate}[label = (\alph*)]
    \item Different proportions of time-varying covariates in the true model (Scenario).
	\item Different signal-to-noise ratios (SNR) labelled as ``High'' and ``Low,'' constructed by choosing different coefficients in the true model. 
    \item Different hazard function settings: a proportional hazards (PH) and a non-proportional hazards (non-PH) setting. 
	\item Different survival relationships between the hazards and covariates: a linear, a nonlinear or an interaction model. 
	\item Different censoring rates: 20\% and 50\%.
	\item Different sample sizes: $N=100,300$ and $500$.
	\item Different amount of knowledge of history of changes in covariates' values: Case I -- When all changes in values of covariates are known, labelled as ``Full,'' and Case II -- When only half of the changes in values of the covariates are known, labelled as ``Half''.
\end{enumerate}

\paragraph{Scenario} 
We consider two different proportions of time-varying covariates in the true model: 2TI $+$ 1TV, and 2TI $+$ 4TV; see Table \ref{tbl:scenario}. Only the first six covariates are given in the table since $X_7$ to $X_{20}$ are never involved in the true DGP. 
\begin{table}[!t]
    \centering
    \caption{Scenario: Numbers of time-invariant and time-varying covariates in the true model.}
    % \begin{tabular}{|L{0.5cm}|C{1cm}C{1cm}|C{1cm}C{1cm}C{1cm}C{1cm}|}
    \begin{tabular}{|l|cc|cccc|}
        \hline
        \multirow{2}{*}{Scenario}&  \multicolumn{2}{c|}{Time-invariant} & \multicolumn{4}{c|}{Time-varying}\\
        \cline{2-7}
        & $X_1$ & $X_2$ & $X_3$ & $X_4$ & $X_5$ & $X_6$\\ 
        \hline
        2TI $+$ 1TV  &$\checkmark$&$\checkmark$&&&$\checkmark$&\\
        \hline
        2TI $+$ 4TV  &$\checkmark$&$\checkmark$&$\checkmark$&$\checkmark$&$\checkmark$&$\checkmark$\\
        \hline
    \end{tabular}
    \label{tbl:scenario}
\end{table}

\paragraph{Survival relationships} Given a survival relationship, the survival time $T$ depends on $\vartheta(t)=\vartheta(\bm{X}(t))$. 
Here we use scenario 2TI $+$ 4TV for illustration. 

\begin{figure}[t]
	\centering
	\includegraphics[scale=0.35]{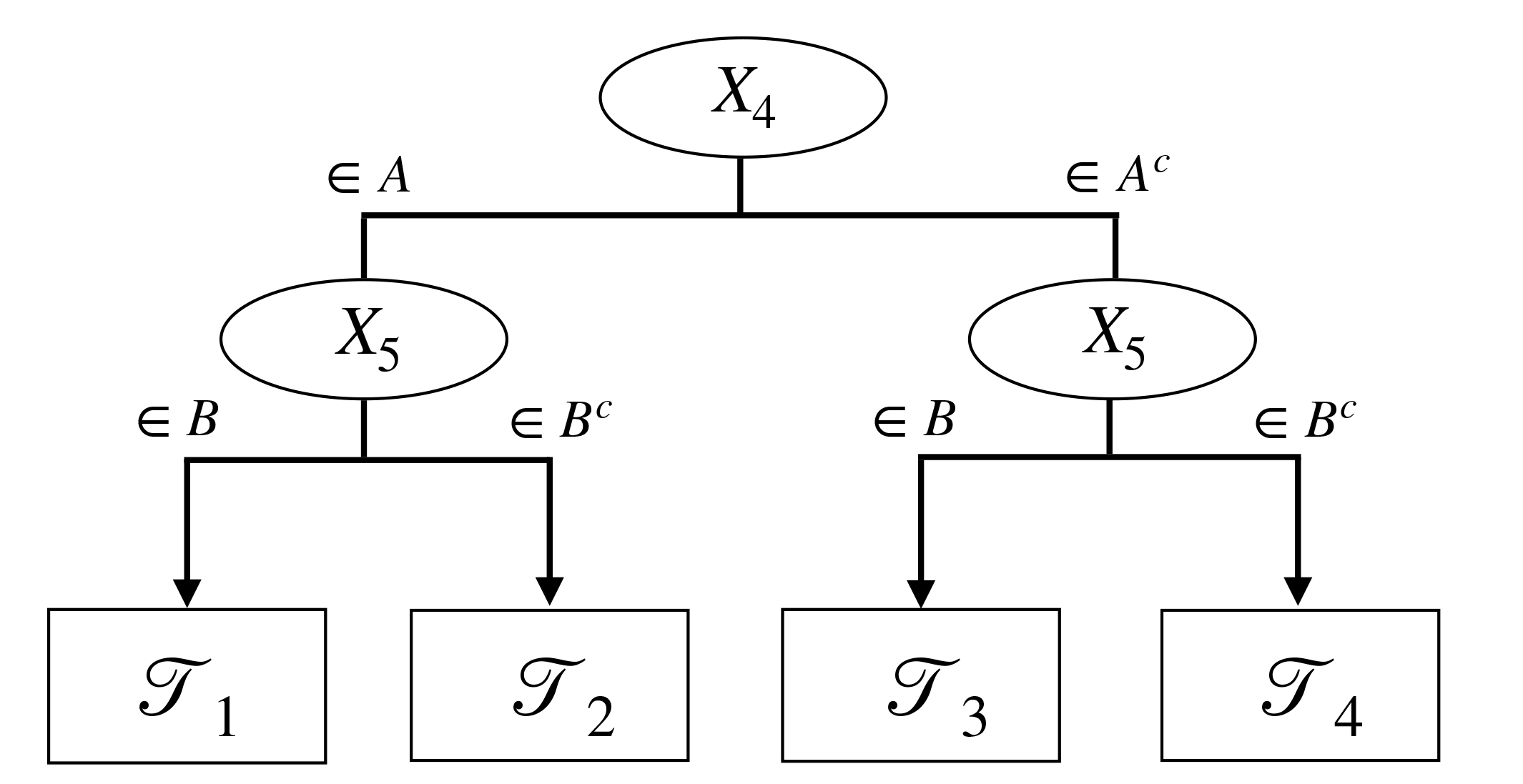}
	\caption{An example of the structure of the covariates driving the Interaction survival relationship, where set $A$ and $B$ are some sets of values of $X_4$ and $X_5$, respectively. }\label{fig:interaction}
\end{figure}

For a linear survival relationship, $\vartheta(t) = \beta_0+\sum_{k=1}^6\beta_kX_{k}(t)$ with constants $\beta_k$, $k=0,\ldots,6$. For a nonlinear survival relationship,
$\vartheta(t) = \phi_1\cos(\sum_{k=1}^{6}X_{k}(t))+\phi_2\log(\psi_0+\sum_{k=1}^6\psi_kX_{k}(t))+\phi_3X_{1}(t)(2X_{2}(t))^{4X_{4}(t)}$
with some constants $\phi_1, \phi_2, \phi_3$, and $\psi_k$, $k=0,\ldots,6$. For an interaction model,
$\vartheta$ is determined by the value of time-varying covariate $X_4$ and the value of time-varying covariate $X_5$. 
Figure \ref{fig:interaction} gives an example of the structure of the covariates driving the interaction survival relationship,
where $\mathscr{T}_1,\mathscr{T}_2,\mathscr{T}_3,\mathscr{T}_4$ correspond to
\begin{enumerate}[label=(\roman*)]
	\item $\vartheta(t) = \eta_1\big[X_1(t)X_2(t)-\log(X_3(t)+X_4(t))-X_6(t)/X_5(t)\big]+\eta_2$
	\item $\vartheta(t) = \gamma_0+\sum_{k=1}^6\gamma_kX_k(t)$
	\item $\vartheta(t) = \eta_3\big[\cos(\pi(X_1(t)+X_5(t)))+\sqrt{X_2(t)+X_6(t)}-X_3(t)\big]+\eta_4$
	\item $\vartheta(t) = \alpha_0+\sum_{k=1}^6\alpha_kX_k(t)$
\end{enumerate}
with some constants $\{\alpha_k\}_{k=0}^6$, $\{\gamma_k\}_{k=0}^6$ and $\{\eta_k\}_{k=1}^4$.

\paragraph{Survival distributions under the PH and the non-PH setting} Given a survival relationship model, the survival time $T$ depends on $\vartheta$ via a Weibull distribution. 

For proportional hazards models, a closed-form solution can be derived to generate survival times with time-varying covariates for the Weibull distribution \citep{survivaltimes}. 
For non-proportional hazards models, a closed-form solution exists for the Weibull distribution, with its nonconstant shape term a function of the covariates (note that the proportional hazards relationship is on the scale parameter for the Weibull distribution). 

To be more specific, for the proportional hazards setting, we consider the underlying hazard function 
\begin{align}
    h(t)=h_0(t)\exp(\vartheta(t)),
\end{align}
where the baseline hazard function is given by $h_0(t) = \lambda\nu t^{\nu-1}$ with $\lambda>0$ and $\nu > 0$.
For the non-proportional hazards setting, the hazard function is set to be
\begin{align}
    h(t) = \lambda\exp(\vartheta(t))(\lambda t)^{\exp(\vartheta(t))-1},
\end{align}
where $\lambda >0$. Values of $\vartheta(t)$ have been scaled to be between $-3$ and $3$.
Note that, compared with the Weibull distribution under the PH setting, now the time-varying effects appear in the shape term instead of the scale term. 
The survival function is then given by $S(t)=\exp (-\int_{0}^{t}h_0(s)\exp (\vartheta(s) )ds )$.  
Further details of simulating the survival time $T$ can be found in Section S1.1 in the Supplemental Material.

Histograms of survival times for typical samples with the number of subjects $N=500$ in each scenario are provided in Section S1.3 in the Supplemental Material to illustrate the data generating processes. The parameters set in the simulation study can be found in Section S1.4 in the Supplemental Material.

\paragraph{Knowledge of history of changes in covariates' values} In practice, it is likely that not all of the changes in the covariates' values are known to the data analyst. 
For example, suppose that a patient's blood pressure is to be measured at regularly scheduled examination times. 
If a patient obeys the schedule then, from the doctor's point of view, all changes in blood pressure are known. 
However, if a patient skips some scheduled examination times, then not all changes in the blood pressure are known. 
In the latter case, this means that whatever modeling method is used to estimate the survival curves, it is operating with incorrect values as inputs and therefore its performance would be expected to deteriorate. 
Of course, the fact that blood pressure is actually changing continuously is an extreme example of this phenomenon; in these simulations we limit ourselves to changes at a finite number of time points. 
The simulations are designed to investigate the performance of different modeling methods in this situation under the following two circumstances:
\begin{itemize}
	\item[-] Case I. When all changes in covariate values are known;
	\item[-] Case II. When only half of the changes in covariate values are known. 
\end{itemize}
The missing changes are selected completely at random. To generate a dataset under Case II, one can start with the dataset generated under Case I. The following example is given to illustrate how to construct such datasets. Suppose the baseline covariates' values of the subject is $\bm{X}(t_0)=\bm{x}_0$ and the covariates values $\bm{X}$ change $J-1$ times at time $t_1,\ldots,t_{J-1}$, before the subject is censored or the event occurs at $t_J=\tilde{T}$. For $J=3$, $\bm{X}(t_1)=\bm{x}_1$ and $\bm{X}(t_2)=\bm{x}_2$. The counting process approach assumes
\begin{equation}
	\begin{aligned} \label{eq:covfull}
	\bm{X}(t) = \bm{x}_{j-1},\quad t_{j-1}\le t<t_{j},\quad j=1,2,3.
	\end{aligned}
\end{equation}
The information of the subject under Case I, displays exactly as in (\ref{eq:covfull}). For a dataset under Case II when only half of the changes are known, only one of $\{t_1,t_2\}$ is known. If only the change at $t_k$ ($k=1,2$) is known, the observed information for the same subject is then
\begin{equation}
	\begin{aligned} \label{eq:covpart}
	\bm{X}(t) &= \bm{x}_0,\quad 0\le t< t_k; \\
	\bm{X}(t) &= \bm{x}_k,\quad t_k \le t \le \tilde{T}.
	\end{aligned}
\end{equation}
Note that for both (\ref{eq:covfull}) and (\ref{eq:covpart}), the true survival curve is constructed using the information as in (\ref{eq:covfull}), when all history of changes in values are known. 

\subsection{Evaluation measures}
Since the goal is to estimate the survival function, we evaluate estimation performance using the average integrated $L_2$ difference between the true and the estimated survival curves $\widehat{S}$. 
Given a dataset $\mathcal{D}$, containing $N$ subjects, each with pseudo-subject information up to the survival/censored time $\tilde{T}^{(i)}$, $\mathcal{X}^{(i)} (\tilde{T}^{(i)} )$, $i=1,2,\ldots,N$,  
\begin{align}
   L_2&(\widehat{S}) = \frac{1}{N}\sum_{i\in \mathcal{D}}\frac{1}{\tilde{T}^{(i)}}\int_0^{\tilde{T}^{(i)}}[S^{(i)}(t)-\widehat{S} (t | \mathcal{X}^{(i)}(t))]^2dt. \label{eq:L2}
\end{align}

Note that we evaluate the integrated $L_2$ difference only up to $\tilde{T}^{(i)}$, the last time point where the survival status is known.
In the simulations, as we generate the true survival time $T^{(i)}$, we have the trajectory of covariate values up to time $T^{(i)}$ for any given subject $i$ even when it is censored at time $C^{(i)} < T^{(i)}$. However, here we intend to match the scenario in real world applications where the covariate information is usually no longer recorded after the event occurs (e.g. the patient dies) or the subject is censored (e.g. lost contact). Thus, we define the best modeling method to be the one that gives us the lowest integrated $L_2$ difference, which is an average value from all subjects; for each subject, the difference between an estimated survival curve and the true survival curve up to its last observed time is measured.

\subsection{Simulation results}
The extended Cox model is included as a benchmark method, since it is one of the most commonly used methods in practice. Another benchmark method used in this paper is the Kaplan-Meier method, which uses no covariates' values to construct the survival function estimate; this helps illustrate the improvement in estimation from incorporating the covariate information.

In this section, we present simulation results based on $500$ simulation trials. The number of trees for bootstrap samples is set to be 100 for all forest methods. We also only focus on the Weibull-Increasing distribution, and omit discussion of the Weibull-Decreasing distribution, since results for the latter distribution are similar. Detailed results are given in Section S1.7 of the Supplemental Material.

\subsubsection{Regulating the construction of trees in forests}
\begin{figure*}[t]
	\centering
	\includegraphics[scale=0.49]{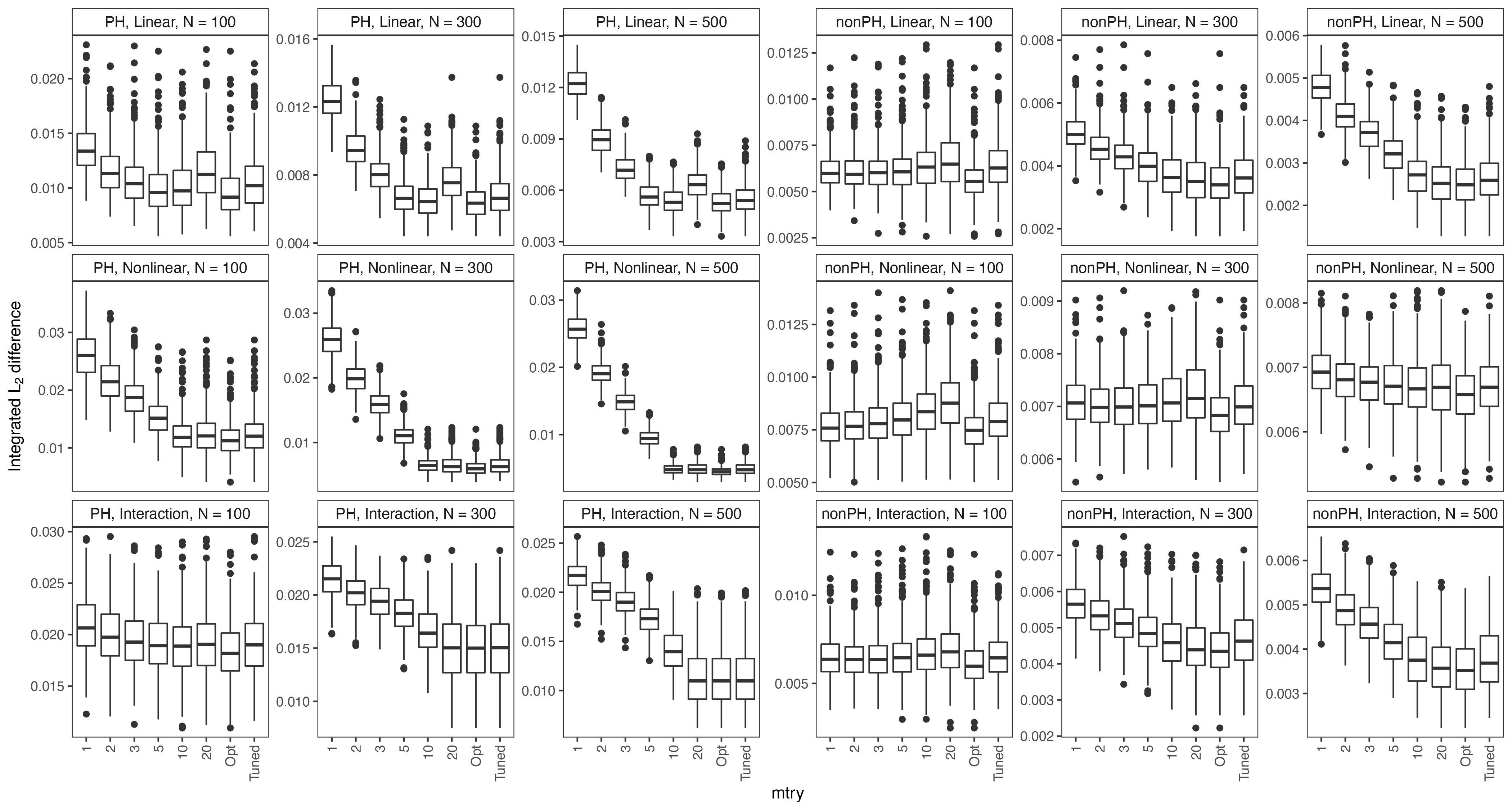}
	\caption{Integrated $L_2$ difference of CIF-TV with different $mtry$ values. Datasets are generated with a light right-censoring rate (20\%), survival times following a Weibull-Increasing distribution. 
	From the top row to the bottom, are given results for the linear, nonlinear and interaction survival relationship; the first three columns show results under the PH setting for the number of subjects $N=100,300,500$, respectively, and the last three columns for results under the non-PH setting. 
	In each plot, 1--CIF-TV with $mtry=1$; 2--CIF-TV with $mtry=2$; 3--CIF-TV with $mtry=3$; 5--CIF-TV with $mtry=5$; 10--CIF-TV with $mtry=10$; 20--CIF-TV with $mtry=20$; Opt--CIF-TV with value of \textit{mtry} that gives the smallest Integrated $L_2$ difference in each round; Tuned--CIF-TV with the value of $mtry$ tuned by the ``out-of-bag'' tuning procedure. The default value in conditional inference forest is $mtry=5$.}
	\label{fig:mtryCIF}
\end{figure*}

Figure \ref{fig:mtryCIF} gives an example of how CIF-TV performs with different values of \textit{mtry} in the scenario 2TI $+$ 4TV, when the censoring rate is 20\%, and the signal-to-noise ratio is low.  
The \textit{mtry} values are tuned based on the ``out-of-bag observations''. 
Similar results for RRF-TV and TSF-TV can be found in Section S1.6 in the Supplemental Material. 

In these examples, one can see that the forests using the ``out-of-bag'' tuning procedure give relatively good performance overall. 
In fact, results from other model setups are broadly similar, in the sense that this tuning procedure provides a relatively reliable choice of \textit{mtry} and it gains in accuracy as the number of subjects $N$ increases. In contrast, the default value of \textit{mtry} does not always perform well, and choosing a different value can sometimes significantly improve performance. 

Table \ref{tbl:defaultvsproposed_PHnonPHm2} gives examples under the PH setting to show the performance comparison between each forest with its default parameter settings and with the proposed parameter settings. 
 
\begin{table*}[t]
    \caption{Comparison between forests with default (D) and proposed parameter settings (P) across different numbers of subjects $N$ in the scenario 2TI $+$ 4TV, when the censoring rate is 20\%, and the signal-to-noise ratio is low.}
    	\label{tbl:defaultvsproposed_PHnonPHm2}
    	\centering
    	\begin{tabular}{cccccccc}
        % \begin{tabular}{C{0.5cm}C{2cm}C{1.6cm}C{1.6cm}C{1.6cm}C{1.6cm}C{1.6cm}C{1.6cm}}
        \hline
        \multicolumn{8}{l}{\textit{Proportional hazards setting}}\\
        \hline
        \multirow{2}{*}{$N$}&\multicolumn{7}{c}{Case I. All changes in covariates' values are known}\\
        \cline{2-8}
        &Extended Cox&CIF-TV(D)&CIF-TV(P)&RRF-TV(D)&RRF-TV(P)&TSF-TV(D)&TSF-TV(P)\\
        \hline
        100 & {\tablenum[table-format=1.2(2),separate-uncertainty]{0.57\pm0.15}} & {\tablenum[table-format=1.2(2),separate-uncertainty]{0.17\pm0.26}} & {\tablenum[table-format=1.2(2),separate-uncertainty]{0.46\pm0.15}} & {\tablenum[table-format=1.2(2),separate-uncertainty]{0.31\pm0.20}} & {\tablenum[table-format=1.2(2),separate-uncertainty]{0.43\pm0.16}} & {\tablenum[table-format=1.2(2),separate-uncertainty]{0.12\pm0.27}} & {\tablenum[table-format=1.2(2),separate-uncertainty]{0.36\pm0.13}} \\
        300 & {\tablenum[table-format=1.2(2),separate-uncertainty]{0.86\pm0.04}} & {\tablenum[table-format=1.2(2),separate-uncertainty]{0.24\pm0.16}} & {\tablenum[table-format=1.2(2),separate-uncertainty]{0.65\pm0.07}} & {\tablenum[table-format=1.2(2),separate-uncertainty]{0.37\pm0.13}} & {\tablenum[table-format=1.2(2),separate-uncertainty]{0.63\pm0.08}} & {\tablenum[table-format=1.2(2),separate-uncertainty]{0.15\pm0.18}} & {\tablenum[table-format=1.2(2),separate-uncertainty]{0.56\pm0.07}} \\
        500 & {\tablenum[table-format=1.2(2),separate-uncertainty]{0.92\pm0.02}} & {\tablenum[table-format=1.2(2),separate-uncertainty]{0.27\pm0.12}} & {\tablenum[table-format=1.2(2),separate-uncertainty]{0.71\pm0.05}} & {\tablenum[table-format=1.2(2),separate-uncertainty]{0.38\pm0.10}} & {\tablenum[table-format=1.2(2),separate-uncertainty]{0.69\pm0.05}} & {\tablenum[table-format=1.2(2),separate-uncertainty]{0.23\pm0.16}} & {\tablenum[table-format=1.2(2),separate-uncertainty]{0.63\pm0.05}} \\
        \hline
        \multirow{2}{*}{$N$}&\multicolumn{7}{c}{Case II. Half of changes in covariates' values are unknown}\\
        \cline{2-8}
        &Extended Cox&CIF-TV(D)&CIF-TV(P)&RRF-TV(D)&RRF-TV(P)&TSF-TV(D)&TSF-TV(P)\\
        \hline
        100 & {\tablenum[table-format=1.2(2),separate-uncertainty]{0.30\pm0.17}} & {\tablenum[table-format=1.2(2),separate-uncertainty]{0.30\pm0.19}} & {\tablenum[table-format=1.2(2),separate-uncertainty]{0.37\pm0.14}} & {\tablenum[table-format=1.2(2),separate-uncertainty]{0.34\pm0.17}} & {\tablenum[table-format=1.2(2),separate-uncertainty]{0.35\pm0.16}} & {\tablenum[table-format=1.2(2),separate-uncertainty]{0.26\pm0.17}} & {\tablenum[table-format=1.2(2),separate-uncertainty]{0.27\pm0.10}} \\
        300 & {\tablenum[table-format=1.2(2),separate-uncertainty]{0.55\pm0.06}} & {\tablenum[table-format=1.2(2),separate-uncertainty]{0.39\pm0.11}} & {\tablenum[table-format=1.2(2),separate-uncertainty]{0.50\pm0.06}} & {\tablenum[table-format=1.2(2),separate-uncertainty]{0.44\pm0.10}} & {\tablenum[table-format=1.2(2),separate-uncertainty]{0.50\pm0.08}} & {\tablenum[table-format=1.2(2),separate-uncertainty]{0.37\pm0.11}} & {\tablenum[table-format=1.2(2),separate-uncertainty]{0.42\pm0.06}} \\
        500 & {\tablenum[table-format=1.2(2),separate-uncertainty]{0.59\pm0.04}} & {\tablenum[table-format=1.2(2),separate-uncertainty]{0.42\pm0.08}} & {\tablenum[table-format=1.2(2),separate-uncertainty]{0.54\pm0.04}} & {\tablenum[table-format=1.2(2),separate-uncertainty]{0.46\pm0.07}} & {\tablenum[table-format=1.2(2),separate-uncertainty]{0.53\pm0.05}} & {\tablenum[table-format=1.2(2),separate-uncertainty]{0.42\pm0.08}} & {\tablenum[table-format=1.2(2),separate-uncertainty]{0.47\pm0.05}}\\
    	\hline
    	\multicolumn{8}{l}{\textit{Non-proportional hazards setting}}\\
    	\hline
        \multirow{2}{*}{$N$}&\multicolumn{7}{c}{Case I. All changes in covariates' values are known}\\
        \cline{2-8}
        &Extended Cox&CIF-TV(D)&CIF-TV(P)&RRF-TV(D)&RRF-TV(P)&TSF-TV(D)&TSF-TV(P)\\
        \hline
        100 & {\tablenum[table-format=1.2(2),separate-uncertainty]{-0.55\pm0.28}} & {\tablenum[table-format=1.2(2),separate-uncertainty]{-0.39\pm0.38}} & {\tablenum[table-format=1.2(2),separate-uncertainty]{0.11\pm0.21}} & {\tablenum[table-format=1.2(2),separate-uncertainty]{-0.27\pm0.33}} & {\tablenum[table-format=1.2(2),separate-uncertainty]{0.12\pm0.18}} & {\tablenum[table-format=1.2(2),separate-uncertainty]{-0.28\pm0.42}} & {\tablenum[table-format=1.2(2),separate-uncertainty]{0.35\pm0.21}} \\
        300 & {\tablenum[table-format=1.2(2),separate-uncertainty]{-0.27\pm0.10}} & {\tablenum[table-format=1.2(2),separate-uncertainty]{-0.27\pm0.27}} & {\tablenum[table-format=1.2(2),separate-uncertainty]{0.44\pm0.12}} & {\tablenum[table-format=1.2(2),separate-uncertainty]{-0.14\pm0.23}} & {\tablenum[table-format=1.2(2),separate-uncertainty]{0.33\pm0.12}} & {\tablenum[table-format=1.2(2),separate-uncertainty]{-0.40\pm0.31}} & {\tablenum[table-format=1.2(2),separate-uncertainty]{0.62\pm0.12}} \\
        500 & {\tablenum[table-format=1.2(2),separate-uncertainty]{-0.23\pm0.07}} & {\tablenum[table-format=1.2(2),separate-uncertainty]{-0.24\pm0.21}} & {\tablenum[table-format=1.2(2),separate-uncertainty]{0.59\pm0.09}} & {\tablenum[table-format=1.2(2),separate-uncertainty]{-0.08\pm0.18}} & {\tablenum[table-format=1.2(2),separate-uncertainty]{0.47\pm0.11}} & {\tablenum[table-format=1.2(2),separate-uncertainty]{-0.30\pm0.28}} & {\tablenum[table-format=1.2(2),separate-uncertainty]{0.72\pm0.08}} \\
        \hline
        \multirow{2}{*}{$N$}&\multicolumn{7}{c}{Case II. Half of changes in covariates' values are unknown}\\
        \cline{2-8}
        &Extended Cox&CIF-TV(D)&CIF-TV(P)&RRF-TV(D)&RRF-TV(P)&TSF-TV(D)&TSF-TV(P)\\
        \hline
        100 & {\tablenum[table-format=1.2(2),separate-uncertainty]{-0.51\pm0.28}} & {\tablenum[table-format=1.2(2),separate-uncertainty]{-0.10\pm0.26}} & {\tablenum[table-format=1.2(2),separate-uncertainty]{0.08\pm0.16}} & {\tablenum[table-format=1.2(2),separate-uncertainty]{-0.11\pm0.26}} & {\tablenum[table-format=1.2(2),separate-uncertainty]{0.09\pm0.18}} & {\tablenum[table-format=1.2(2),separate-uncertainty]{0.04\pm0.29}} & {\tablenum[table-format=1.2(2),separate-uncertainty]{0.21\pm0.17}} \\
        300 & {\tablenum[table-format=1.2(2),separate-uncertainty]{-0.16\pm0.10}} & {\tablenum[table-format=1.2(2),separate-uncertainty]{0.00\pm0.19}} & {\tablenum[table-format=1.2(2),separate-uncertainty]{0.22\pm0.10}} & {\tablenum[table-format=1.2(2),separate-uncertainty]{0.01\pm0.18}} & {\tablenum[table-format=1.2(2),separate-uncertainty]{0.20\pm0.08}} & {\tablenum[table-format=1.2(2),separate-uncertainty]{0.09\pm0.22}} & {\tablenum[table-format=1.2(2),separate-uncertainty]{0.37\pm0.13}} \\
        500 & {\tablenum[table-format=1.2(2),separate-uncertainty]{-0.11\pm0.07}} & {\tablenum[table-format=1.2(2),separate-uncertainty]{0.05\pm0.16}} & {\tablenum[table-format=1.2(2),separate-uncertainty]{0.28\pm0.10}} & {\tablenum[table-format=1.2(2),separate-uncertainty]{0.05\pm0.15}} & {\tablenum[table-format=1.2(2),separate-uncertainty]{0.24\pm0.08}} & {\tablenum[table-format=1.2(2),separate-uncertainty]{0.15\pm0.17}} & {\tablenum[table-format=1.2(2),separate-uncertainty]{0.43\pm0.12}}\\
        \hline 
        \multicolumn{8}{l}{\footnotesize{$^\ast$ Given a method $A$, each cell value are given as mean $\pm$ one standard deviation of  $\big(L_2(\mathrm{KM})-L_2(\mathrm{A})\big)/L_2(\mathrm{KM})$}}\\
        \multicolumn{8}{l}{\footnotesize{$\,\,\,$ based on all simulations.}}\\
        \multicolumn{8}{l}{\footnotesize{$^\ast$ For similar results under other model setups, please refer to Section S1.6 in the Supplemental Material.}}\\
    	\end{tabular}
    \end{table*}	 

In Table \ref{tbl:defaultvsproposed_PHnonPHm2}, positive numbers indicate a decrease in integrated $L_2$ difference compared to a Kaplan-Meier fit on the dataset, while negative numbers indicate an increase. 
The absolute value of the numbers represents the size of the difference between the integrated $L_2$ difference of the candidate and that of a Kaplan-Meier fit. 
The table shows that forests with the proposed parameter settings can provide improved performance over those with default parameter settings across all different numbers of subjects $N$ by a substantial amount. 
Note that, under the non-PH setting, for datasets with all of the changes in covariates' history known, the negative numbers indicate the poor performance of forests with default parameter settings even compared to a simple Kaplan-Meier curve, showing that the default methods can fail miserably. 
In contrast, for all forests with the proposed parameter settings, as $N$ increases, the change in sign and in the absolute value of the numbers indicates better and better performance in general. 
Overall, the performance of the proposed parameter setting is relatively stable and better than that of the default values.

In the following discussion, we therefore only focus on the forest methods with the proposed parameter settings.

\subsubsection{Properties of the proposed forest methods}
\begin{figure}[t]
    \centering
    \includegraphics[scale = 0.17]{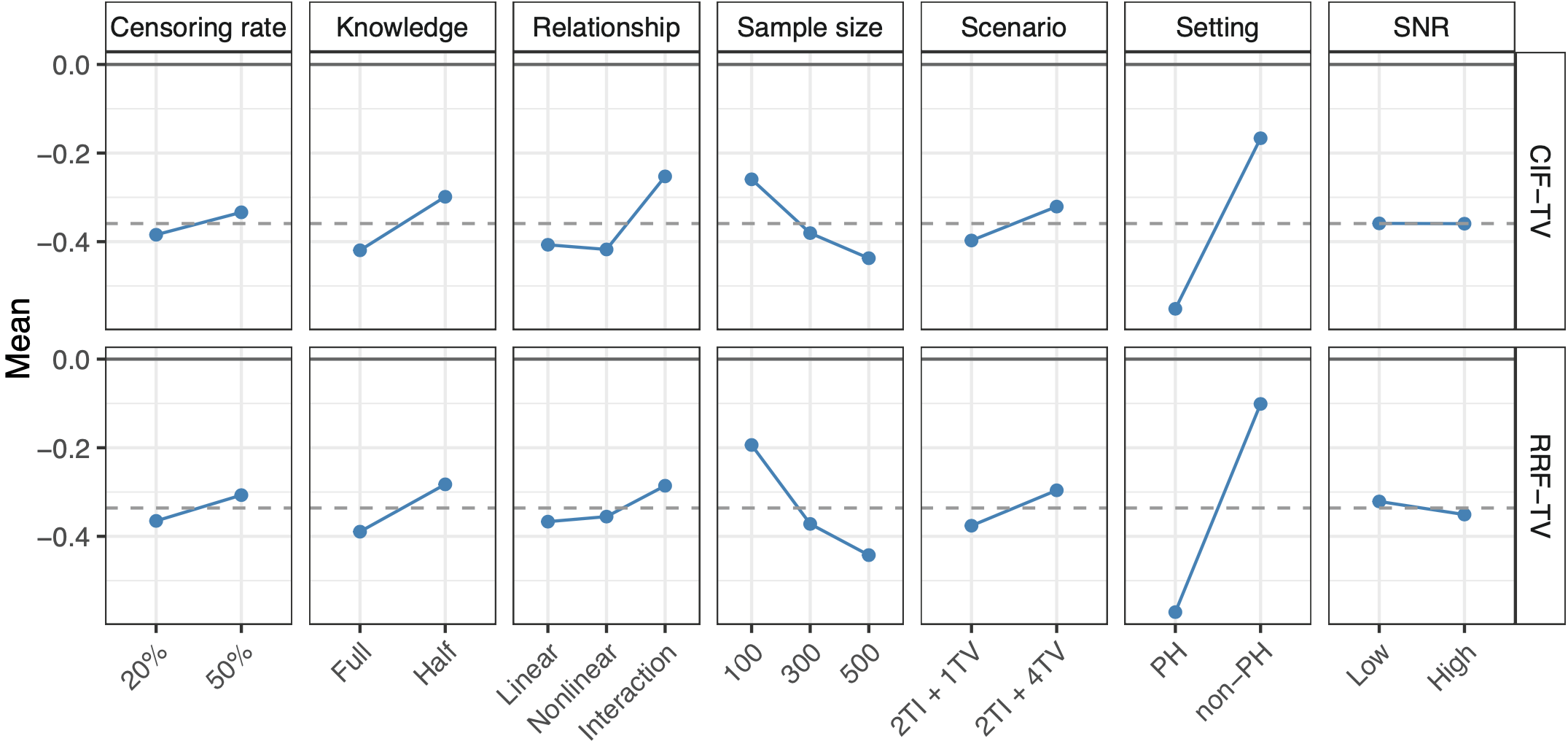}
    \caption{Main effects plots of integrated $L_2$ difference improvement from the proposed forests over a simple Kaplan-Meier fit. Given a method $A$ (CIF-TV or RRF-TV), the difference improvement is computed as $(L_2(\mathrm{A}) - L_2(\mathrm{KM}))/L_2(\mathrm{KM})$. The solid line gives the zero value and the dashed line gives the mean value over all effects for reference.}
    \label{fig:maineffects}
\end{figure}

Using factorial designs, we study the difference between each of the proposed forest 
methods and a simple Kaplan-Meier fit under the effects of the following factors: censoring 
rate, amount of knowledge, survival relationship, training sample size, scenario, hazards 
setting and SNR. The effects are estimated based on an analysis of variance model fit with 
these factors as main effects. Figure \ref{fig:maineffects} provides main effects plots 
for the integrated $L_2$ difference improvement from the proposed forest methods over a 
simple Kaplan-Meier fit. 

In Figure \ref{fig:maineffects}, the overall center of location is negative, 
highlighting that both of the proposed forest methods perform better than a simple 
Kaplan-Meier fit. The overall mean integrated $L_2$ difference of CIF-TV is slightly 
smaller than that of RRF-TV. 
The relative performance of the proposed forest methods can vary with changes in factors. 
Note that the $p$-values of the hypothesis testing on the simple main effect of SNR 
is insignificant at a 0.10 level, which suggests that the impact of the change of its 
level on the performance of the proposed forest methods over a Kaplan-Meier fit 
is negligible (more details can be found in Section S1.8 in the Supplemental Material).
 
For the other factors, the fewer the number of changes in values of covariates 
that are known, higher censoring rate, smaller training sample size, larger portion 
of covariates being time-varying, more relaxed assumption on the hazard setting, 
and more complicated structure of the survival relationship (all reflecting more 
difficult estimation tasks), the less the proposed forest methods improve over a 
simple Kaplan-Meier fit. Conversely, in the opposite situations, the stronger the 
ability of the proposed forest methods to estimate the underlying survival relationship 
and therefore bring a greater improvement.

In particular, the two proposed forest methods win by a much larger margin under 
the proportional hazards setting, which is expected since the log-rank-type splitting 
procedures used in the proposed forest methods rely to some extent on the proportional 
hazards assumption.

It is also clear that the difference between the number of time-invariant (TI) 
and the number of time-varying (TV) covariates is driving the scenario effect. 
When \#TV $-$ \#TI increases, the relative performance of the proposed forest methods 
deteriorates. Presumably, this is because the increasing level of local time-varying 
effects makes the underlying relationship more difficult to estimate.
 
Note that the improvement from the proposed forest methods over a Kaplan-Meier fit remains relatively stable to the change of levels in both the censoring rate and the scenario factors compared to the change in other factors.
In the following discussion, we mainly focus on the factors that are more influential based on our previous study: the number of changes in values of covariates that are known, the underlying survival relationship, the sample size, and the hazard function setting. 
The simulation results presented are based on the datasets generated under the scenario 2TI $+$ 4TV, the lower signal-to-noise ratio, with 20\% censoring rate.

\subsubsection{Estimation performance comparison}
\begin{figure*}[t]
    \centering
	\includegraphics[scale=0.45]{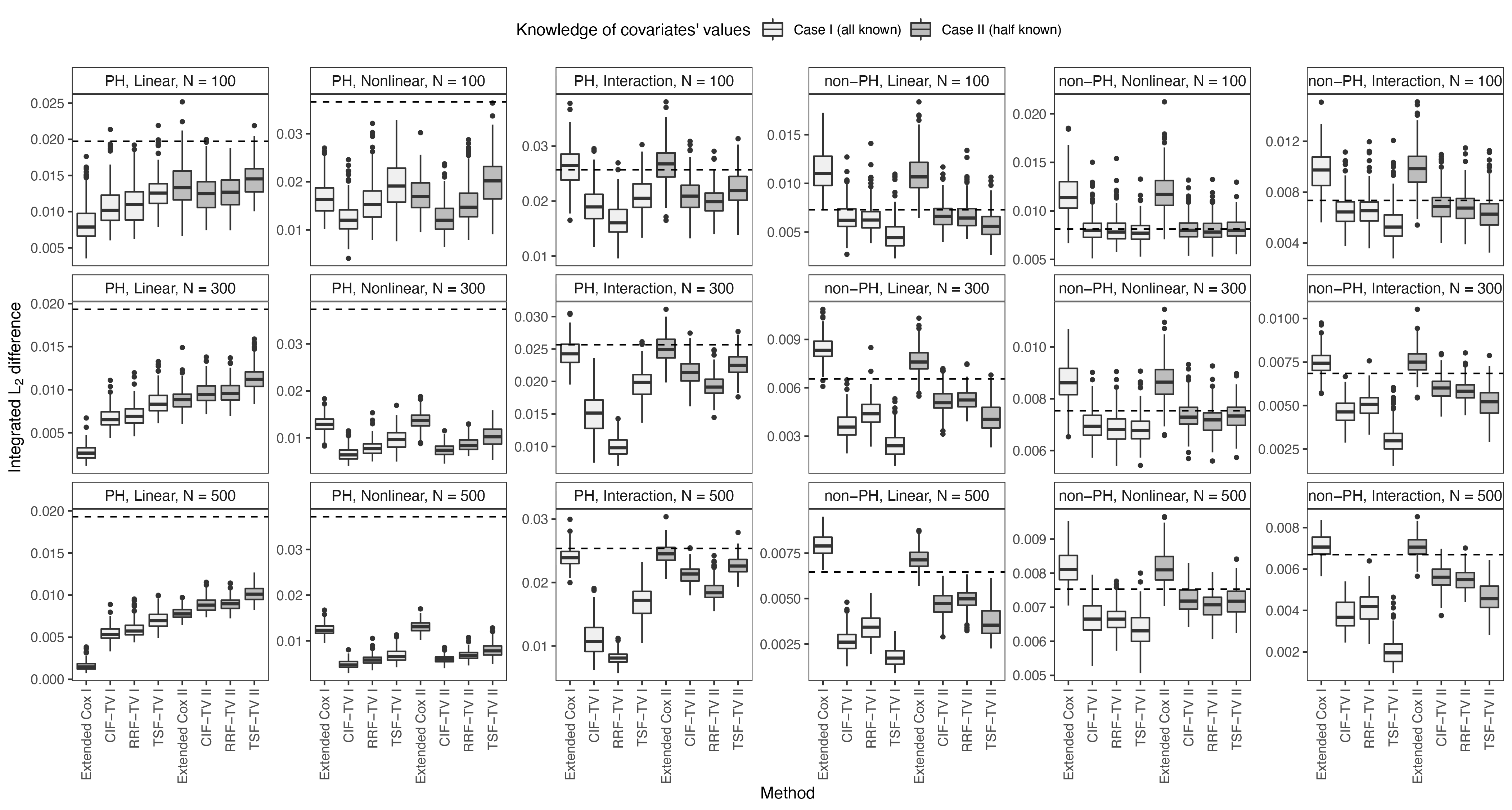}
	\caption{Boxplots of integrated $L_2$ difference for performance comparison. Datasets are generated with survival times following a Weibull distribution, light right-censoring rate (20\%). The three rows show results for the number of subjects $N=100,300,500$, respectively; the first three columns show results under the PH setting for survival relationship linear, nonlinear and interaction, respectively, and the last three columns for results under the non-PH setting. The horizontal dashed line shows the median integrated $L_2$ difference of a Kaplan-Meier fit on the datasets. 
	In each of the plots, the set of boxplots lightly-shaded shows the performance of different methods on datasets with history of changes in covariates' values known; the set heavily-shaded shows the performance on datasets with half of the changes in covariates' values unknown. }
	\label{fig:evaperf}
\end{figure*}

Figure \ref{fig:evaperf} gives side-by-side boxplots of integrated $L_2$ difference defined in (\ref{eq:L2}) 
for performance comparison under different model setups. 
It shows that for the linear survival relationship under the PH setting, the extended Cox model performs the best. This is expected as the extended Cox model relies exactly on the assumption of proportional hazards and a log-linear relationship between the hazard function and covariates. 
For nonlinear and interaction survival relationships, all forests outperform the extended Cox model, showing their advantage in dealing with a relatively complex situation. 
More specifically, for nonlinear setups, CIF-TV performs the best and RRF-TV the second, while for interaction model setups, RRF-TV performs the best and CIF-TV the second. 
In addition, CIF-TV and RRF-TV outperform TSF-TV across all different numbers of subjects and survival relationships. 
Under the non-PH setting, the extended Cox model cannot even outperform a simple Kaplan-Meier fit on the dataset, whether all changes in values of covariates are known or not. 
As discussed, the presence of non-proportional hazards settings poses great challenges to modeling methods that assume proportional hazards; not just Cox, but also the survival forests like CIF-TV and RRF-TV that use a log-rank splitting rule. 
On the other hand, TSF-TV, which is specifically designed to detect non-proportional hazards deviations, performs the best across all different setups under the non-PH setting.

It is not surprising that having all changes in values known gives increasingly better performance compared to only having half of the changes known as the sample size increases. 
As $N$ increases, false information due to unknown changes has a negative effect on performance of all modeling methods. In particular, this affects the extended Cox model more than the forest methods when the underlying survival relationship is linear under the PH setting, while it affects the extended Cox model less for all other cases. 
This is simply because the extended Cox model already performs poorly in nonlinear and non-PH situations, so the misleading information from incorrect knowledge of covariates' values cannot hurt performance very much. 
 
Generally, if the true underlying model setup is known, one should choose CIF-TV or RRF-TV under the PH setting, and TSF-TV under the non-PH setting. However, none of the forest methods can perform well all of the time. In the next section, we provide guidance on how to choose among these forest methods.

\subsubsection{Guidance for choosing the modeling method}
Cross-validation methods have been used in the past for the error estimation  of survival models \citep{predictionmeasure}. 
We propose to use one of the most common methods, $K$-fold cross-validation, implemented with integrated Brier scores for survival data, to select the ``best'' modeling method for a given dataset $\mathcal{D}$ as follows.

For a given survival curve estimate $\widehat{S}$,
\begin{enumerate}
	\item[i.] Split the dataset into $K$ non-overlapping subsets $\mathcal{D}_k$ $(k = 1, 2, \ldots, K)$, each 
	containing (roughly) equal number of subjects;
	\item[ii.] For each $k=1,2,\ldots, K$
	\begin{enumerate}
		\item Modeling methods $\widehat{S}_k$ are then trained with the data $\mathcal{D} \backslash \mathcal{D}_k$ where the $k$-th subset is removed;
		\item Test $\widehat{S}_k$ on data in the $k$-th test set $\mathcal{D}_k$ and compute the corresponding integrated Brier score $\widehat{\mathrm{IBS}}\big(\widehat{S}_k;\mathcal{D}_k\big)$ as given in (\ref{eq:IBSrewrite});
	\end{enumerate}
	\item[iii.] Average over all $K$ subsets and obtain 
	\begin{align}
	    \mathrm{IBSCVErr}(\widehat{S})=\frac{1}{K}\sum_{k=1}^{K}\widehat{\mathrm{IBS}}(\widehat{S}_k;\mathcal{D}_k).\label{eq:IBSCVErr}
	\end{align}
\end{enumerate}
We then choose the modeling method that gives the smallest $\mathrm{IBSCVErr}(\widehat{S})$ in (\ref{eq:IBSCVErr}). 
 	
For the simulated data sets, we use $10$-fold cross-validation to choose between modeling methods. 
The measures $p_{\mathrm{B}}$, $r_{\mathrm{B}}$ and $r_{\mathrm{W}}$ are used to evaluate the performance,
\begin{align}
    p_{\mathrm{B}} &=\#\{x_{\mathrm{CV}}=x_{\min}\}/n_{\mathrm{rep}} \label{eq:p_B}\\
    r_{\mathrm{B}} &= \vert x_{\min}-x_{\mathrm{CV}}\vert/x_{\min}\label{eq:r_B}\\
    r_{\mathrm{W}} &= \vert x_{\max}-x_{\mathrm{CV}}\vert/x_{\max}\label{eq:r_W}
\end{align}
where $n_{\mathrm{rep}}$ denotes the number of simulations ($n_{\mathrm{rep}}=500$), $x_{\mathrm{CV}}$ denotes the integrated $L_2$ difference of the method chosen by cross-validation, and $x_{\min}$ and $x_{\max}$ denote the lowest and highest integrated $L_2$ difference among all modeling methods, respectively. In each round of simulation, we call the method that gives $x_{\min}$ the best modeling method and the method that gives $x_{\max}$ the worst modeling method. By definition, $p_{\mathrm{B}}$ provides the proportion of the times IBS-based $10$-fold CV selects the best modeling method, and $r_{\mathrm{B}}$ and $r_{\mathrm{W}}$ compute the relative errors from the best and the worst modeling method, respectively. The smaller $r_{\mathrm{B}}$ is, or the larger $r_{\mathrm{W}}$ is, the better IBS-based $10$-fold CV works.
 
\begin{table*}[t]
	\centering
	\caption{Summary of the performance of IBS-based 10-fold CV rule. 
	}\label{tbl:IBSCV3measures}
	\begin{tabular}{lllcccccc}
		\hline
		\multirow{2}{*}{Sample size}&\multirow{2}{*}{Setting}&\multirow{2}{*}{Relationship} &\multicolumn{3}{c}{Case I}&\multicolumn{3}{c}{Case II}\\
		\cline{4-9}
		&&&$p_{\mathrm{B}}$\footnotemark[1]&$r_{\mathrm{B}}$\footnotemark[2] &$r_{\mathrm{W}}$\footnotemark[2]&$p_{\mathrm{B}}$\footnotemark[1]&$r_{\mathrm{B}}$\footnotemark[2] &$r_{\mathrm{W}}$\footnotemark[2]\\
		\hline
		\multirow{6}{*}{$N=100$}& \multirow{3}{*}{PH}& Linear & {\tablenum[table-format = 1.2]{0.34}} & {\tablenum[table-format = 1.2]{0.35}}$\;\pm\;0.44$ & {\tablenum[table-format = 1.2]{0.21}}$\;\pm\;0.18$ & {\tablenum[table-format = 1.2]{0.35}} & {\tablenum[table-format = 1.2]{0.14}}$\;\pm\;0.16$ & {\tablenum[table-format = 1.2]{0.16}}$\;\pm\;0.12$\\
        && Nonlinear & {\tablenum[table-format = 1.2]{0.66}} & {\tablenum[table-format = 1.2]{0.11}}$\;\pm\;0.22$ & {\tablenum[table-format = 1.2]{0.34}}$\;\pm\;0.18$ & {\tablenum[table-format = 1.2]{0.69}} & {\tablenum[table-format = 1.2]{0.09}}$\;\pm\;0.21$ & {\tablenum[table-format = 1.2]{0.35}}$\;\pm\;0.16$\\
        && Interaction & {\tablenum[table-format = 1.2]{0.71}} & {\tablenum[table-format = 1.2]{0.03}}$\;\pm\;0.08$ & {\tablenum[table-format = 1.2]{0.33}}$\;\pm\;0.11$ & {\tablenum[table-format = 1.2]{0.33}} & {\tablenum[table-format = 1.2]{0.08}}$\;\pm\;0.10$ & {\tablenum[table-format = 1.2]{0.21}}$\;\pm\;0.09$\\
        \cline{2-9}
        & \multirow{3}{*}{non-PH}& Linear & {\tablenum[table-format = 1.2]{0.87}} & {\tablenum[table-format = 1.2]{0.04}}$\;\pm\;0.16$ & {\tablenum[table-format = 1.2]{0.56}}$\;\pm\;0.14$ & {\tablenum[table-format = 1.2]{0.63}} & {\tablenum[table-format = 1.2]{0.06}}$\;\pm\;0.13$ & {\tablenum[table-format = 1.2]{0.45}}$\;\pm\;0.14$\\
        && Nonlinear & {\tablenum[table-format = 1.2]{0.53}} & {\tablenum[table-format = 1.2]{0.04}}$\;\pm\;0.07$ & {\tablenum[table-format = 1.2]{0.33}}$\;\pm\;0.10$ & {\tablenum[table-format = 1.2]{0.47}} & {\tablenum[table-format = 1.2]{0.04}}$\;\pm\;0.07$ & {\tablenum[table-format = 1.2]{0.32}}$\;\pm\;0.10$\\
        && Interaction & {\tablenum[table-format = 1.2]{0.79}} & {\tablenum[table-format = 1.2]{0.04}}$\;\pm\;0.12$ & {\tablenum[table-format = 1.2]{0.42}}$\;\pm\;0.14$ & {\tablenum[table-format = 1.2]{0.55}} & {\tablenum[table-format = 1.2]{0.05}}$\;\pm\;0.10$ & {\tablenum[table-format = 1.2]{0.34}}$\;\pm\;0.12$\\
        \hline 
        \multirow{6}{*}{$N=300$}& \multirow{3}{*}{PH}& Linear & {\tablenum[table-format = 1.2]{0.99}} & {\tablenum[table-format = 1.2]{0.03}}$\;\pm\;0.40$ & {\tablenum[table-format = 1.2]{0.69}}$\;\pm\;0.11$ & {\tablenum[table-format = 1.2]{0.49}} & {\tablenum[table-format = 1.2]{0.07}}$\;\pm\;0.11$ & {\tablenum[table-format = 1.2]{0.19}}$\;\pm\;0.10$\\
        && Nonlinear & {\tablenum[table-format = 1.2]{0.83}} & {\tablenum[table-format = 1.2]{0.03}}$\;\pm\;0.14$ & {\tablenum[table-format = 1.2]{0.49}}$\;\pm\;0.12$ & {\tablenum[table-format = 1.2]{0.76}} & {\tablenum[table-format = 1.2]{0.04}}$\;\pm\;0.12$ & {\tablenum[table-format = 1.2]{0.45}}$\;\pm\;0.12$\\
        && Interaction & {\tablenum[table-format = 1.2]{0.95}} & {\tablenum[table-format = 1.2]{0.00}}$\;\pm\;0.02$ & {\tablenum[table-format = 1.2]{0.58}}$\;\pm\;0.07$ & {\tablenum[table-format = 1.2]{0.58}} & {\tablenum[table-format = 1.2]{0.06}}$\;\pm\;0.09$ & {\tablenum[table-format = 1.2]{0.19}}$\;\pm\;0.08$\\
        \cline{2-9}
        & \multirow{3}{*}{non-PH}& Linear & {\tablenum[table-format = 1.2]{0.95}} & {\tablenum[table-format = 1.2]{0.01}}$\;\pm\;0.05$ & {\tablenum[table-format = 1.2]{0.70}}$\;\pm\;0.09$ & {\tablenum[table-format = 1.2]{0.77}} & {\tablenum[table-format = 1.2]{0.04}}$\;\pm\;0.10$ & {\tablenum[table-format = 1.2]{0.44}}$\;\pm\;0.13$\\
        && Nonlinear & {\tablenum[table-format = 1.2]{0.59}} & {\tablenum[table-format = 1.2]{0.02}}$\;\pm\;0.03$ & {\tablenum[table-format = 1.2]{0.21}}$\;\pm\;0.06$ & {\tablenum[table-format = 1.2]{0.42}} & {\tablenum[table-format = 1.2]{0.02}}$\;\pm\;0.03$ & {\tablenum[table-format = 1.2]{0.16}}$\;\pm\;0.05$\\
        && Interaction & {\tablenum[table-format = 1.2]{0.97}} & {\tablenum[table-format = 1.2]{0.01}}$\;\pm\;0.05$ & {\tablenum[table-format = 1.2]{0.59}}$\;\pm\;0.11$ & {\tablenum[table-format = 1.2]{0.73}} & {\tablenum[table-format = 1.2]{0.03}}$\;\pm\;0.08$ & {\tablenum[table-format = 1.2]{0.30}}$\;\pm\;0.11$\\
        \hline 
        \multirow{6}{*}{$N=500$}& \multirow{3}{*}{PH}& Linear & {\tablenum[table-format = 1.2]{1.00}} & {\tablenum[table-format = 1.2]{0.00}}$\;\pm\;0.00$ & {\tablenum[table-format = 1.2]{0.78}}$\;\pm\;0.07$ & {\tablenum[table-format = 1.2]{0.79}} & {\tablenum[table-format = 1.2]{0.03}}$\;\pm\;0.06$ & {\tablenum[table-format = 1.2]{0.22}}$\;\pm\;0.08$\\
        && Nonlinear & {\tablenum[table-format = 1.2]{0.80}} & {\tablenum[table-format = 1.2]{0.02}}$\;\pm\;0.07$ & {\tablenum[table-format = 1.2]{0.60}}$\;\pm\;0.08$ & {\tablenum[table-format = 1.2]{0.78}} & {\tablenum[table-format = 1.2]{0.03}}$\;\pm\;0.07$ & {\tablenum[table-format = 1.2]{0.54}}$\;\pm\;0.07$\\
        && Interaction & {\tablenum[table-format = 1.2]{0.84}} & {\tablenum[table-format = 1.2]{0.01}}$\;\pm\;0.05$ & {\tablenum[table-format = 1.2]{0.65}}$\;\pm\;0.05$ & {\tablenum[table-format = 1.2]{0.86}} & {\tablenum[table-format = 1.2]{0.03}}$\;\pm\;0.07$ & {\tablenum[table-format = 1.2]{0.22}}$\;\pm\;0.07$\\
        \cline{2-9}
        & \multirow{3}{*}{non-PH}& Linear & {\tablenum[table-format = 1.2]{0.94}} & {\tablenum[table-format = 1.2]{0.01}}$\;\pm\;0.06$ & {\tablenum[table-format = 1.2]{0.77}}$\;\pm\;0.06$ & {\tablenum[table-format = 1.2]{0.80}} & {\tablenum[table-format = 1.2]{0.03}}$\;\pm\;0.10$ & {\tablenum[table-format = 1.2]{0.48}}$\;\pm\;0.12$\\
        && Nonlinear & {\tablenum[table-format = 1.2]{0.74}} & {\tablenum[table-format = 1.2]{0.01}}$\;\pm\;0.02$ & {\tablenum[table-format = 1.2]{0.22}}$\;\pm\;0.06$ & {\tablenum[table-format = 1.2]{0.34}} & {\tablenum[table-format = 1.2]{0.02}}$\;\pm\;0.02$ & {\tablenum[table-format = 1.2]{0.12}}$\;\pm\;0.04$\\
        && Interaction & {\tablenum[table-format = 1.2]{0.99}} & {\tablenum[table-format = 1.2]{0.00}}$\;\pm\;0.01$ & {\tablenum[table-format = 1.2]{0.71}}$\;\pm\;0.09$ & {\tablenum[table-format = 1.2]{0.82}} & {\tablenum[table-format = 1.2]{0.02}}$\;\pm\;0.06$ & {\tablenum[table-format = 1.2]{0.34}}$\;\pm\;0.11$\\
		\hline
		\multicolumn{9}{l}{\scriptsize{$^1$ $p_{\mathrm{B}}$ is computed as in (\ref{eq:p_B}), with mean value over all simulations.}}\\
		\multicolumn{9}{l}{\scriptsize{$^2$ $r_{\mathrm{B}}$ and $r_{\mathrm{W}}$ are computed as in (\ref{eq:r_B}) and (\ref{eq:r_W}), respectively, with mean value $\pm$ one standard deviation over all simulations.}}\\
	\end{tabular}
\end{table*}
    
Table \ref{tbl:IBSCV3measures} presents the summary of the performance of the IBS-based 10-fold CV Rule. 
It is not surprising that IBS-based 10-fold CV works better under Case I where all changes in covariate values are known in general, with larger values of $p_{\mathrm{B}}$, smaller values of $r_{\mathrm{B}}$ and larger values of $r_{\mathrm{W}}$. That is, the incorrect knowledge of covariate values also hurts the performance of the selection procedure. 
In general, as the number of subjects $N$ increases, 
the value of $p_{\mathrm{B}}$ gets larger for most of the scenarios, 
indicating IBS-based 10-fold CV is able to pick up the best modeling 
method at a higher frequency; even under those scenarios where 
$p_{\mathrm{B}}$ is lower than 50\%, the relative error 
from the best modeling method, $r_{\mathrm{B}}$, remains 
within 10\% for most of the cases. 
Note that more than half of the cases for $N=100$ have the 
relative error from the best modeling method, $r_{\mathrm{B}}$, 
less than 10\%, and almost all of the cases for $N=500$ have 
$r_{\mathrm{B}}$ under 5\%. 
That means that even when the IBS-based 10-fold CV does 
not pick the best modeling method, it is still able to pick 
a method that works reasonably well, resulting in the 
integrated $L_2$ difference being not much higher than 
that of the best method.

\subsection{Proposed forest methods for time-invariant covariate data}
We have focused on ensemble methods for survival data with time-varying covariates, as we feel that this is a very common and important situation that has been understudied in the past. Having said that, there are certainly many situations in which only time-invariant (baseline) covariate information is available, and understanding the properties of different methods in that situation is important. Section S2 in the Supplemental Material describes the results of simulations related to this question.
In those simulations, datasets with left-truncated and right-censored survival times are generated based on time-invariant covariates.
% , which is the situation for which the proposed forest methods are designed.

In fact, the simulation results of all comparative estimation performance in the case of time-invariant covariates are broadly similar to those in the time-varying covariates cases. That is,
\begin{enumerate}
	\item The ``out-of-bag'' tuning procedure can provide a reliable choice of \textit{mtry} that gives relatively good performance in general. One should also consider adjusting other tuning parameters such as \textit{minsplit}, \textit{minbucket} in the conditional inference forest and the transformation forest, and \textit{nodesize} in random survival forest, as the size of dataset grows;
	\item Taking into an account all other factors, under the PH setting, the best method is always one of the two proposed forests, while under the non-PH setting, it is the transformation forest method.  
	\item The IBS-based CV rule is a good option for choosing among the various methods, as the comparative performance of methods appears to be different from setting to setting.
\end{enumerate}
% This guidance is thus appropriate for survival data with both time-invariant and time-varying covariates.

\subsection{Real data application}
\begin{figure}[!t]
    \centering
    \includegraphics[scale=0.55]{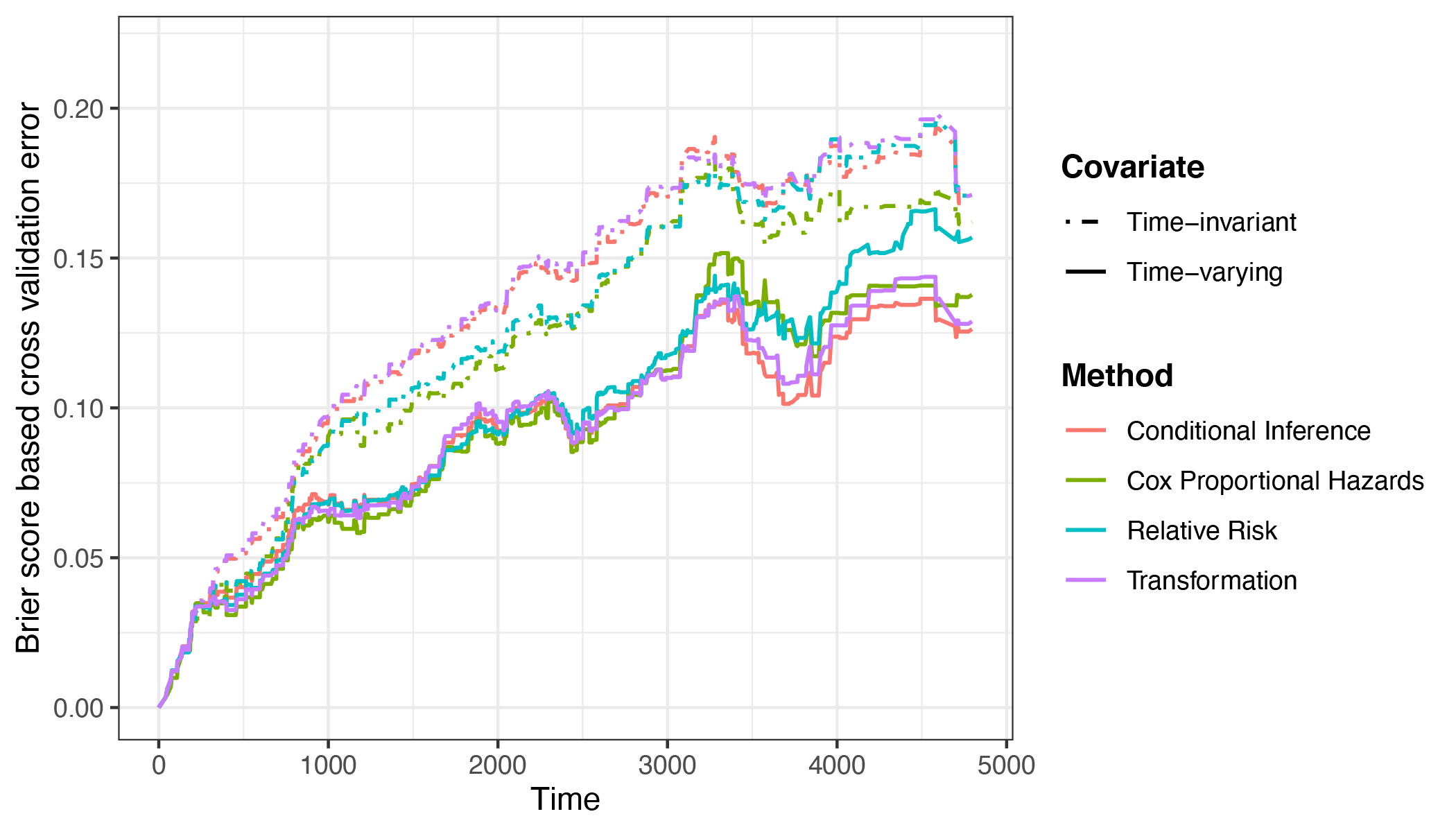}
    \caption{Brier score-based $10$-fold cross-validation errors at $t$-th recording day provided for (1) the extended Cox model, CIF-TV, RRF-TV and TSF-TV on the PBC data with time-varying covariate values obtained in the extended study; (2) Cox model, CIF, RRF and TSF on the PBC dataset with only the initial covariate values (i.e. with time-invariant covariates). The results are shown up to time point where only 5\% of the subjects are still at risk.}
	\label{fig:realset}
\end{figure}

We now illustrate application of the proposed time-varying covariates forests to a real data set, the Mayo Clinic Primary Biliary Cirrhosis Data, available in the R package \textsf{survival}. To study the effectiveness of using $D$-penicillamine as treatment, 312 patients with primary biliary cirrhosis (PBC) were enrolled in a randomized medical trial at the Mayo Clinic from January in 1974 to May in 1984 \citep{PBC1}. 
The outcome of interest is the time to death for these patients. In this study, medical measurements and other patient information were recorded as covariates' values at entry and at yearly intervals. 
The Cox model was used to estimate the survival function for patients with primary biliary cirrhosis based on twelve noninvasive, easily collected covariates that require only a blood sample and clinical evaluation \citep{PBC1}. 
These twelve covariates include age at entry, alkaline phosphotase (U/liter), logarithm of serum albumin (g/dl), presence of ascites, aspartate aminotransferase (U/ml), logarithm of serum bilirubin (mg/dl), serum cholesterol (mg/dl), condition of edema, presence of hepatomegaly or enlarged liver, platelet count, logarithm of prothrombin time and presence or absence of spiders. As the study was extended for another four years, a total of 1945 visits were generated. 
All of the twelve covariates except age become time-varying covariates in the follow-up data. 
At the end of the follow-up study, 169 of the 312 patients were still alive, 140 had died, and three had been lost contact with.
The extended study allows the researchers to study the effects of the changes in the prognostic variables, as time-varying covariates \citep{PBC2}. 
We therefore fit the proposed forest methods as well as the extended Cox model \citep{PBC2} on the dataset with time-varying covariates from the extended study. 
Note that estimates of this kind for different sets of covariate values over time can be useful in providing guidance from a public policy point of view, as they highlight the different average survival experiences of different subpopulations with different covariate paths.
To better illustrate the effects of the time-varying covariates, we also consider the corresponding time-invariant dataset where the covariate values are never updated after the initial observation. 
For performance comparison, the Brier score-based $10$-fold cross-validation error at the $t$-th recording day is computed for each method on both the time-varying and time-invariant covariate datasets, shown in Figure \ref{fig:realset}. The corresponding integrated Brier score cross-validation results are given in Table \ref{tbl:PBC_IBSCVerr}. 

\begin{table}[ht]
    \caption{IBS-Based $10$-fold cross-validation errors provided for (1) the extended Cox model, CIF-TV, RRF-TV and TSF-TV on the PBC data with time-varying covariate values obtained in the extended study; (2) Cox model, CIF, RRF and TSF on the corresponding time-invariant covariate dataset.}
    	\centering 
        \begin{tabular}{lcccc}% {L{1.7cm}C{1cm}C{1cm}C{1cm}C{1cm} }
        \hline
         Covariate & Cox & CIF  & RRF  & TSF  \\
        % \multicolumn{5}{l}{\textit{Proportional hazards setting}}\\
        \hline
        % Errors & 0.4745 & 0.4620 & 0.5073 & 0.4675\\
        Time-invariant & 0.1245 & 0.1354 & 0.1302 & 0.1371 \\
        Time-varying$^\ast$ & 0.0984 & 0.0952 & 0.1049 & 0.0961 \\
        \hline
        \multicolumn{5}{l}{$^{\ast}${\scriptsize For time-varying covariate data, the results are shown for the extended Cox}}\\
        \multicolumn{5}{l}{{\scriptsize $\;\;$model, CIF-TV, RRF-TV and TSF-TV, respectively.}}
        \end{tabular}
        \label{tbl:PBC_IBSCVerr}
\end{table}

Figure \ref{fig:realset} shows that the cross-validation errors from all methods on the 
time-varying covariate dataset are lower than the corresponding ones in the time-invariant 
covariate dataset after $t > 300$, which suggests that the updated covariate information 
can significantly improve performance. Given the results on the time-varying covariate 
data in Figure \ref{fig:realset}, one can see that the differences in performance are 
relatively small. The extended Cox model slightly outperforms the others before $t = 3000$, 
where around 30\% of the subjects are still at risk. CIF-TV and TSF-TV give the best performance 
between $t = 3000$ and $t = 4200$. After $t = 4200$, since the results are based on very few 
failures (given only 10\% of subjects are still at risk), the Brier scores are highly variable 
and not as reliable. 
Note that the Brier score results also suggest that the survival relationship between 
the hazards and covariates is relatively (log-)linear for $t < 3000$, but not for $t > 3000$. 
Based on these Brier score-based cross-validation results, we would recommend an extended Cox model 
for $0 < t < 3000$, and CIF-TV for $t > 3000$. Table \ref{tbl:PBC_IBSCVerr} shows that CIF-TV 
gives the lowest corresponding integrated Brier score cross-validation errors, suggesting 
that CIF-TV provides the best estimated accuracy overall. In practice, where the underlying 
true survival distribution may possess a complex structure with time-varying features, 
a plot of the cross-validation-based Brier scores as in Figure \ref{fig:realset} can 
potentially help data analysts make better choices for survival estimation at different 
time points, compared to a universal choice of method for all time points.

\section{Discussion}
The estimation of a population-level survival probabilities for time-varying covariate data 
are useful in many survival analysis settings, causal inference research being an important 
example. Additional challenges arise when considering time-varying treatment regimens.  
To draw real-world evidence about the effectiveness of such regimens on patient survival, 
the key is to account for the time-varying confounding effects and one way to address 
this issue is by using the inverse probability of treatment weighting 
\citep{Barber2004causaltv,Robins2000causaltv}. For survival data, the estimation of 
the time-varying weights can potentially be improved by using flexible tree-based 
methods allowing time-varying covariates (confounders). A recent work \citep{Hu2022Causal} 
that uses LTRC forests shows that the use of more flexible models for the estimation of 
time-varying weights can lead to more accurate treatment effect estimation.

In this paper, we propose extending the relative risk, conditional inference, and 
transformation forests to provide survival function estimation with time-varying covariates. 
The extension of random survival forests \citep{RSF} for cumulative hazard function 
estimation with time-varying covariates based on log-rank-type splitting rules requires 
reformulation of the log-rank test statistics \citep{LeBlanc1993logrank,Segal1988logrank} 
and/or the log-rank scores \citep{HOTHORN2003logrankscore}. For a log-rank score rule, one can make use of the log-rank score specified in (\ref{eq:logrank_score_ltrc}). 
However, in this case one cannot construct the extension by modifying the custom splitting 
rule feature and employing the fast \textsf{C} code from \textsf{randomForestSRC}, 
since its basic structure only allows for right-censored survival. A fast and efficient 
algorithm to construct random survival forests for left-truncated right-censored 
survival data is needed, and is a goal of future work. 
 
\section{Conclusion}
In this paper, we have proposed two new ensemble algorithms, CIF-TV and RRF-TV, and adapted the transformation algorithm, TSF-TV. These three forest algorithms can handle (left-truncated) right-censored survival data with time-varying covariates and provide dynamic estimation.  
The tuning parameters in the proposed forest methods for survival data with time-varying covariates affect their overall performance. 
Guidance on how to choose those parameters is provided to improve on the potentially poor performance of forests with the default parameter settings. 

The estimation performance of the proposed forest methods is investigated to understand how the improvement over a Kaplan-Meier fit is related to changes in different factors.
Focusing on the more influential factors, the estimation performance comparison against other methods shows that the proposed forest methods outperform others under certain circumstances, while no method can dominate in all cases. 
We then provide guidance for choosing the modeling method in practice, showing that cross-validation is able to pick the best modeling method most of the time, or at least select a method that performs not much worse than the best method. 

The estimation of a population-level survival probabilities for time-varying 
covariate data are useful in many survival analysis settings, causal inference 
research being an important example. Additional challenges arise when considering 
time-varying treatment regimens.  To draw real-world evidence about the effectiveness 
of such regimens on patient survival, the key is to account for the time-varying 
confounding effects and one way to address this issue is by using the inverse probability 
of treatment weighting \citep{Barber2004causaltv,Robins2000causaltv}. For survival data, 
the estimation of the time-varying weights can potentially be improved by using flexible 
tree-based methods allowing time-varying covariates (confounders). A recent work 
\citep{Hu2022Causal} that uses the proposed survival forests shows that the use of more flexible models 
for the estimation of time-varying weights can lead to more accurate treatment effect estimation.

Our developed methodology and algorithms allow for estimation using the proposed forests for (left-truncated) right-censored data with time-invariant covariates.
The same data-driven guidance for tuning the parameters or selecting a modeling method also applies to the time-invariant covariates case (for both left-truncated right-censored survival data and right-censored survival data), which implies its broad effectiveness regardless of additional left-truncation and regardless of the presence of time-varying effects.

\section{Code availability}
R scripts for reproducibility of the simulations and real dataset illustrative example analysis are available from the github repository,
\url{https://github.com/WeichiYao/TimeVaryingData\_LTRCforests}. 

An \texttt{R} package, \textsf{LTRCforests}, which implements CIF-TV and RRF-TV for LTRC data with application to time-varying data, is available on CRAN.     

\appendices
\section{Dynamically Adjusted Survival Function for the Hypothetical COVID-19 Example}  
The time-to-event is the time to a positive COVID test. There is a time varying covariate, $X(t)$, which describes the vaccination status of a subject with values in $\{0,1,2,3\}$ where $0=$ unvaccinated, $1=$ vaccinated with a single dose and $2=$ vaccinated with two doses, and $3=$ vaccinated with two doses and a booster. 
The covariate may change values at discrete time points; in this example at times $1$ and $2$.

\begin{figure*}[t]
    \centering
    \includegraphics[scale=0.5]{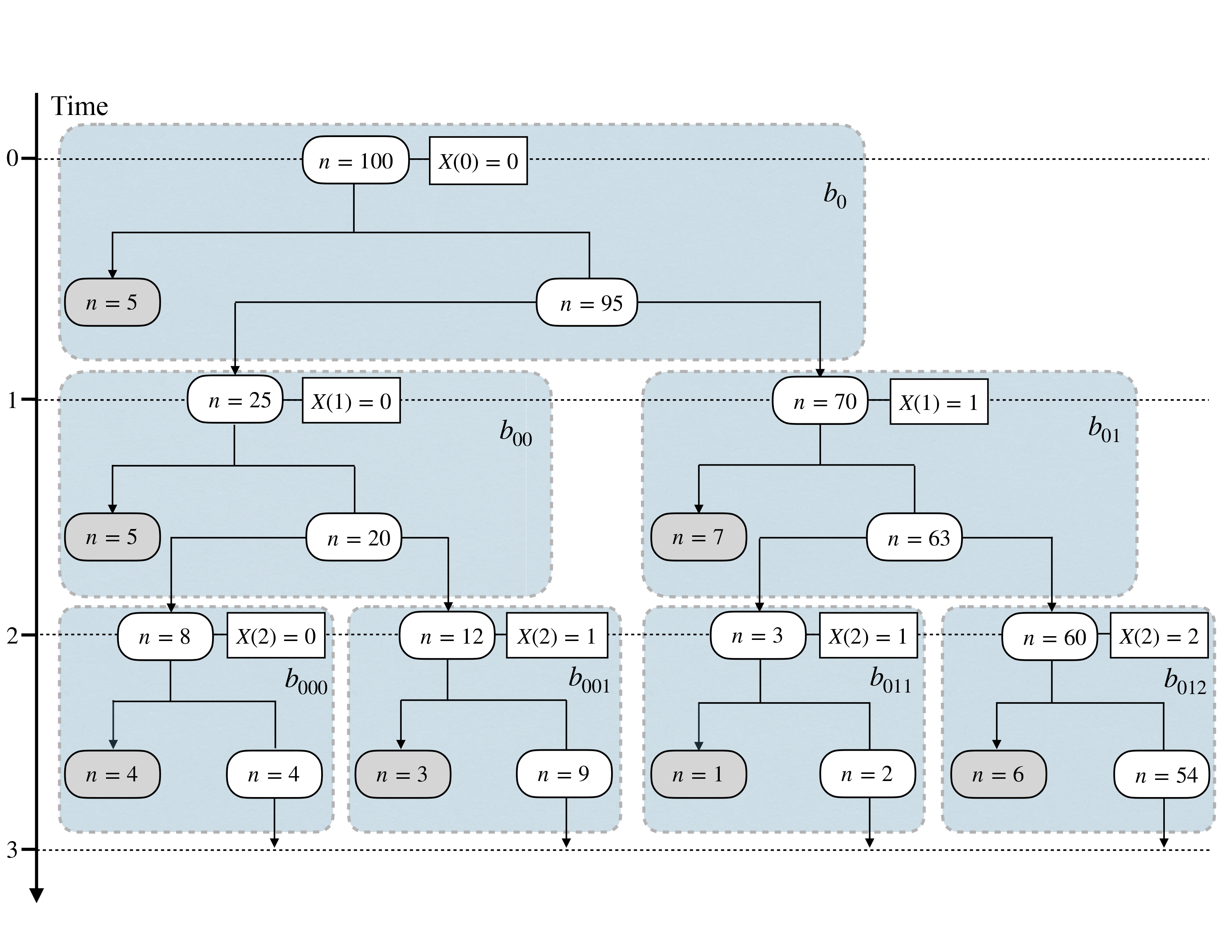}
    \caption{The hypothetical tree for COVID progression during $[0,3)$. The tree nodes are shown as ovals with values indicating the number of subjects in the corresponding group. 
    Darker nodes stand for the groups of subjects infected with COVID, and the lighter ones for COVID-free.
    At time $t=0$, $1$ and $2$, the updated vaccination status $X(t)$ are shown in the squares alongside with the nodes. 
    Each of the seven gray shaded areas corresponds to the group with a different vaccination status. For example, the $b_{0}$ area indicates the group unvaccinated in time interval $[0,1)$, $b_{01}$ the group vaccinated at time $1$, and $b_{012}$ the group vaccinated at times $t=1$ and $2$. Similar interpretations hold for the other gray areas. } \label{fig:covid_example}
\end{figure*}

We construct the hypothetical tree for COVID progression, see Figure \ref{fig:covid_example}. At time $t=0$, there is a sample of $n=100$ unvaccinated and COVID-free subjects at the root of the tree, i.e., every subject has $X(0)=0$. Thus, the estimated survival function at time $t=0$, $\widehat{S}(0)=1$. Each of the branches on the tree are labeled $b_{0}$, $b_{00}$, etc. The initial branch is $b_0$; it represents a subject not vaccinated in the time interval $[0,1)$. Similarly, branch $b_{00}$ represents a subject unvaccinated over the time interval $[0,2)$; branch $b_{01}$, a subject vaccinated with at single dose at time $t=1$; $b_{011}$, a subject vaccinated with a single dose at time $t=1$ and not receiving second dose at time $t=2$, etc. For simplicity, we will assume that there is no censoring. 

At time $t=1$, there are five subjects infected with COVID and 95 COVID-free. Hence, the estimated survival function at time $t=1$ is 
\begin{align*}
    \widehat{S}(1)=\widehat{S}(1\mid X(u)=0,\,0\leq u<1)=\widehat{S}(1\mid b_0)=0.95.
\end{align*}

Of the 95 COVID-free subjects at time $t=1$, 70 get their first dose of vaccine and 25 remain unvaccinated. At time $t=2$ of the 25 unvaccinated subjects, 20 are COVID-free and of 70 subjects who received a single vaccine, 63 are COVID-free. Thus, the estimated survival function at time $t=2$ on $b_{00}$ $(X(u)=0,\;0\leq u<2)$ is    
\begin{align*}
    \widehat{S}(2\mid b_{00})=\widehat{S}(1\mid b_0)(20/25)=0.95(0.8)=0.76,
\end{align*}
and on $b_{01}$ is
\begin{align*}
    \widehat{S}(2\mid b_{01})=\widehat{S}(1\mid b_0)(63/70)=0.95(0.9)=0.855.
\end{align*}

Of the 63 COVID-free vaccinated subjects at time $1$, 60 receive the second dose of vaccine at time $t=2$ and $3$ do not. Of those 60, 54 are COVID-free at time $t=3$. Thus, the estimated survival probability at $t=3$ for this group is 
\begin{align*}
    \widehat{S}(3\mid b_{012})=\widehat{S}(2\mid b_{01})(54/60)=0.855(0.9)=0.7695.
\end{align*}
For the group of three subjects with no second vaccine at time $t=2$, the estimated survival probability at time $t=3$ is lower: 
\begin{align*}
    \widehat{S}(3\mid b_{011})=\widehat{S}(2\mid b_{01})(2/3)=0.855(2/3)=0.57.
\end{align*}

At time $t=2$, of the 20 unvaccinated COVID-free subjects, 12 receive their first dose of vaccine and eight do not. Of those 12 subjects, nine are COVID-free at time $t=3$. For this group, which gets a single vaccine at time $t=2$, the estimated survival probability at $t=3$ is   
\begin{align*}
    \widehat{S}(3\mid b_{001})=\widehat{S}(2 \mid b_{00})(9/12)=0.76(0.75)=0.57,
\end{align*}
which happens in this example to be the same as the estimated survival probability of the group that got the single vaccine at time $t=1$.

Finally, for the non-vaccinated group ($b_{000}$), the estimated survival probability at time $t=3$ is the lowest: 
\begin{align*}
    \widehat{S}(3\mid b_{000})=\widehat{S}(2\mid b_{00})\left( 4/8\right) =0.76(0.5)=0.38.
\end{align*} 

The subjects who did or did not receive a booster at the next time period would be handled in a similar way.

\section{Derivation of the Survival Estimate} % \label{appendix:derivation_estimate}
Recall that by definition of survival functions, $S (t | \mathcal{X}^\ast(t) ) = \mathbb{P}(T>t | \mathcal{X}^\ast(t))$. At given time $t\in  [t_j^\ast, t_{j+1}^\ast )$, note that $\mathbb{P}(T>t | \mathcal{X}^\ast(t)) = \mathbb{P} (T>t, \;T> t_j^\ast\mid \mathcal{X}^\ast(t) )$, we apply the conditional probability and obtain 
\small
\begin{align}
    S (t | \mathcal{X}^\ast(t) )
    = \mathbb{P}\big(T>t\mid T> t_j^\ast,\;\mathcal{X}^\ast(t)\big)S\big(t_j^\ast\mid \mathcal{X}^\ast(t_j^\ast)\big).
    \label{eq:SurvEstimate_def}
\end{align}
\normalsize

In constructing the survival function estimate, we assume that the hazard at time $t$ is a function only of the current covariate values at time $t$ (but these covariates can include lagged values of some covariates). 
This allows us to construct the estimate at time $t$ using any subjects in the population with the specified value at that precise time point;
that is, we estimate $\mathbb{P}(T>t\mid T> t_j^\ast,\;\mathcal{X}^\ast(t))$ by computing
\small
\begin{align*}
    \widehat{\mathbb{P}}\big(T>t\mid T> t_j^\ast,\;\bm{x}_j^\ast\big)
    = \frac{\widehat{\mathbb{P}}\big(T>t\mid\bm{x}_j^\ast\big)}{\widehat{\mathbb{P}}\big(T>t_j^\ast\mid \bm{x}_j^\ast\big)},
\end{align*}
\normalsize
where both the numerator and the denominator are the values of the estimated survival function in the hypothetical case with the covariate $\bm{x}_j^\ast$ at $t$ and $t_j^\ast$, respectively. 
The risk sets that are used to compute these two quantities consider all subjects with covariate values $\bm{x}_j^\ast$ at $t_j^\ast$, regardless of their covariate paths before $t_j^\ast$. Note that this hypothetical estimated survival function is in fact the output of the  algorithm for the input with covariate value $\bm{x}_j^\ast$. 

Therefore, by substituting $\widehat{S}_{A,j}(t)\triangleq\widehat{\mathbb{P}}(T>t\mid \bm{x}_j^\ast)$, we can then approximate (\ref{eq:SurvEstimate_def}) as
\begin{align*}
    \widehat{S}(t \mid  \mathcal{X}^\ast(t) )=\frac{\widehat{S}_{A,j}(t)}{\widehat{S}_{A,j}(t^\ast)}\widehat{S}(t_j^\ast \mid  \mathcal{X}^\ast(t_j^\ast)),
\end{align*}
which gives us the formula in (\ref{eq:SurvEstimate_recursive}).

\bibliographystyle{chicago}
\bibliography{reference.bib}  



\section*{Supplementary material}

# Supplemental Material for "Ensemble Methods for Survival Function Estimation with Time-Varying Covariates"

Weichi Yao[1]*, Halina Frydman[1], Denis Larocque[2], and Jeffrey S. Simonoff[1]

*Correspondence: wyao@stern.nyu.edu
[1]Department of Technology, Operations, and Statistics, Stern School of
Business, New York University, New York City, USA
[2]Department of Decision Sciences, HEC Montréal, Montréal, CA

# Contents

# S1 Proposed forests for time-varying covariate survival data

## S1.1 Generating the survival times

Suppose the covariates $\boldsymbol{X}$ are initially observed at $t_0 = 0$, and change values at $t_1, \ldots, t_{m-1}$, before the subject is censored or the event occurs at $t_m$. That means there are in total $m$ intervals, within which the values of all covariates remain the same, $\boldsymbol{X}(t) = \boldsymbol{x}_i$, for $t \in [t_i, t_{i+1})$, $i = 0, 1, \ldots, m-1$. Consider the Weibull distribution with coefficient $(\lambda, v)$ and a survival relationship $\vartheta(t) = \vartheta(\boldsymbol{X}(t))$. Denote $\vartheta_i = \vartheta(\boldsymbol{x}_i)$, $i = 0, \ldots, m-1$.

The survival times are generated as follows:

1. Compute the values of cumulative hazard function $H(t)$ at $t_1, t_2, \ldots, t_{m-1}$ from

$$
H(t) = \begin{cases}
\lambda \exp(\vartheta_0) t^v & \text{if } 0 \leq t \leq t_1; \\
\lambda \exp(\vartheta_i) \left(t^v - t_i^v\right) + H(t_i, \vartheta_{i-1}) & \text{if } t_i < t \leq t_{i+1},\ i = 1, \ldots, m-2; \\
\lambda \exp(\vartheta_{m-1}) \left(t^v - t_{m-1}^v\right) + H(t_{m-1}, \vartheta_{m-2}) & \text{if } t_{m-1} < t.
\end{cases}
$$

2. Divide time lines into the following regions

$$
R_i = \begin{cases}
[H(t_{i-1}), H(t_i)] & i = 1, \ldots, m-1; \\
[H(t_{i-1}), \infty) & i = m.
\end{cases}
$$

3. Randomly generate $u \in \mathcal{U}(0, 1)$, the survival time is determined by which region $-\log(u)$ falls in:

$$
T = \begin{cases}
\left[\frac{-\log(u)}{\lambda \exp(\vartheta_0)}\right]^{1/v} & \text{if } -\log(u) \in R_1; \\
\left[\frac{-\log(u) - H(t_{i-1})}{\lambda \exp(\vartheta_{i-1})} + t_{i-1}^v\right]^{1/v} & \text{if } -\log(u) \in R_i,\ i = 2, \ldots, m.
\end{cases}
$$

4. The corresponding survival probability function is

$$
S(t) = \exp(-H(t)),
$$

where $t$ falls into mutually exclusive intervals $D_1 = [0, t_1), D_2 = [t_1, t_2), \ldots, D_m = [t_{m-1}, \infty)$.

## S1.2 Changing pattern of the time-varying covariates in the simulations

This section provide further details on the changing pattern of the time-varying covariates, specifically, on how to generate the observed values of time-varying covariates $X_6$, $X_{13}$, $X_{16}$, $X_{18}$, $X_{20}$ at $t_0, \ldots, t_{m-1}$: $x_{j,k}$, for $j = 0, \ldots, m-1$ and $k = 6, 13, 16, 18, 20$.

- $X_6$, whose initial value is randomly generated from $\{0, 1, 2\}$ with equal probability:

  - if the initial value is 2 then it stays at 2 for all the following time points: $x_{0,6} = \cdots = x_{m-1,6} = 2$;

- – if the initial value is less than 2, randomly sample the number of time points from $\{1, \ldots, m\}$ that it will stay at the initial value, $\tilde{n}_0$,
  - ∗ if the initial value is 1,
    - · if $\tilde{n}_0 < m$, then the values at the rest $m - \tilde{n}_0$ time points are all $2's$: $x_{0,6} = \cdots = x_{\tilde{n}_0-1,6} = 1$ and $x_{\tilde{n}_0,6} = \cdots = x_{m-1,6} = 2$;
    - · otherwise, it stays at 1 for all time points: $x_{0,6} = x_{1,6} = \cdots = x_{m-1,6} = 1$;
  - ∗ if the initial value is 0, randomly sample the number of time points from $\{1, \ldots, m - \tilde{n}_0\}$ that its value becomes 1, $\tilde{n}_1$,
    - · if $\tilde{n}_1 < m - \tilde{n}_0$, then its values at the rest $m - \tilde{n}_0 - \tilde{n}_1$ time points are $2's$: $x_{0,6} = \cdots = x_{\tilde{n}_0-1,6} = 0$, $x_{\tilde{n}_0,6} = \cdots = x_{\tilde{n}_0+\tilde{n}_1-1,6} = 1$ and $x_{\tilde{n}_0+\tilde{n}_1,6} = \cdots = x_{m-1,6} = 2$;
    - · otherwise its values at the rest $m - \tilde{n}_0$ time points are 1: $x_{0,6} = \cdots = x_{\tilde{n}_0-1,6} = 0$ and $x_{\tilde{n}_0,6} = \cdots = x_{m-1,6} = 1$.

- $X_{13}$, whose changing pattern is always $0 \to 1$: randomly sample the number of time points from $\{1, \ldots, m-1\}$ that it stays at the initial value 0, $\tilde{n}_0$, and it then changes value to 1 for the rest of the time points: $x_{0,13} = \cdots = x_{\tilde{n}_0-1,13} = 0$ and $x_{\tilde{n}_0,13} = \cdots = x_{m-1,13} = 1$;

- $X_{16}$, whose changing pattern is either $0 \to 1$ or $1 \to 2$: first randomly generated its initial value from $\text{Bern}(0.5)$, then randomly sample the number of time points from $\{1, \ldots, m-1\}$ that it stays at the initial value, $\tilde{n}_0$, and it then increases its value by 1 for the rest of the time points: $x_{0,16} = \cdots = x_{\tilde{n}_0-1,16}$ and $x_{\tilde{n}_0,16} = \cdots = x_{m-1,16} = x_{0,16} + 1$;

- $X_{18}$, whose changing pattern is $0 \to 1 \to 2$: randomly sample two numbers of time points from $\{1, \ldots, m-1\}$, $\tilde{n}_0$ and $\tilde{n}_1$, then $x_{0,18} = \cdots = x_{\tilde{n}_0-1,18} = 1$, $x_{\tilde{n}_0,18} = \cdots = x_{\tilde{n}_1-1,18} = 1$ and $x_{\tilde{n}_1,18} = \cdots = x_{m-1,18} = 2$;

- $X_{20}$, which is a linear function of the left-truncated time point of the interval with slope and intercept follows $\text{Unif}(0,1)$: first generate $c_1, c_2 \sim \text{Unif}(0,1)$, then $x_{j,20} = c_1 t_j + c_2$, $j = 0, \ldots, m-1$.

## S1.3 Histogram of simulated survival times

Histograms of survival times for typical samples with the number of subjects $N = 500$ in each scenario are provided in Figure S1.1.

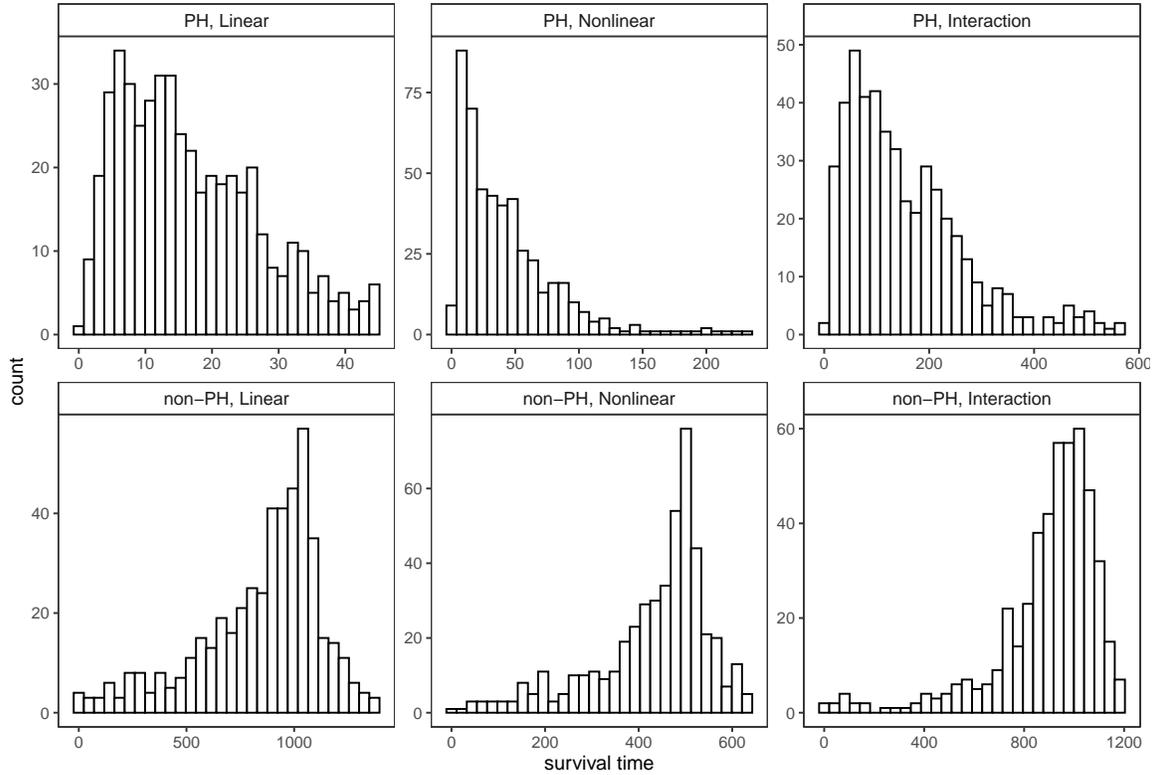

Figure S1.1: Histograms of the lower 95% of the survival times generated in typical samples with no right-censoring, with the number of subjects $N = 500$. The first row gives results under the PH setting, the second under the non-PH setting. The first column gives results for the linear survival relationship, second column for the nonlinear survival relationship, and the last column for the interaction survival relationship.

## S1.4 Parameters set in the simulation study

In the simulation study, we set the parameters for the basic scenario "2TI + 4TV" as follows.

- Under the PH setting:

  - When the survival relationship is linear: set $\lambda = 0.012$, $\nu = 0.8$ if the underlying survival distribution is Weibull-Decreasing, and $\lambda = 0.001$, $\nu = 2$ for Weibull-Increasing.

    * When the signal-to-noise ratio is "Low," $\beta_0 = 0$, $\beta_1 = 1$, $\beta_2 = 1$, $\beta_3 = -1$, $\beta_4 = -1$, $\beta_5 = 0.5$, $\beta_6 = -0.5$;
    * When the signal-to-noise ratio is "High," $\beta_0 = -4$, $\beta_1 = 5$, $\beta_2 = 5$, $\beta_3 = -5$, $\beta_4 = -5$, $\beta_5 = 2.5$, $\beta_6 = -2.5$;

  - When the survival relationship is nonlinear: set $\lambda = 0.15$, $\nu = 0.8$ if the underlying survival distribution is Weibull-Decreasing, and $\lambda = 0.0025$, $\nu = 1.8$ for Weibull-Increasing.

    * When the signal-to-noise ratio is "Low," $\phi_1 = \phi_4 = 0$, $\phi_2 = \phi_3 = -1$, $\psi_0 = 0.2$, $\psi_3 = 0.1$, $\psi_5 = 0.05$, $\psi_6 = 1$;
    * When the signal-to-noise ratio is "High," $\phi_1 = 0$, $\phi_2 = \phi_3 = -5$, $\phi_4 = 5$, $\psi_0 = 0.2$, $\psi_3 = 0.1$, $\psi_5 = 0.05$, $\psi_6 = 1$;

  - When the survival relationship is interaction, set $A = [0.7, 1]$, $B = \{4, 5\}$, and $\lambda = 0.14$, $\nu = 0.5$ if the underlying survival distribution is Weibull-Decreasing, $\lambda = 0.0001$, $\nu = 1.8$ for Weibull-Increasing.

    * When the signal-to-noise ratio is "Low," $\eta_1 = \eta_3 = 1$, $\eta_2 = \eta_4 = 0$, $\gamma_0 = 0$, $\gamma_1 = 1$, $\gamma_2 = 1$, $\gamma_3 = -1$, $\gamma_4 = -1$, $\gamma_5 = 0.5$, $\gamma_6 = -0.5$, $\alpha_0 = 0$, $\alpha_1 = -1$, $\alpha_2 = 1$, $\alpha_3 = 1$, $\alpha_4 = -1$, $\alpha_5 = -0.5$, $\alpha_6 = -0.5$;
    * When the signal-to-noise ratio is "High," $\eta_1 = \eta_3 = 5$, $\eta_2 = 1.5$, $\eta_4 = 1.101$, $\gamma_0 = -7$, $\gamma_1 = 5$, $\gamma_2 = 5$, $\gamma_3 = -5$, $\gamma_4 = -5$, $\gamma_5 = 2.5$, $\gamma_6 = -2.5$, $\alpha_0 = -10$, $\alpha_1 = -5$, $\alpha_2 = 5$, $\alpha_3 = 5$, $\alpha_4 = -5$, $\alpha_5 = -2.5$, $\alpha_6 = -2.5$;

- Under the non-PH setting:

  - When the survival relationship is linear, set $\lambda = 0.001$ for Weibull-Increasing.

    * When the signal-to-noise ratio is "Low," $\beta_0 = 0$, $\beta_1 = 1$, $\beta_2 = 1$, $\beta_3 = 1$, $\beta_4 = 1$, $\beta_5 = 10$, $\beta_6 = 1$;
    * When the signal-to-noise ratio is "High," $\beta_0 = -64.5$, $\beta_1 = 3$, $\beta_2 = 3$, $\beta_3 = 3$, $\beta_4 = 3$, $\beta_5 = 30$, $\beta_6 = 3$;

  - When the survival relationship is nonlinear, set $\lambda = 0.002$ for Weibull-Increasing.

    * When the signal-to-noise ratio is "Low," $\phi_1 = 1$, $\phi_2 = \phi_3 = \phi_4 = 0$;
    * When the signal-to-noise ratio is "High," $\phi_1 = 5$, $\phi_2 = \phi_3 = 0$, $\phi_4 = -2.835$;

  - When the survival relationship is interaction, set $A = [0.7, 1]$, $B = \{5\}$, and $\lambda = 0.001$ for Weibull-Increasing as the underlying survival function.

    * When the signal-to-noise ratio is "Low," $\eta_1 = \eta_3 = 1$, $\eta_2 = \eta_4 = 0$, $\gamma_0 = 0$, $\gamma_1 = -1$, $\gamma_2 = -1$, $\gamma_3 = -1$, $\gamma_4 = -1$, $\gamma_5 = -10$, $\gamma_6 = -1$, $\alpha_0 = 0$, $\alpha_1 = 1$, $\alpha_2 = 1$, $\alpha_3 = 1$, $\alpha_4 = 1$, $\alpha_5 = 10$, $\alpha_6 = 1$;
    * When the signal-to-noise ratio is "High," $\eta_1 = \eta_3 = 5$, $\eta_2 = 0.6$, $\eta_4 = 1.101$, $\gamma_0 = 64.5$, $\gamma_1 = -3$, $\gamma_2 = -3$, $\gamma_3 = -3$, $\gamma_4 = -3$, $\gamma_5 = -30$, $\gamma_6 = -3$, $\alpha_0 = -64.5$, $\alpha_1 = 3$, $\alpha_2 = 3$, $\alpha_3 = 3$, $\alpha_4 = 3$, $\alpha_5 = 30$, $\alpha_6 = 3$.

## S1.5 Bootstrapping subjects vs. bootstrapping pseudo-subjects

Figures S1.2 and S1.3 give side-by-side boxplots of integrated $L_2$ difference, showing the performance comparison between bootstrapping pseudo-subjects and bootstrapping subjects for each type of the forests under the PH setting, and under the non-PH setting, respectively. The results are provided for data generated under the true model 2TI + 4TV.

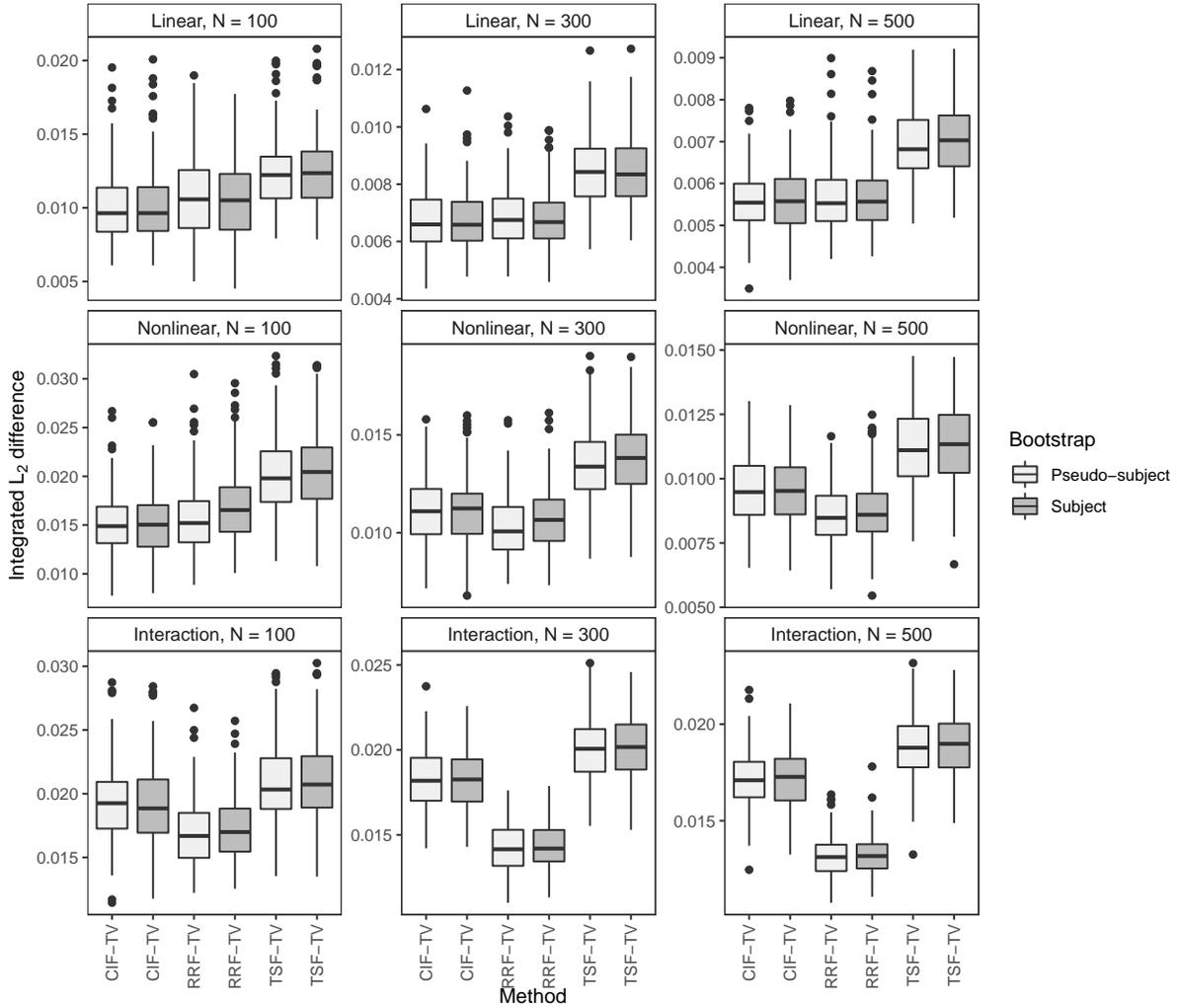

Figure S1.2: Integrated $L_2$ difference of three forest methods with different bootstrap mechanisms, on datasets having 2TI + 4TV in the true model, survival times generated from a Weibull-Increasing distribution under the PH setting, light right-censoring rate (20%). All forest methods are trained with $mtry = 5$ by default, and tuning parameters set to be $\sqrt{n}$. From the top row to the bottom, it gives results for the linear, nonlinear and the interaction survival relationship. From the first column to the third, it gives results for the number of subjects $N = 100, 300, 500$.

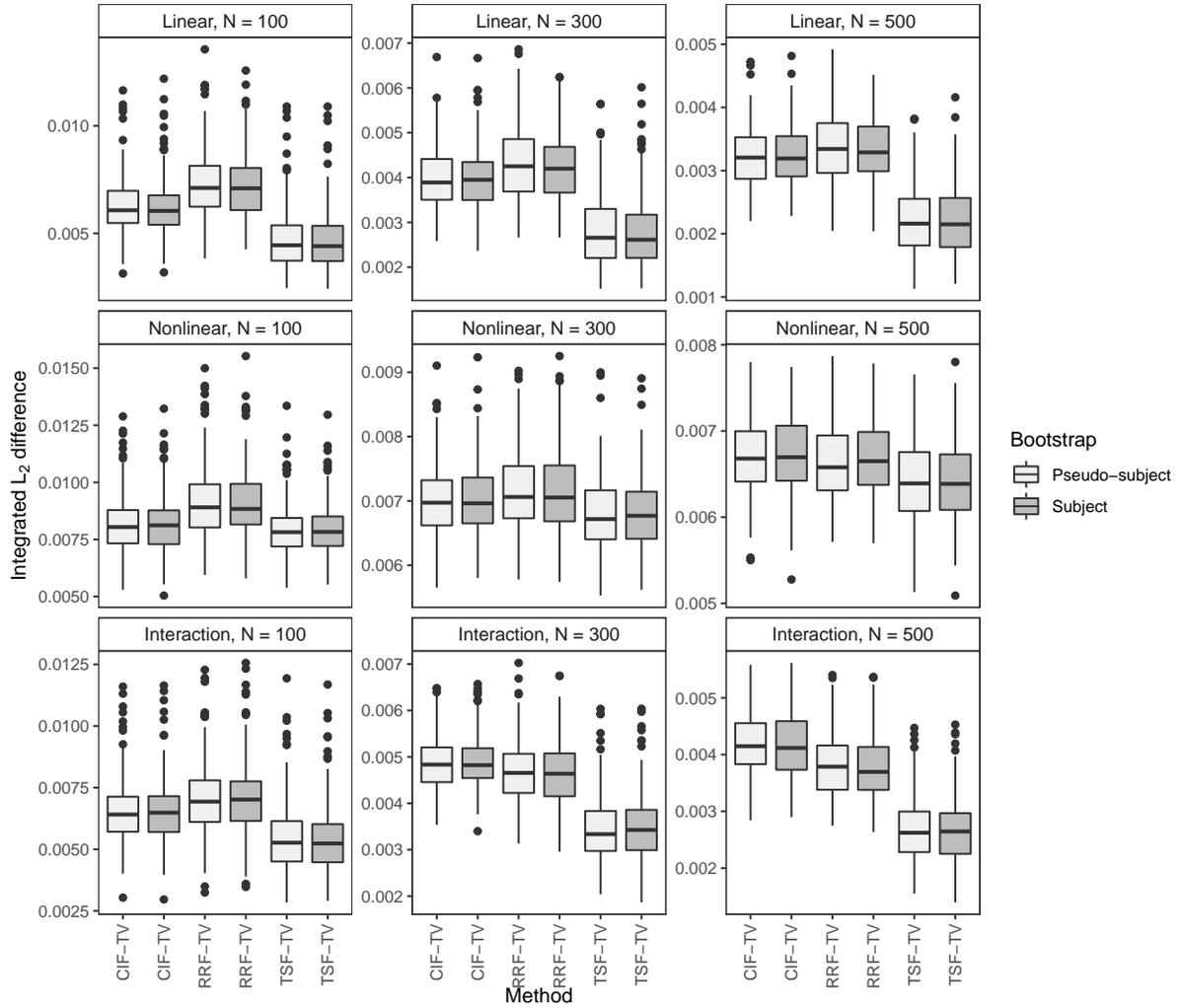

Figure S1.3: Integrated $L_2$ difference of three forest methods with different bootstrap mechanisms, on datasets having 2TI + 4TV in the true model, survival times generated from a Weibull-Increasing distribution under the non-PH setting, light right-censoring rate (20%). All forest methods are trained with $mtry = 5$ by default, and tuning parameters set to be $\sqrt{n}$. From the top row to the bottom, it gives results for the linear, nonlinear and the interaction survival relationship. From the first column to the third, it gives results for the number of subjects $N = 100, 300, 500$.

One can see that for each type of forest, the results when bootstrapping pseudo-subjects are very similar to those when bootstrapping the subjects. That is, the two different bootstrapping mechanisms do not result in fundamentally different levels of performance.

## S1.6   Regulating the construction of trees in forests

Figures S1.4 to S1.7 give examples of how RRF-TV and TSF-TV perform with different values of *mtry* under the PH setting and the non-PH setting, respectively. The *mtry* values are chosen by the "out-of-bag" tuning procedure. The results are very similar compared to those for CIF-TV given in the manuscript.

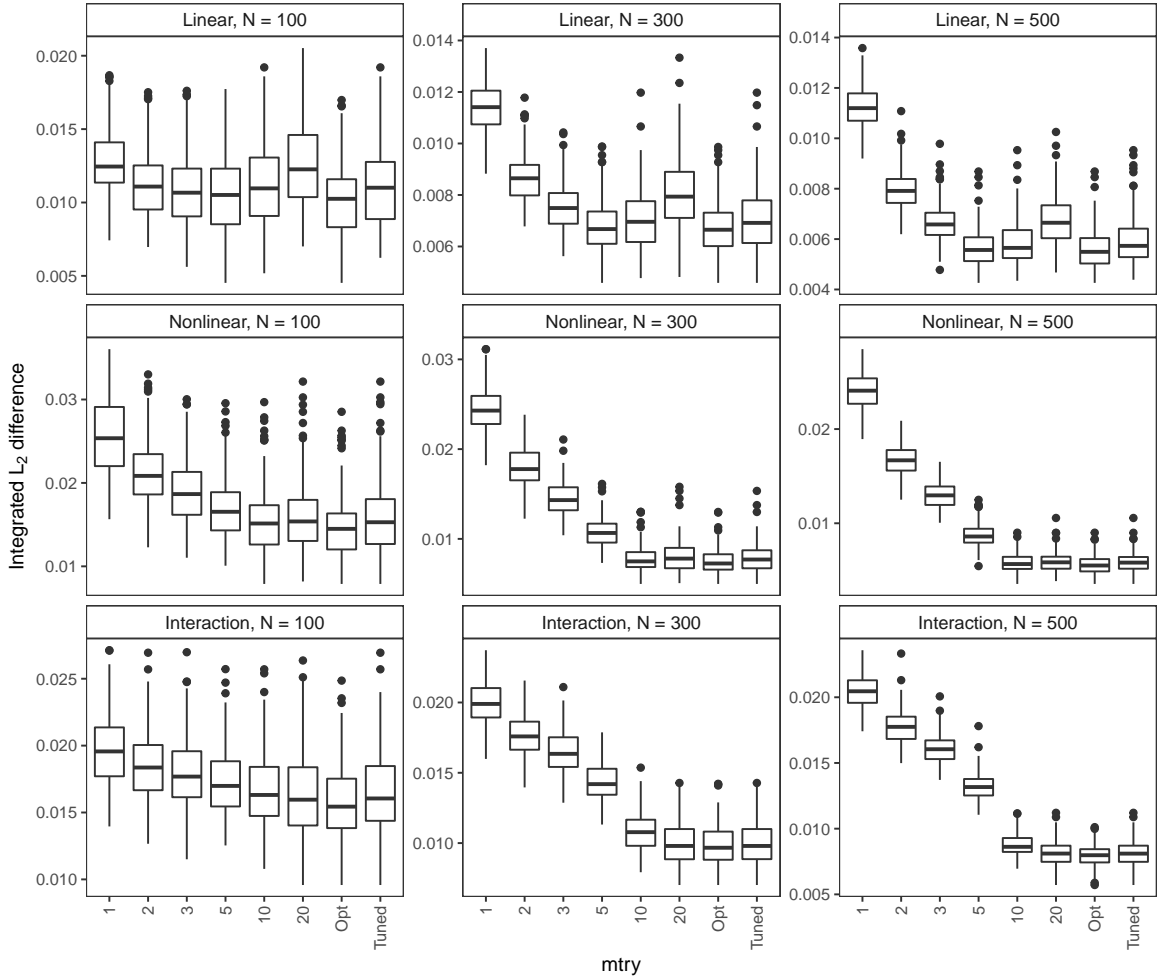

Figure S1.4: Integrated $L_2$ difference of RRF-TV with different *mtry* values under the PH setting. Datasets are generated with light right-censoring rate (20%), survival times generated from a Weibull-Increasing distribution. From the top row to the bottom, it gives results for the linear, nonlinear and the interaction survival relationship. From the first column to the last, it gives results for the number of subjects $N = 100, 300, 500$. In each plot, 1–RRF-TV with $mtry = 1$; 2–RRF-TV with $mtry = 2$; 3–RRF-TV with $mtry = 3$; 5–RRF-TV with $mtry = 5$; 10–RRF-TV with $mtry = 10$; 20–RRF-TV with $mtry = 20$; Opt–RRF-TV with value of $mtry$ that gives the smallest Integrated $L_2$ difference in each round; Tuned–RRF-TV with the value of $mtry$ tuned by the "out-of-bag" tuning procedure.

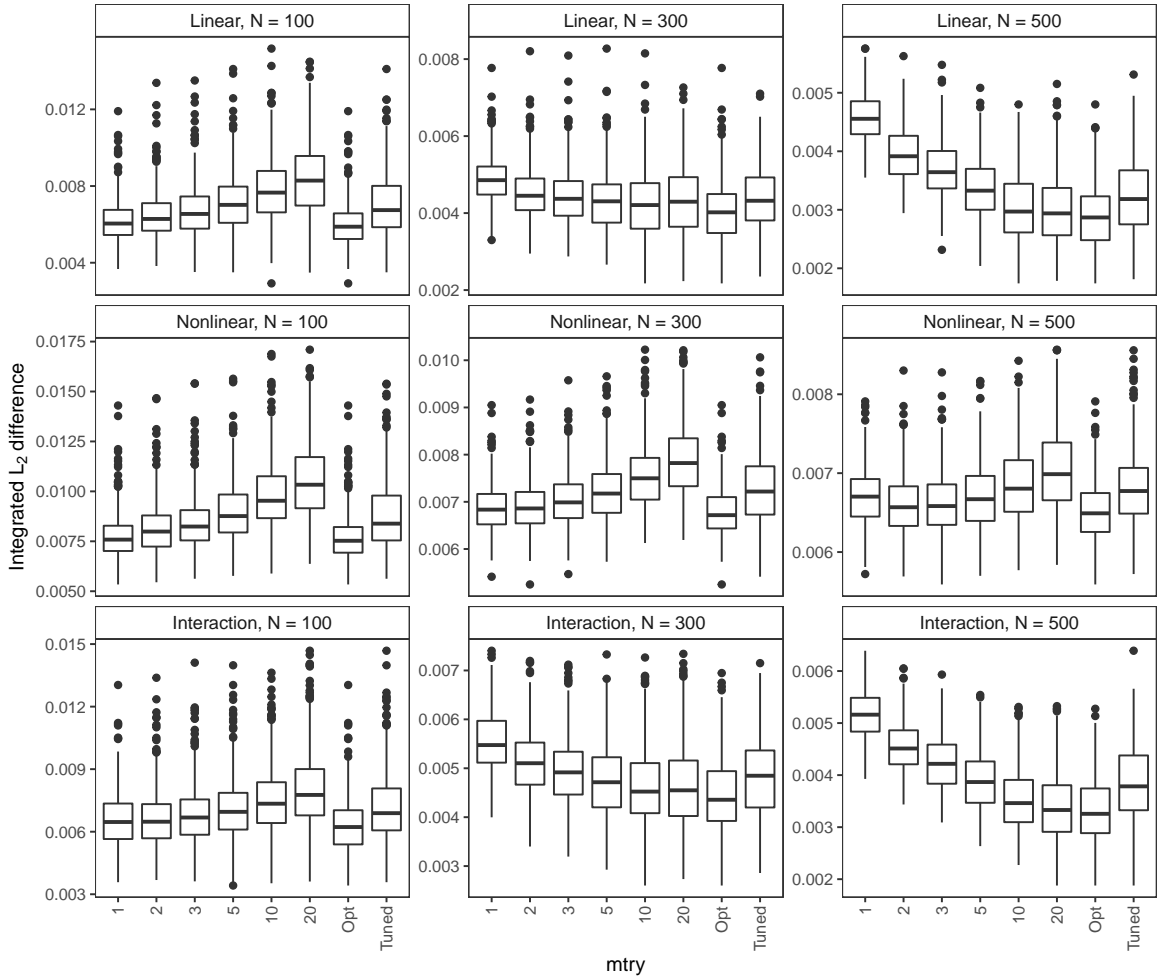

Figure S1.5: Integrated $L_2$ difference of RRF-TV with different *mtry* values under the non-PH setting. Datasets are generated with light right-censoring rate (20%), survival times generated from a Weibull-Increasing distribution. From the top row to the bottom, it gives results for the linear, nonlinear and the interaction survival relationship. From the first column to the last, it gives results for the number of subjects $N = 100, 300, 500$. In each plot, 1–RRF-TV with *mtry* = 1; 2–RRF-TV with *mtry* = 2; 3–RRF-TV with *mtry* = 3; 5–RRF-TV with *mtry* = 5; 10–RRF-TV with *mtry* = 10; 20–RRF-TV with *mtry* = 20; Opt–RRF-TV with value of *mtry* that gives the smallest Integrated $L_2$ difference in each round; Tuned–RRF-TV with the value of *mtry* tuned by the "out-of-bag" tuning procedure.

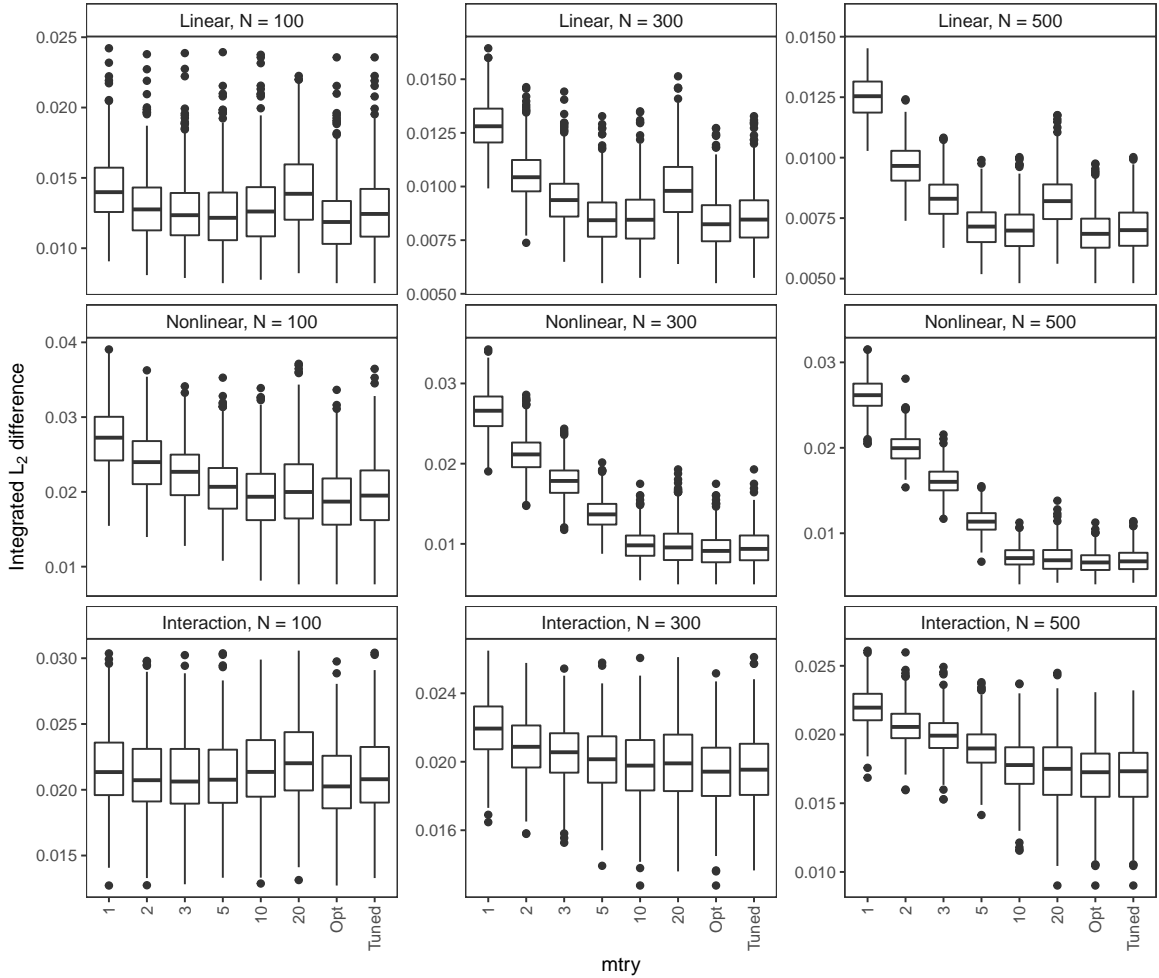

Figure S1.6: Integrated $L_2$ difference of TSF-TV with different *mtry* values under the PH setting. Datasets are generated with light right-censoring rate (20%), survival times generated from a Weibull-Increasing distribution. From the top row to the bottom, it gives results for the linear, nonlinear and the interaction survival relationship. From the first column to the last, it gives results for the number of subjects $N = 100, 300, 500$. In each plot, 1–TSF-TV with $mtry = 1$; 2–TSF-TV with $mtry = 2$; 3–TSF-TV with $mtry = 3$; 5–TSF-TV with $mtry = 5$; 10–TSF-TV with $mtry = 10$; 20–TSF-TV with $mtry = 20$; Opt–TSF-TV with value of *mtry* that gives the smallest Integrated $L_2$ difference in each round; Tuned–TSF-TV with the value of *mtry* tuned by the "out-of-bag" tuning procedure.

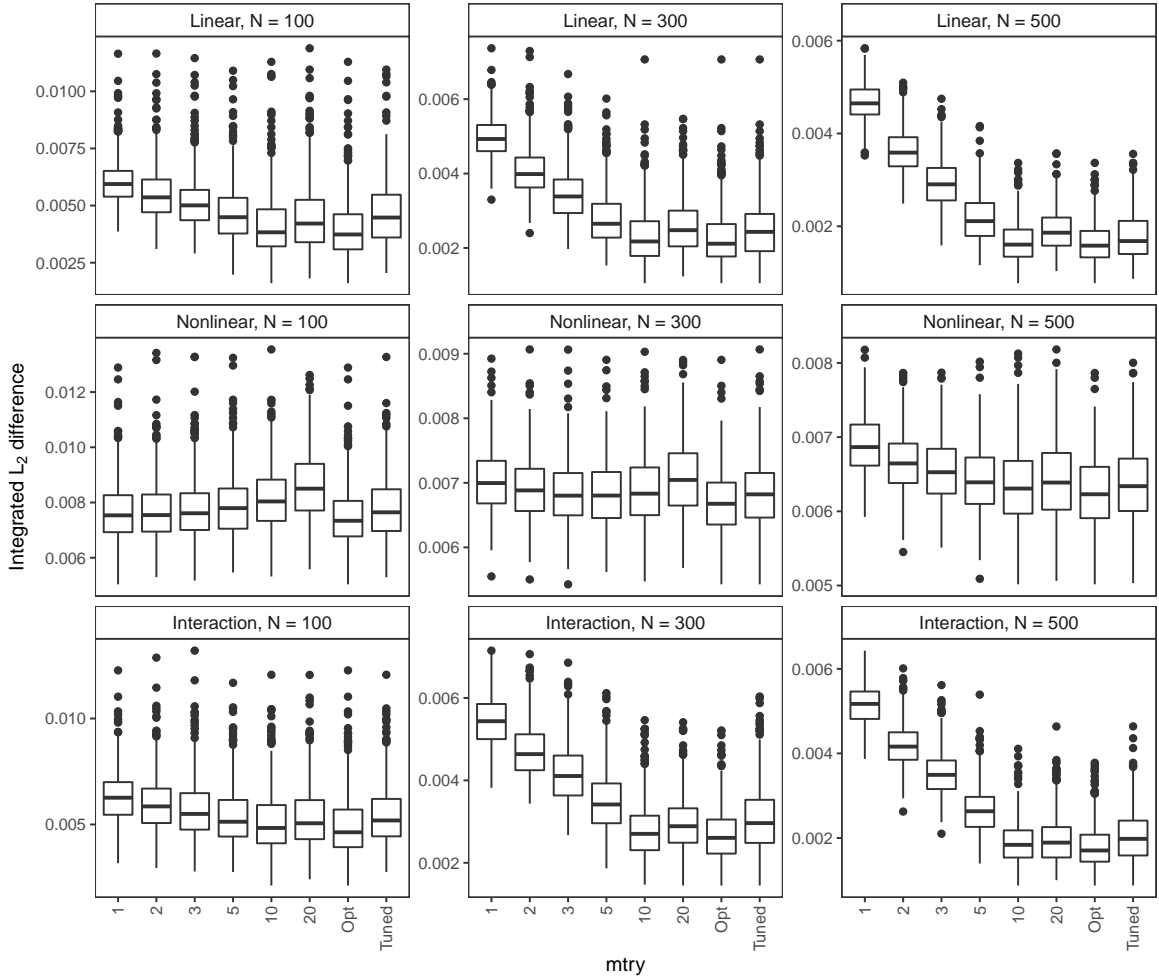

Figure S1.7: Integrated $L_2$ difference of TSF-TV with different *mtry* values under the non-PH setting. Datasets are generated with light right-censoring rate (20%), survival times generated from a Weibull-Increasing distribution. From the top row to the bottom, it gives results for the linear, nonlinear and the interaction survival relationship. From the first column to the last, it gives results for the number of subjects $N = 100, 300, 500$. In each plot, 1–TSF-TV with $mtry = 1$; 2–TSF-TV with $mtry = 2$; 3–TSF-TV with $mtry = 3$; 5–TSF-TV with $mtry = 5$; 10–TSF-TV with $mtry = 10$; 20–TSF-TV with $mtry = 20$; Opt–TSF-TV with value of *mtry* that gives the smallest Integrated $L_2$ difference in each round; Tuned–TSF-TV with the value of *mtry* tuned by the "out-of-bag" tuning procedure.

Table S1.1 gives the performance comparison between each forest method with its default parameter settings and with the proposed parameter settings for the nonlinear survival relationship under the PH setting and the non-PH setting, respectively. Table S1.2 gives the performance comparison between each forest method with its default parameter settings and with the proposed parameter settings for the interaction survival relationship under the PH setting and the non-PH setting, respectively.

Figure S1.8 shows the effect of different numbers of trees for bootstrap samples on CIF-TV, RRF-TV and TSF-TV. The number of trees increases from 100 to 500. The results show that the improvement in integrated $L_2$ difference from more trees is negligible at 100.

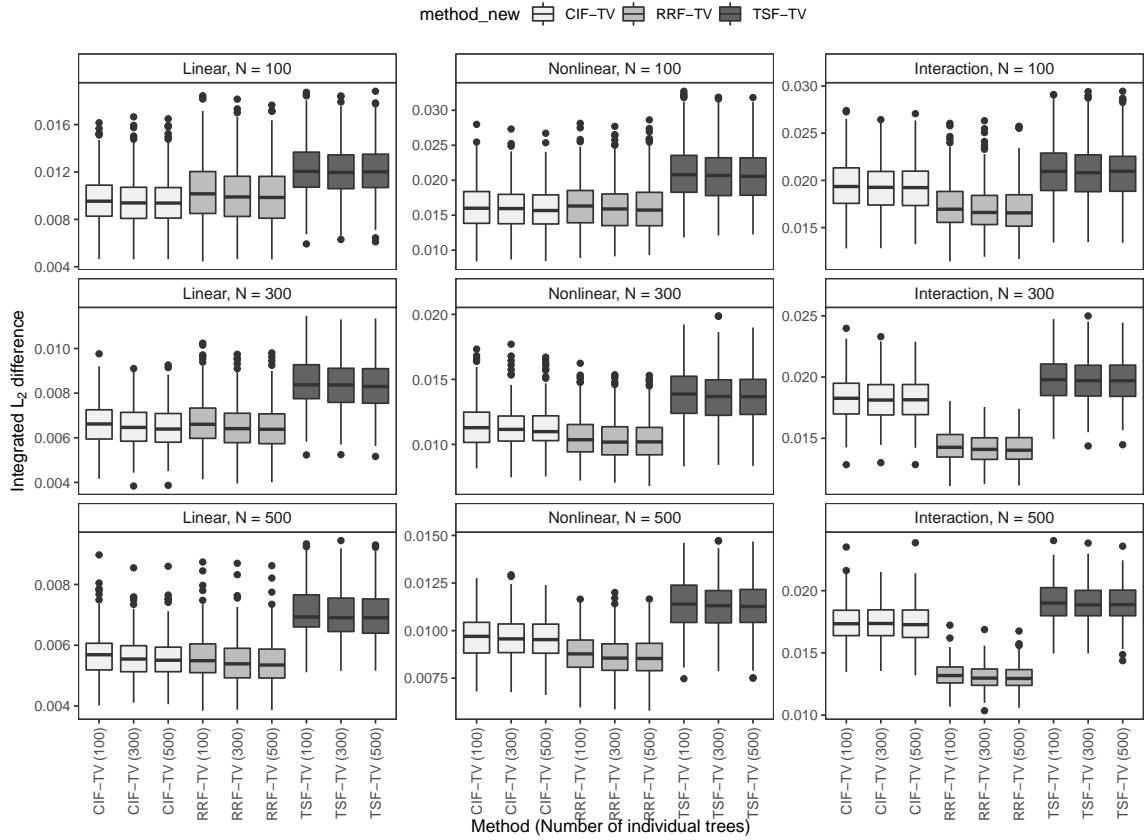

Figure S1.8: Effects of different numbers of trees for bootstrap samples on CIF-TV, RRF-TV and TSF-TV. Datasets are generated with survival times following a Weibull-Increasing distribution, light right-censoring rate 20% under the PH setting. The first column shows results for the number of subjects $N = 100$, second column for $N = 300$, bottom column for $N = 500$; the top row shows results for linear survival relationship, middle row for nonlinear, the bottom for interaction. In each of the plots, the set of boxplots lightly shaded shows the performance of CIF-TV with number of trees $100, 300, 500$ from left to right; the set moderately shaded shows the performance of RRF-TV with number of trees $100, 300, 500$ from left to right; the set heavily shaded shows the performance of TSF-TV with number of trees $100, 300, 500$ from left to right.

Figure S1.9 gives side-by-side integrated $L_2$ difference boxplots on datasets with survival times generated following a Weibull-Decreasing distribution.

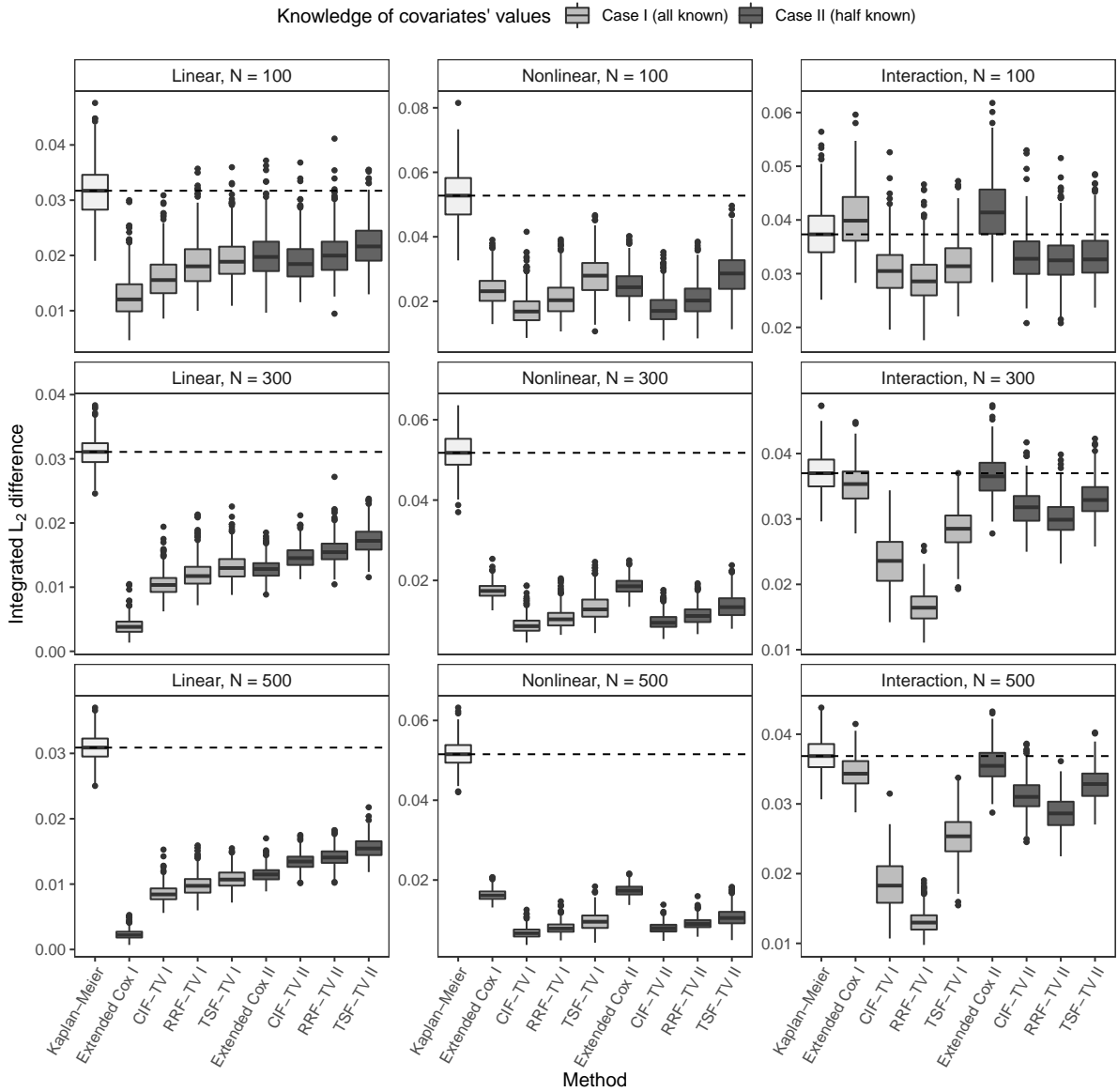

Figure S1.9: Boxplots of integrated $L_2$ difference for performance comparison under the PH setting. Datasets are generated with survival times following a Weibull-Decreasing distribution, light right-censoring rate 20%. The first row shows results for the number of subjects $N = 100$, second row for $N = 300$, bottom row for $N = 500$; the first column shows results for linear survival relationship, second column for nonlinear, the third column for interaction. The horizontal dashed line shows the median integrated $L_2$ difference of a Kaplan-Meier fit on the datasets. In each of the plots, the set of boxplots lightly shaded shows the performance of different methods on datasets with history of changes in covariates' values known; the set heavily shaded shows the performance on datasets with part of history of changes in covariates' values unknown.

## S1.8 Analysis of variance

Table S1.3: ANOVA table for comparing group means of CIF-TV/RRF-TV improvement over a simple Kaplan-Meier fit.

| Response: $(L_2(\text{Method}) - L_2(\text{KM}))/L_2(\text{KM})$ | | | | | | |
|---|---|---|---|---|---|---|
| Method: | CIF-TV | | | RRF-TV | | |
| | Sum Sq | Df | Pr(>F) | Sum Sq | Df | Pr(>F) |
| Censoring rate | 0.1849 | 1 | 0.0046 | 0.2443 | 1 | 0.0015 |
| Knowledge | 1.0512 | 1 | < 0.0001 | 0.8215 | 1 | < 0.0001 |
| Relationship | 1.6353 | 2 | < 0.0001 | 0.3697 | 2 | 0.0005 |
| Sample size | 1.5899 | 2 | < 0.0001 | 3.1521 | 2 | < 0.0001 |
| Scenario | 0.4215 | 1 | < 0.0001 | 0.4597 | 1 | < 0.0001 |
| Setting | 10.6751 | 1 | < 0.0001 | 15.9043 | 1 | < 0.0001 |
| SNR | 0.0000 | 1 | 0.9729 | 0.0633 | 1 | 0.1039 |

# S2 Proposed forests for time-invariant covariate survival data

The section is organized as follows. Section S2.1 describes the model setup, Section S2.2 provides the performance evaluation of the "out-of-bag" tuning procedure for *mtry* and gives comparative performance between the proposed parameter settings and the default settings. The overall performance comparison among the Cox model, the three forest methods with the proposed parameter settings, as well as the best method and the method selected by the IBS-based 10-fold CV rule are given in Section S2.3. All results with integrated $L_2$ difference are computed with $\tau_i = \widetilde{T}_i$. The comparative performance of the different methods was broadly similar when using $\tau_i = \widetilde{\tau} = \max_j \widetilde{T}_j$.

Here we use the proposed forest methods for left-truncated right-censored survival data with time-invariant covariates, the proposed forest methods and the transformation forest method are referred to as LTRC CIF, LTRC RRF and LTRC TSF, respectively.

## S2.1 Model setup

We generate left-truncated right-censored survival time data with time-invariant covariates as follows. The left-truncation time $T_0$ is generated as a $\mathcal{U}(0, L)$ random variable with some constant $L > 0$. The event time $T$ is randomly generated with a Weibull distribution. If the generated $T < T_0$, i.e. the event time is less than the left-truncation time, then this observation is discarded. Otherwise, the observation is retained, with censoring time $C = T_0 + D$, where $D \sim weibull(\text{shape} = 2, \text{scale} = \lambda_D)$ has an weibull distribution. The parameter $\lambda_D$ is selected to ensure 20% censoring rate. If $C < T$, then this observation is censored ($\Delta = 0$), otherwise the survival time $T$ is observed ($\Delta = 1$). Note that $D$ and $T_0$ are both independently generated from $T$ and from each other. The observed response for each observation is a triplet $(T_0, \widetilde{T}, \Delta)$, where $Y = \min(T, C)$.

There are 20 covariates in total, with the first six determining the survival times. $X_1, X_3 \sim$ Bern(0.5), $X_2, X_4 \sim$ Unif(0, 1), $X_5$ follows a categorical distribution with possible outcomes $\{1, 2, 3, 4, 5\}$ with equal probability, $X_6$ follows a categorical distribution with possible outcomes $\{0, 1, 2\}$ with equal probability. Among the rest of the 14 noise covariates, $X_7$, $X_{10}$, $X_{15}$, $X_{17}$, $X_{20} \sim$ Unif(0, 1), $X_8 \sim$ Unif(1, 2), $X_{11}$, $X_{13}$, $X_{16}$, $X_{19} \sim$ Bern(0.5), $X_{12}$ and $X_{18}$ both follow a categorical distribution with possible outcomes $\{0, 1, 2\}$ with equal probability, $X_9$ and $X_{14}$ both follow a categorical distribution with possible outcomes $\{1, 2, 3, 4, 5\}$ with equal probability. The survival times generating schemes with linear, nonlinear and interaction survival relationships are the same as in the time-varying cases, described in Section 3.2 in the manuscript.

## S2.2  Regulating the construction of trees in forests

In the simulations the value of *mtry* is tuned based on the "out-of-bag observations," and the values of *minprob* and *minbucket* are set to be the maximum of the default value and the square root of the number of pseudo-subject observations $n$.

Figures S2.1 and S2.2 show how LTRC CIF performs with different values of *mtry* under the PH setting and non-PH setting, respectively. The datasets are generated with survival times following a Weibull-Increasing distribution, light (right-)censoring rate. The results for LTRC RRF and LTRC TSF are similar, as given in Figures S2.3 to S2.6.

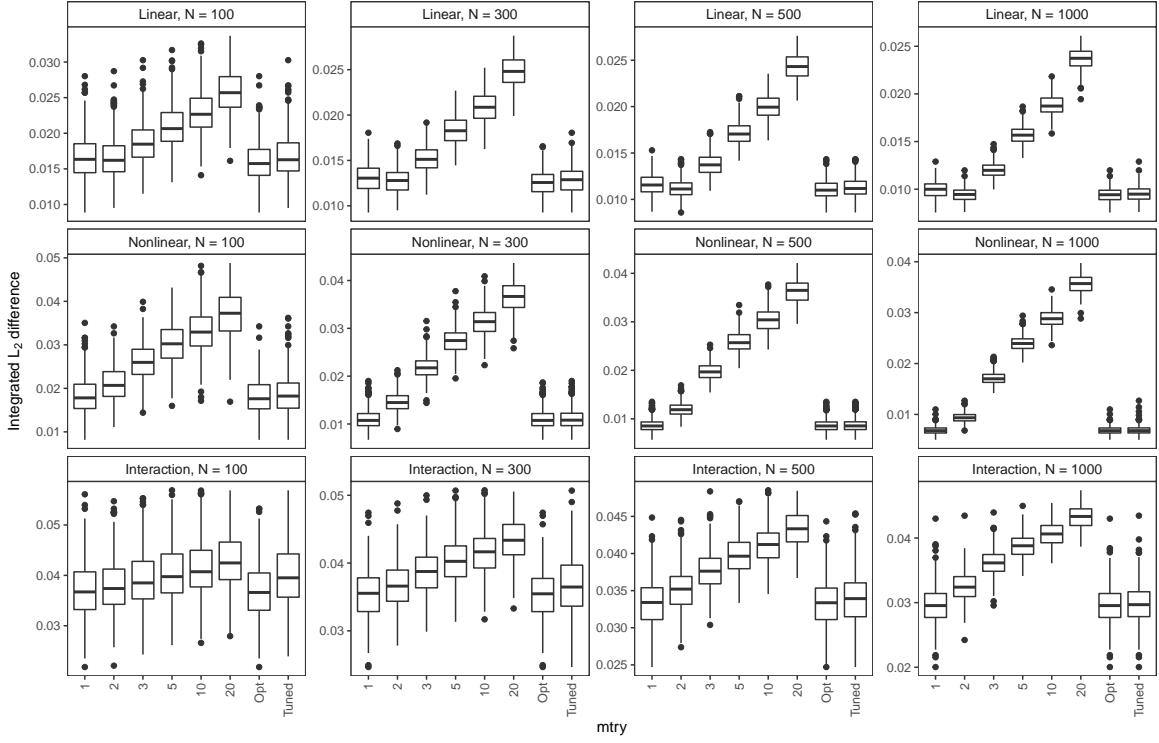

Figure S2.1: Integrated $L_2$ difference of LTRC CIF with different *mtry* values distribution under the PH setting. Datasets are generated with time-invariant covariates, light right-censoring rate (20%), left-truncated and right-censored survival times following a Weibull-Increasing. From the top row to the bottom, are given results for the linear, nonlinear and interaction survival relationship. From the first column to the last, are given results for the number of subjects $N = 100, 300, 500, 1000$. In each plot, 1–LTRC CIF with *mtry* = 1; 2–LTRC CIF with *mtry* = 2; 3–LTRC CIF with *mtry* = 3; 5–LTRC CIF with *mtry* = 5; 10–LTRC CIF with *mtry* = 10; 20–LTRC CIF with *mtry* = 20; Opt–LTRC CIF with value of *mtry* that gives the smallest Integrated $L_2$ difference in each round; Tuned–LTRC CIF with the value of *mtry* tuned by the "out-of-bag" tuning procedure. The default value in LTRC CIF is $mtry = 5$.

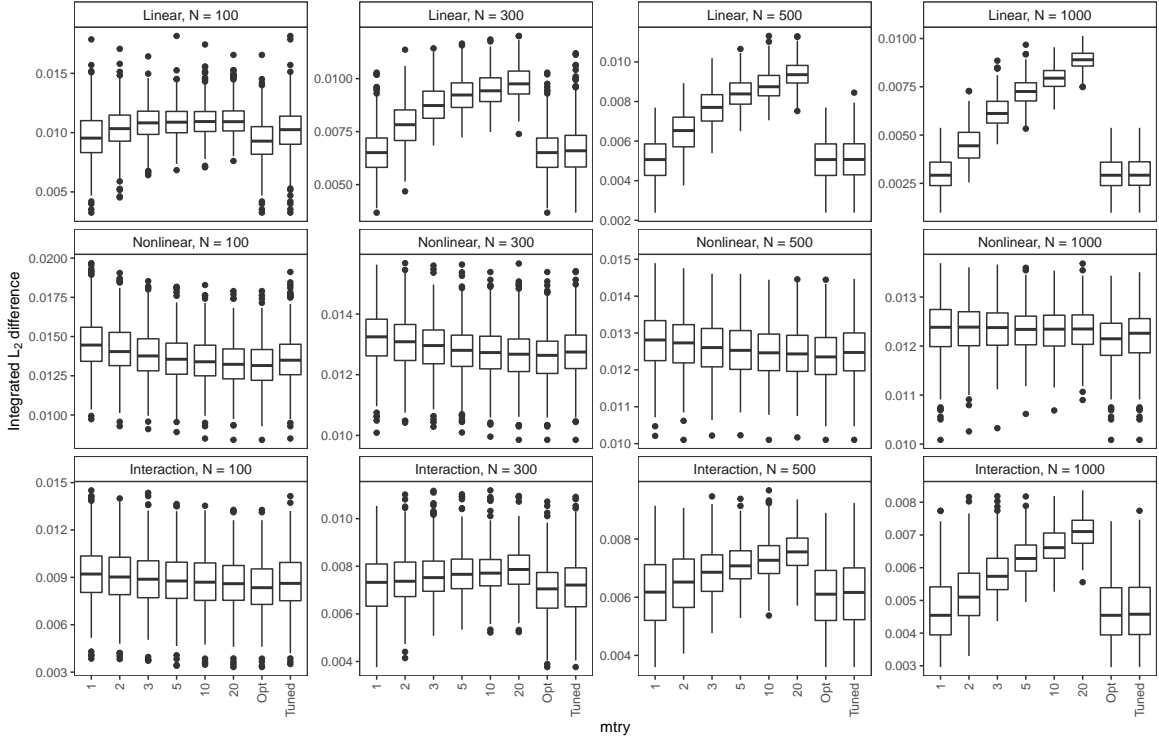

Figure S2.2: Integrated $L_2$ difference of LTRC CIF with different *mtry* values under the non-PH setting. Datasets are generated with time-invariant covariates, light right-censoring rate (20%), left-truncated and right-censored survival times following a Weibull-Increasing distribution. From the top row to the bottom, are given results for the linear, nonlinear and interaction survival relationship. From the first column to the last, are given results for the number of subjects $N = 100, 300, 500, 1000$. In each plot, 1–LTRC CIF with *mtry* = 1; 2–LTRC CIF with *mtry* = 2; 3–LTRC CIF with *mtry* = 3; 5–LTRC CIF with *mtry* = 5; 10–LTRC CIF with *mtry* = 10; 20–LTRC CIF with *mtry* = 20; Opt–LTRC CIF with value of *mtry* that gives the smallest Integrated $L_2$ difference in each round; Tuned–LTRC CIF with the value of *mtry* tuned by the "out-of-bag" tuning procedure. The default value in LTRC CIF is *mtry* = 5.

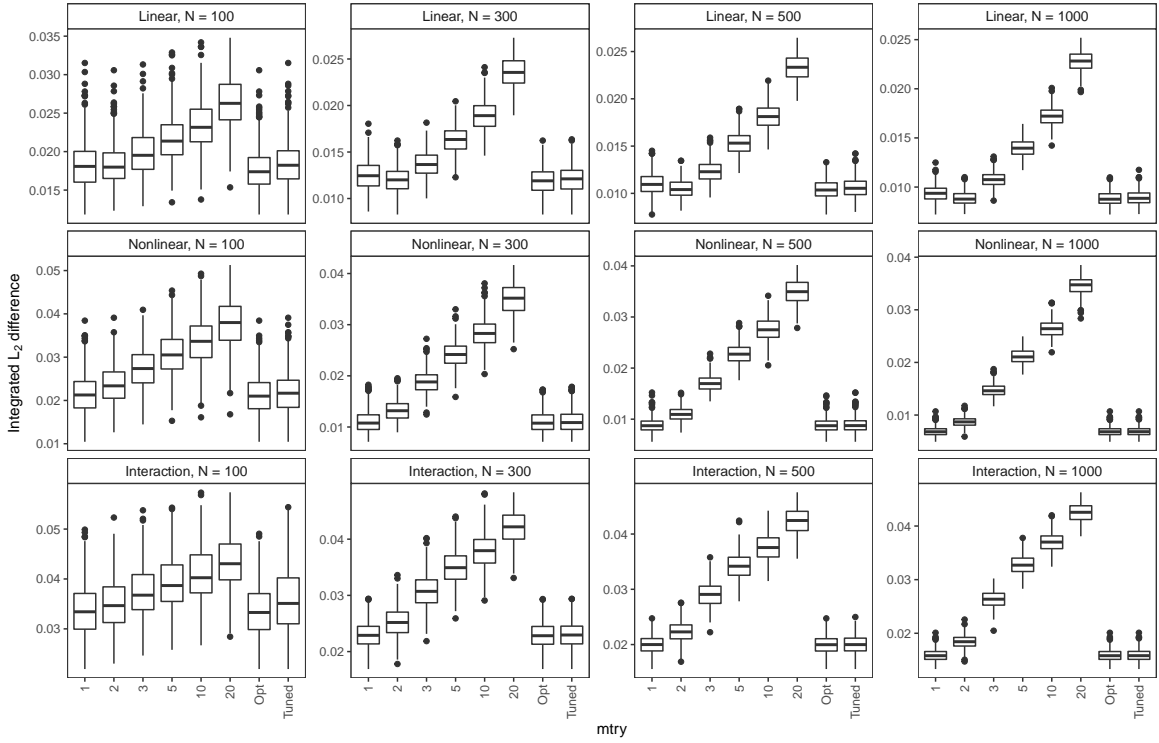

Figure S2.3: Integrated $L_2$ difference of LTRC RRF with different *mtry* values under the PH setting. Datasets are generated with time-invariant covariates, light right-censoring rate (20%), left-truncated and right-censored survival times following a Weibull-Increasing distribution. From the top row to the bottom, are given results for the linear, nonlinear and interaction survival relationship. From the first column to the last, are given results for the number of subjects $N = 100, 300, 500, 1000$. In each plot, 1–LTRC RRF with *mtry* = 1; 2–LTRC RRF with *mtry* = 2; 3–LTRC RRF with *mtry* = 3; 5–LTRC RRF with *mtry* = 5; 10–LTRC RRF with *mtry* = 10; 20–LTRC RRF with *mtry* = 20; Opt–LTRC RRF with value of *mtry* that gives the smallest Integrated $L_2$ difference in each round; Tuned–LTRC RRF with the value of *mtry* tuned by the "out-of-bag" tuning procedure. The default value in LTRC RRF is *mtry* = 5.

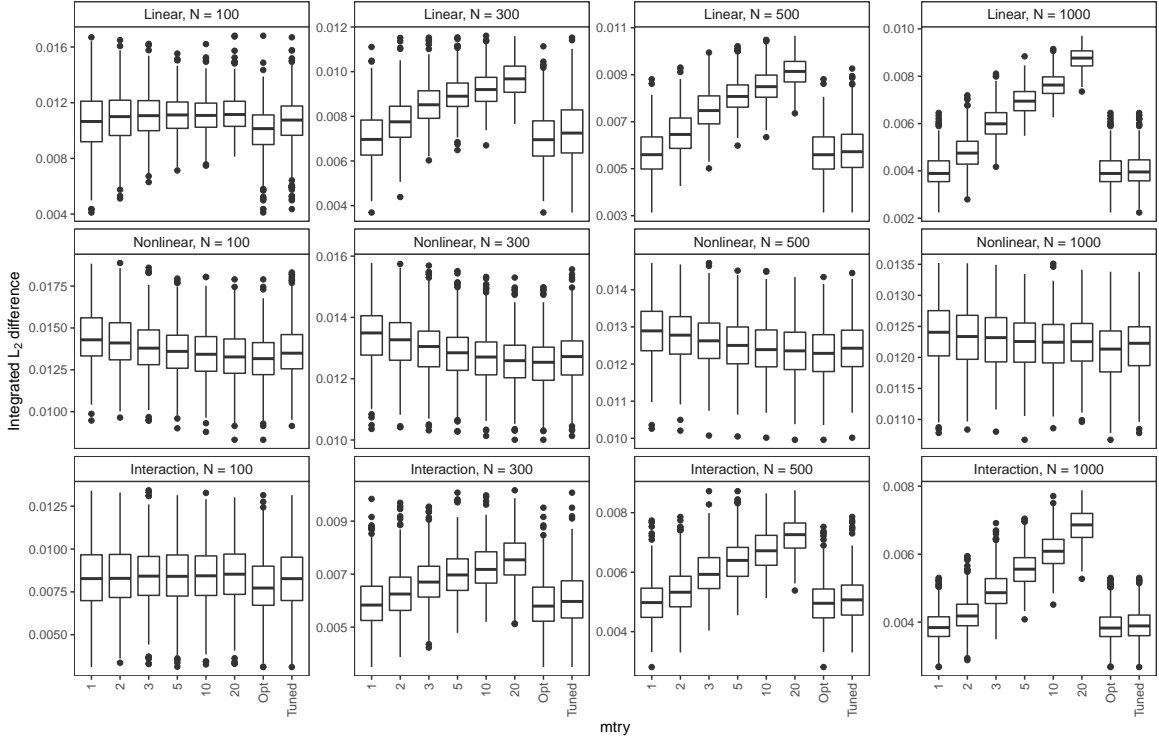

Figure S2.4: Integrated $L_2$ difference of LTRC RRF with different *mtry* values under the non-PH setting. Datasets are generated with time-invariant covariates, light right-censoring rate (20%), left-truncated and right-censored survival times following a Weibull-Increasing distribution. From the top row to the bottom, are given results for the linear, nonlinear and interaction survival relationship. From the first column to the last, are given results for the number of subjects $N = 100, 300, 500, 1000$. In each plot, 1–LTRC RRF with $mtry = 1$; 2–LTRC RRF with $mtry = 2$; 3–LTRC RRF with $mtry = 3$; 5–LTRC RRF with $mtry = 5$; 10–LTRC RRF with $mtry = 10$; 20–LTRC RRF with $mtry = 20$; Opt–LTRC RRF with value of *mtry* that gives the smallest Integrated $L_2$ difference in each round; Tuned–LTRC RRF with the value of *mtry* tuned by the "out-of-bag" tuning procedure. The default value in LTRC RRF is $mtry = 5$.

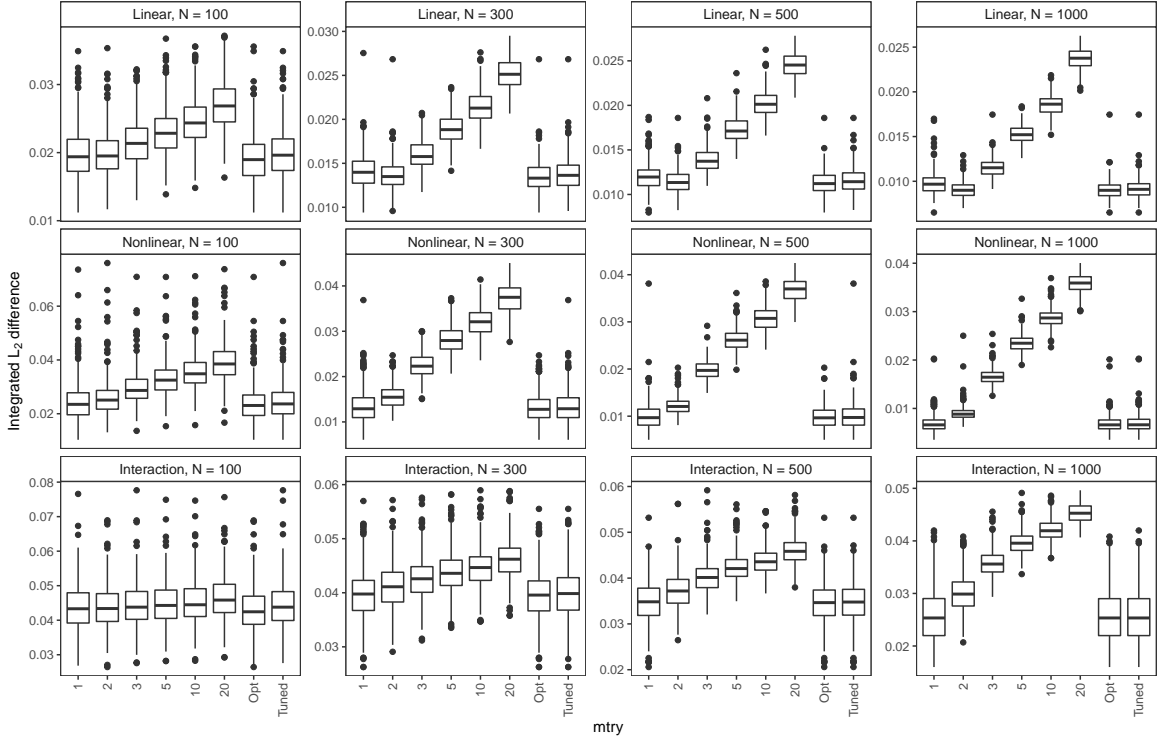

Figure S2.5: Integrated $L_2$ difference of LTRC TSF with different *mtry* values under the PH setting. Datasets are generated with time-invariant covariates, light right-censoring rate (20%), left-truncated and right-censored survival times following a Weibull-Increasing distribution. From the top row to the bottom, are given results for the linear, nonlinear and interaction survival relationship. From the first column to the last, are given results for the number of subjects $N = 100, 300, 500, 1000$. In each plot, 1–LTRC TSF with *mtry* = 1; 2–LTRC TSF with *mtry* = 2; 3–LTRC TSF with *mtry* = 3; 5–LTRC TSF with *mtry* = 5; 10–LTRC TSF with *mtry* = 10; 20–LTRC TSF with *mtry* = 20; Opt–LTRC TSF with value of *mtry* that gives the smallest Integrated $L_2$ difference in each round; Tuned–LTRC TSF with the value of *mtry* tuned by the "out-of-bag" tuning procedure. The default value in LTRC TSF is *mtry* = 5.

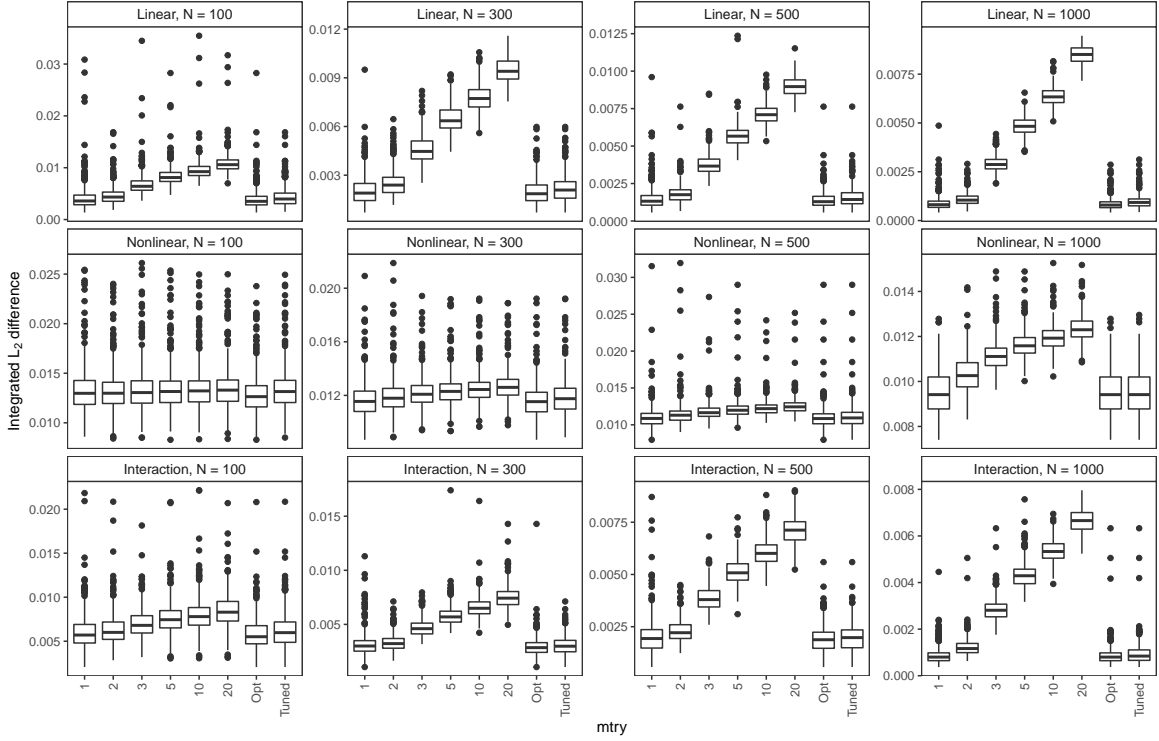

Figure S2.6: Integrated $L_2$ difference of LTRC TSF with different *mtry* values under the non-PH setting. Datasets are generated with time-invariant covariates, light right-censoring rate (20%), left-truncated and right-censored survival times following a Weibull-Increasing distribution. From the top row to the bottom, are given results for the linear, nonlinear and interaction survival relationship. From the first column to the last, are given results for the number of subjects $N = 100, 300, 500, 1000$. In each plot, 1–LTRC TSF with *mtry* = 1; 2–LTRC TSF with *mtry* = 2; 3–LTRC TSF with *mtry* = 3; 5–LTRC TSF with *mtry* = 5; 10–LTRC TSF with *mtry* = 10; 20–LTRC TSF with *mtry* = 20; Opt–LTRC TSF with value of *mtry* that gives the smallest Integrated $L_2$ difference in each round; Tuned–LTRC TSF with the value of *mtry* tuned by the "out-of-bag" tuning procedure. The default value in LTRC TSF is *mtry* = 5.

Table S2.1 gives performance comparison between each forest method with its default parameter settings and with the proposed parameter settings, using datasets generated with survival times following a Weibull-Increasing distribution, nonlinear relationship, with 20% right-censoring rate, as an example.

In Table S2.1, positive numbers indicate a decrease in integrated $L_2$ difference compared to a Cox model on the dataset, while negative numbers indicate an increase. The absolute value of the numbers represents the size of the difference between the integrated $L_2$ difference of the candidate and that of a Cox model. The two tables show that forests with the proposed parameter settings can provide improved performance over those with default parameter settings across almost all different numbers of subjects $N$ by a substantial amount, under both PH and non-PH settings.

## S2.3 Performance comparison and IBS-based CV model selection

Figures S2.7 and S2.8 give side-by-side integrated $L_2$ difference boxplots on datasets with light (right-)censoring rate, survival times generated following a Weibull-I distribution under the

PH setting and the non-PH setting, respectively. These figures compare the performance of the four methods, the best method and the method chosen by IBS-based 10-fold CV rule with $\tau_i = 1.5\widetilde{T}_i$.

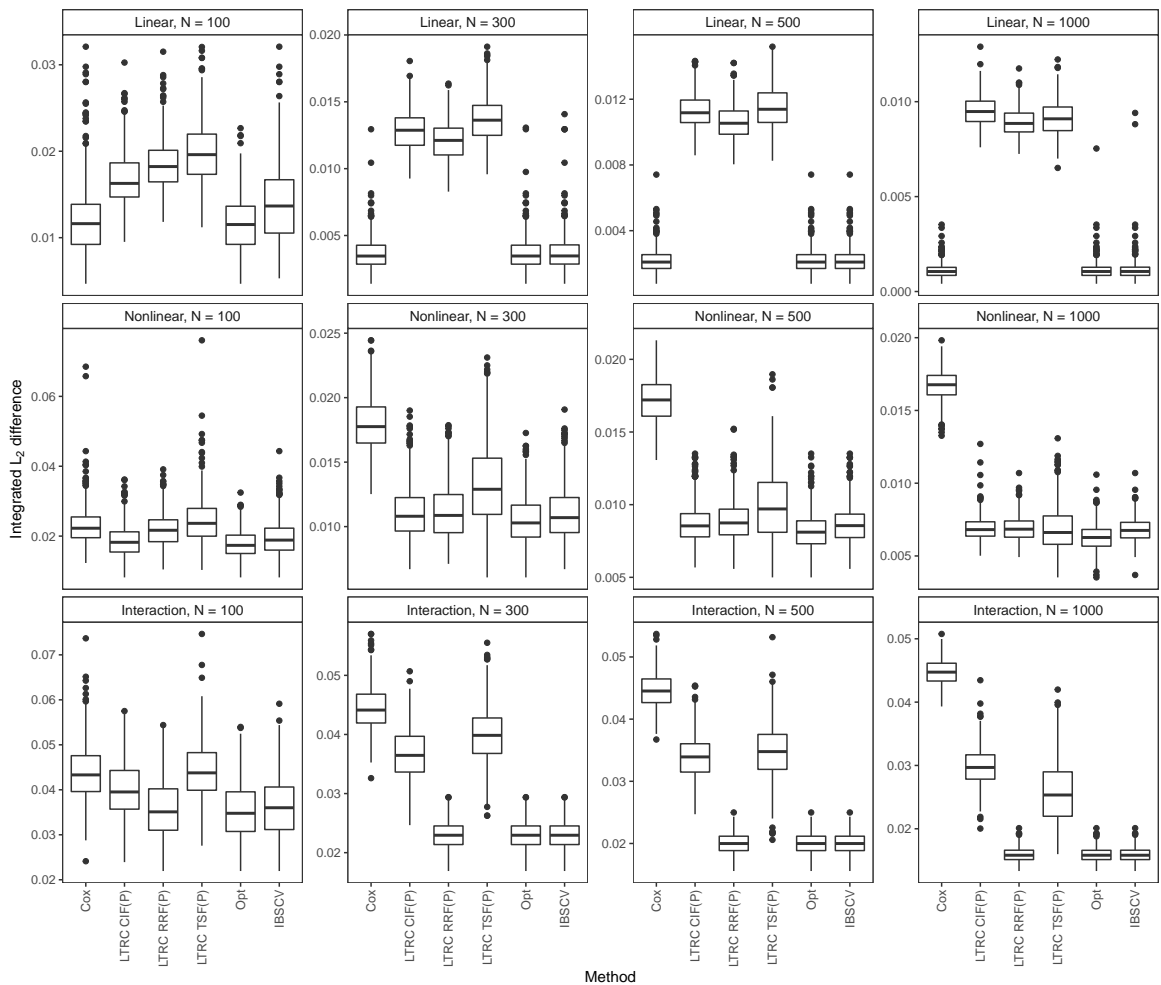

Figure S2.7: Boxplots of integrated $L_2$ difference for performance comparison across different survival relationships and different numbers of subjects $N$, under the PH setting. Datasets are generated with time-invariant covariates, left-truncated right-censored survival times following a Weibull-Increasing distribution. The first row shows results for the number of subjects $N = 100$, second row for $N = 300$, third row for $N = 500$, bottom row for $N = 1000$; the first column shows results for linear survival relationship, second column for nonlinear, the third column for interaction. In each plot, LTRC CIF(P)–LTRC CIF with proposed parameter settings; LTRC RRF(P)–LTRC RRF with proposed parameter settings; LTRC TSF(P)–LTRC TSF with proposed parameter settings; Opt–Best method; IBSCV–Method chosen by IBS-based 10-fold CV.

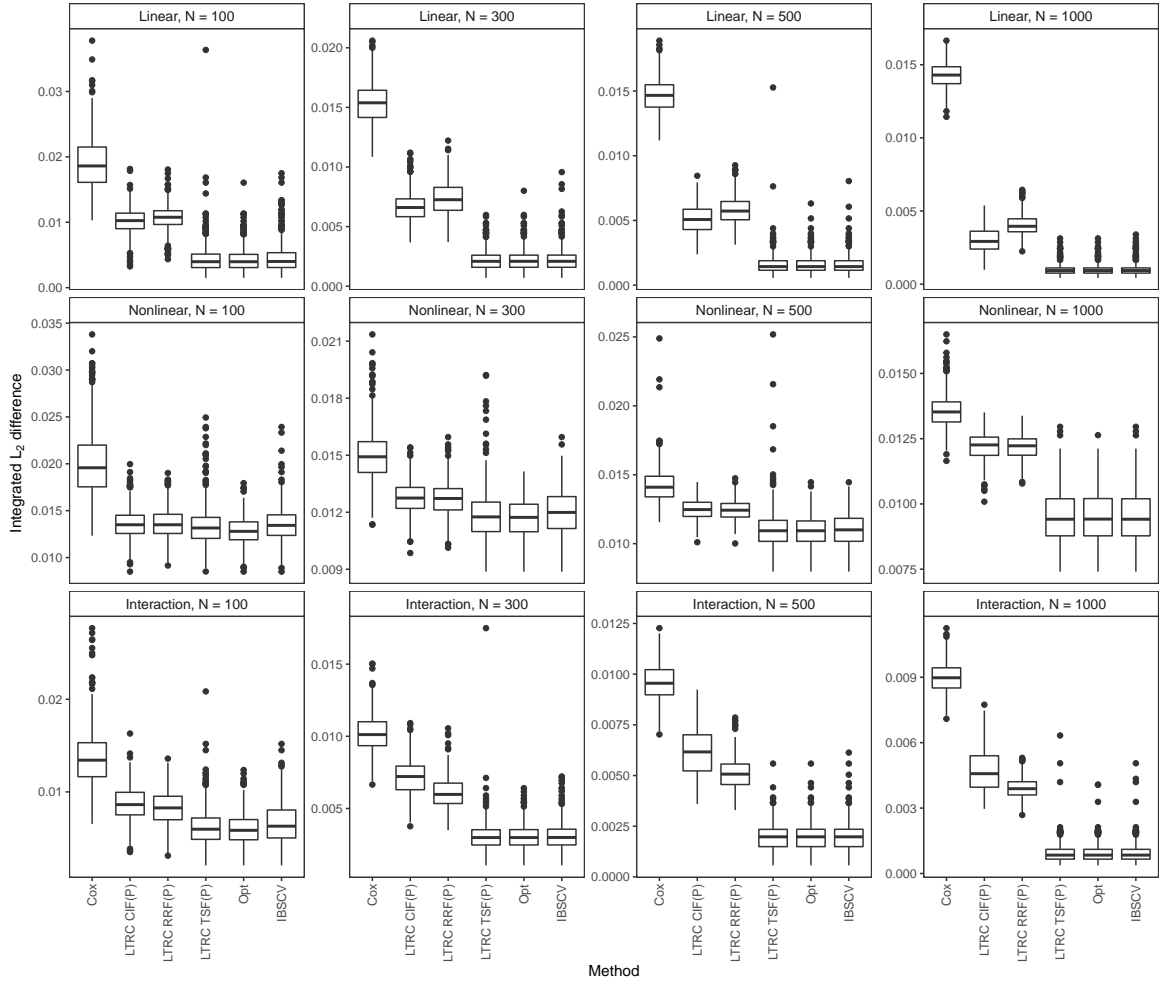

Figure S2.8: Boxplots of integrated $L_2$ difference for performance comparison across different survival relationships and different numbers of subjects $N$, under the non-PH setting. Datasets are generated with time-invariant covariates, left-truncated right-censored survival times following a Weibull-Increasing distribution. The first row shows results for the number of subjects $N = 100$, second row for $N = 300$, third row for $N = 500$, bottom row for $N = 1000$; the first column shows results for linear survival relationship, second column for nonlinear, the third column for interaction. In each plot, LTRC CIF(P)–LTRC CIF with proposed parameter settings; LTRC RRF(P)–LTRC RRF with proposed parameter settings; LTRC TSF(P)–LTRC TSF with proposed parameter settings; Opt–Best method; IBSCV–Method chosen by IBS-based 10-fold CV.

# References

Table S1.1: Comparison between forests with default (D) and proposed parameter settings (P) across different numbers of subjects $N$ given a nonlinear survival relationship. Given a method $A$, each cell value are given as mean $\pm$ one standard deviation of $(L_2(\text{KM}) - L_2(A))/L_2(\text{KM})$ based on all simulations. The mean value is the average % decrease in integrated $L_2$ difference compared to the Kaplan–Meier fit.

*Proportional hazards setting*

Case I. All changes in covariates' values are known

| $N$ | Extended Cox | CIF-TV(D) | CIF-TV(P) | RRF-TV(D) | RRF-TV(P) | TSF-TV(D) | TSF-TV(P) |
|---|---|---|---|---|---|---|---|
| 100 | $0.55 \pm 0.10$ | $0.53 \pm 0.14$ | $0.66 \pm 0.12$ | $0.52 \pm 0.14$ | $0.56 \pm 0.15$ | $0.41 \pm 0.17$ | $0.46 \pm 0.13$ |
| 300 | $0.65 \pm 0.04$ | $0.64 \pm 0.07$ | $0.82 \pm 0.04$ | $0.68 \pm 0.06$ | $0.79 \pm 0.05$ | $0.55 \pm 0.09$ | $0.74 \pm 0.06$ |
| 500 | $0.66 \pm 0.03$ | $0.68 \pm 0.05$ | $0.87 \pm 0.03$ | $0.72 \pm 0.05$ | $0.84 \pm 0.03$ | $0.61 \pm 0.07$ | $0.81 \pm 0.04$ |

Case II. Half of changes in covariates' values are unknown

| $N$ | Extended Cox | CIF-TV(D) | CIF-TV(P) | RRF-TV(D) | RRF-TV(P) | TSF-TV(D) | TSF-TV(P) |
|---|---|---|---|---|---|---|---|
| 100 | $0.52 \pm 0.10$ | $0.59 \pm 0.10$ | $0.65 \pm 0.10$ | $0.54 \pm 0.13$ | $0.57 \pm 0.13$ | $0.47 \pm 0.12$ | $0.45 \pm 0.12$ |
| 300 | $0.63 \pm 0.04$ | $0.71 \pm 0.05$ | $0.80 \pm 0.04$ | $0.71 \pm 0.05$ | $0.77 \pm 0.05$ | $0.66 \pm 0.06$ | $0.72 \pm 0.06$ |
| 500 | $0.64 \pm 0.03$ | $0.75 \pm 0.04$ | $0.84 \pm 0.03$ | $0.75 \pm 0.04$ | $0.81 \pm 0.03$ | $0.72 \pm 0.04$ | $0.78 \pm 0.04$ |

*Non-proportional hazards setting*

Case I. All changes in covariates' values are known

| $N$ | Extended Cox | CIF-TV(D) | CIF-TV(P) | RRF-TV(D) | RRF-TV(P) | TSF-TV(D) | TSF-TV(P) |
|---|---|---|---|---|---|---|---|
| 100 | $-0.42 \pm 0.22$ | $-0.33 \pm 0.30$ | $0.02 \pm 0.14$ | $-0.40 \pm 0.29$ | $0.02 \pm 0.15$ | $-0.20 \pm 0.27$ | $0.06 \pm 0.11$ |
| 300 | $-0.13 \pm 0.06$ | $-0.31 \pm 0.20$ | $0.08 \pm 0.07$ | $-0.35 \pm 0.20$ | $0.10 \pm 0.07$ | $-0.26 \pm 0.19$ | $0.11 \pm 0.06$ |
| 500 | $-0.07 \pm 0.04$ | $-0.25 \pm 0.16$ | $0.12 \pm 0.05$ | $-0.29 \pm 0.15$ | $0.12 \pm 0.07$ | $-0.18 \pm 0.17$ | $0.16 \pm 0.06$ |

Case II. Half of changes in covariates' values are unknown

| $N$ | Extended Cox | CIF-TV(D) | CIF-TV(P) | RRF-TV(D) | RRF-TV(P) | TSF-TV(D) | TSF-TV(P) |
|---|---|---|---|---|---|---|---|
| 100 | $-0.45 \pm 0.23$ | $-0.16 \pm 0.23$ | $0.02 \pm 0.12$ | $-0.22 \pm 0.23$ | $0.02 \pm 0.13$ | $-0.06 \pm 0.19$ | $0.03 \pm 0.10$ |
| 300 | $-0.13 \pm 0.07$ | $-0.15 \pm 0.14$ | $0.04 \pm 0.05$ | $-0.20 \pm 0.16$ | $0.07 \pm 0.05$ | $-0.08 \pm 0.14$ | $0.04 \pm 0.04$ |
| 500 | $-0.08 \pm 0.04$ | $-0.11 \pm 0.12$ | $0.05 \pm 0.03$ | $-0.15 \pm 0.12$ | $0.07 \pm 0.07$ | $-0.04 \pm 0.11$ | $0.05 \pm 0.03$ |

Table S1.2: Comparison between forests with default (D) and proposed parameter settings (P) across different numbers of subjects $N$ given an interaction survival relationship. Given a method $A$, each cell value are given as mean $\pm$ one standard deviation of $(L_2(\text{KM}) - L_2(A))/L_2(\text{KM})$ difference compared to the Kaplan-Meier fit.

*Proportional hazards setting*

Case I. All changes in covariates' values are known

| $N$ | Extended Cox | CIF-TV(D) | CIF-TV(P) | RRF-TV(D) | RRF-TV(P) | TSF-TV(D) | TSF-TV(P) |
|---|---|---|---|---|---|---|---|
| 100 | $-0.02 \pm 0.12$ | $0.21 \pm 0.19$ | $0.26 \pm 0.11$ | $0.29 \pm 0.18$ | $0.35 \pm 0.16$ | $0.15 \pm 0.20$ | $0.19 \pm 0.08$ |
| 300 | $0.05 \pm 0.05$ | $0.29 \pm 0.11$ | $0.41 \pm 0.11$ | $0.40 \pm 0.10$ | $0.61 \pm 0.07$ | $0.17 \pm 0.13$ | $0.23 \pm 0.06$ |
| 500 | $0.06 \pm 0.03$ | $0.33 \pm 0.09$ | $0.55 \pm 0.11$ | $0.43 \pm 0.07$ | $0.68 \pm 0.04$ | $0.22 \pm 0.11$ | $0.33 \pm 0.09$ |

Case II. Half of changes in covariates' values are unknown

| $N$ | Extended Cox | CIF-TV(D) | CIF-TV(P) | RRF-TV(D) | RRF-TV(P) | TSF-TV(D) | TSF-TV(P) |
|---|---|---|---|---|---|---|---|
| 100 | $-0.05 \pm 0.13$ | $0.24 \pm 0.14$ | $0.19 \pm 0.08$ | $0.24 \pm 0.17$ | $0.21 \pm 0.16$ | $0.19 \pm 0.12$ | $0.14 \pm 0.06$ |
| 300 | $0.02 \pm 0.05$ | $0.28 \pm 0.08$ | $0.16 \pm 0.05$ | $0.30 \pm 0.08$ | $0.24 \pm 0.10$ | $0.23 \pm 0.08$ | $0.12 \pm 0.03$ |
| 500 | $0.04 \pm 0.03$ | $0.30 \pm 0.06$ | $0.16 \pm 0.04$ | $0.33 \pm 0.06$ | $0.26 \pm 0.07$ | $0.24 \pm 0.06$ | $0.11 \pm 0.03$ |

*Non-proportional hazards setting*

Case I. All changes in covariates' values are known

| $N$ | Extended Cox | CIF-TV(D) | CIF-TV(P) | RRF-TV(D) | RRF-TV(P) | TSF-TV(D) | TSF-TV(P) |
|---|---|---|---|---|---|---|---|
| 100 | $-0.34 \pm 0.21$ | $-0.27 \pm 0.31$ | $0.09 \pm 0.18$ | $-0.20 \pm 0.27$ | $0.09 \pm 0.14$ | $-0.11 \pm 0.30$ | $0.25 \pm 0.18$ |
| 300 | $-0.09 \pm 0.06$ | $-0.14 \pm 0.19$ | $0.32 \pm 0.12$ | $-0.06 \pm 0.18$ | $0.27 \pm 0.10$ | $-0.11 \pm 0.22$ | $0.55 \pm 0.11$ |
| 500 | $-0.05 \pm 0.03$ | $-0.09 \pm 0.15$ | $0.44 \pm 0.11$ | $0.02 \pm 0.13$ | $0.38 \pm 0.11$ | $0.00 \pm 0.20$ | $0.70 \pm 0.09$ |

Case II. Half of changes in covariates' values are unknown

| $N$ | Extended Cox | CIF-TV(D) | CIF-TV(P) | RRF-TV(D) | RRF-TV(P) | TSF-TV(D) | TSF-TV(P) |
|---|---|---|---|---|---|---|---|
| 100 | $-0.35 \pm 0.22$ | $-0.07 \pm 0.22$ | $0.05 \pm 0.13$ | $-0.08 \pm 0.22$ | $0.07 \pm 0.14$ | $0.05 \pm 0.22$ | $0.13 \pm 0.13$ |
| 300 | $-0.10 \pm 0.05$ | $0.01 \pm 0.14$ | $0.13 \pm 0.07$ | $0.02 \pm 0.14$ | $0.15 \pm 0.06$ | $0.10 \pm 0.16$ | $0.25 \pm 0.11$ |
| 500 | $-0.06 \pm 0.03$ | $0.06 \pm 0.12$ | $0.17 \pm 0.08$ | $0.06 \pm 0.11$ | $0.18 \pm 0.06$ | $0.17 \pm 0.14$ | $0.31 \pm 0.11$ |

Table S2.1: Comparison between forests with default (D) and proposed parameter settings (P) across different numbers of subjects $N$ for left-truncated right-censored survival data. Datasets are generated with time-invariant covariates, under the nonlinear survival relationship. Given a method $A$, each cell value are given as mean ± one standard deviation of $(L_2(\text{Cox}) - L_2(A))/L_2(\text{KM})$ based on all simulations. The mean value is the average % decrease in integrated $L_2$ difference compared to the Cox model.

*Proportional hazards setting*

| | LTRC CIF(D) | LTRC CIF(P) | LTRC RRF(D) | LTRC RRF(P) | LTRC TSF(D) | LTRC TSF(P) |
|---|---|---|---|---|---|---|
| $N = 100$ | $-0.10 \pm 0.22$ | $0.16 \pm 0.22$ | $-0.22 \pm 0.25$ | $0.02 \pm 0.25$ | $-0.27 \pm 0.26$ | $-0.09 \pm 0.29$ |
| $N = 300$ | $0.07 \pm 0.12$ | $0.37 \pm 0.13$ | $0.02 \pm 0.12$ | $0.37 \pm 0.14$ | $-0.00 \pm 0.13$ | $0.25 \pm 0.18$ |
| $N = 500$ | $0.19 \pm 0.09$ | $0.49 \pm 0.09$ | $0.16 \pm 0.09$ | $0.48 \pm 0.10$ | $0.17 \pm 0.09$ | $0.42 \pm 0.15$ |
| $N = 1000$ | $0.34 \pm 0.06$ | $0.59 \pm 0.06$ | $0.33 \pm 0.06$ | $0.59 \pm 0.06$ | $0.37 \pm 0.09$ | $0.58 \pm 0.10$ |

*Non-proportional hazards setting*

| | LTRC CIF(D) | LTRC CIF(P) | LTRC RRF(D) | LTRC RRF(P) | LTRC TSF(D) | LTRC TSF(P) |
|---|---|---|---|---|---|---|
| $N = 100$ | $0.29 \pm 0.10$ | $0.31 \pm 0.10$ | $0.30 \pm 0.10$ | $0.31 \pm 0.10$ | $0.33 \pm 0.09$ | $0.33 \pm 0.09$ |
| $N = 300$ | $0.13 \pm 0.07$ | $0.15 \pm 0.06$ | $0.13 \pm 0.07$ | $0.15 \pm 0.06$ | $0.23 \pm 0.06$ | $0.21 \pm 0.07$ |
| $N = 500$ | $0.10 \pm 0.07$ | $0.12 \pm 0.06$ | $0.10 \pm 0.07$ | $0.12 \pm 0.06$ | $0.22 \pm 0.09$ | $0.22 \pm 0.09$ |
| $N = 1000$ | $0.10 \pm 0.05$ | $0.10 \pm 0.04$ | $0.09 \pm 0.05$ | $0.10 \pm 0.04$ | $0.27 \pm 0.05$ | $0.30 \pm 0.07$ |


\end{document}